\documentclass{article}

\usepackage{microtype}
\usepackage{graphicx}
\usepackage{subcaption}
\usepackage{booktabs} 

\usepackage{url}
\usepackage{svg}
\usepackage{tabularx}
\usepackage{enumitem}
\usepackage{comment}
\usepackage{stfloats}
\usepackage{adjustbox}

\usepackage{hyperref}

\usepackage[preprint]{icml2026}

\usepackage{amsmath}
\usepackage{amssymb}
\usepackage{mathtools}
\usepackage{amsthm}

\theoremstyle{plain}

\theoremstyle{definition}

\theoremstyle{remark}

\usepackage[textsize=tiny]{todonotes}

\icmltitlerunning{LLM-Mediated Guidance of MARL Systems}

\begin{document}

\twocolumn[
  \icmltitle{LLM-Mediated Guidance of MARL Systems}

  \begin{icmlauthorlist}
    \icmlauthor{Philipp D. Siedler}{yyy}
    \icmlauthor{Ian Gemp}{xxx}
  \end{icmlauthorlist}

  \icmlaffiliation{yyy}{Aleph Alpha Research, Heidelberg, Germany}
  \icmlaffiliation{xxx}{Google DeepMind, London, United Kingdom}

  \icmlcorrespondingauthor{Philipp D. Siedler}{p.d.siedler@gmail.com}
  \icmlcorrespondingauthor{Ian Gemp}{imgemp@google.com}

  \icmlkeywords{MARL, LLMs, Reinforcement Learning, Multi-Agent Reinforcement Learning}

  \vskip 0.3in
]



\printAffiliationsAndNotice{}  

\begin{abstract}
In complex multi-agent environments, achieving efficient learning and effective coordination remains a major challenge for Multi-Agent Reinforcement Learning (MARL) systems. This work explores the integration of MARL with Large Language Model (LLM)-mediated guidance to steer agents toward improved collective behaviour. We investigate a centralized intervention framework in which an LLM temporarily overrides agent actions during training to provide sparse, high-level guidance. We evaluate two types of controllers: a Rule-Based (RB) Controller and a Natural Language (NL) Controller that uses a small (7B/8B) instruction-tuned LLM to simulate human-like guidance. Experiments in the Aerial Wildfire Suppression (AWS) environment show that both controllers outperform a no-intervention baseline, with RB interventions yielding the strongest gains. Notably, interventions applied during early training stages significantly accelerate learning and lead to higher final performance. Our results demonstrate that LLM-mediated interventions can improve coordination and training efficiency in challenging MARL settings without modifying reward functions or policies, highlighting their potential as a practical mechanism for human-inspired steering of multi-agent systems.
\end{abstract}

\section{Introduction} 

Cooperative MARL research has developed techniques to effectively optimize collective return in simulated environments \citep{rashid_monotonic_2020,yuan2023survey,albrecht2024multi}. This enables the deployment of multi-agent systems (MAS) that can efficiently solve complex problems, particularly those that factorize into parallel subtasks and/or occur in the physical world (e.g., robotics), where spatially distributed agents are beneficial \citep{calvaresi_real-time_2021}.
However, what if the reward function is misspecified? This can happen because the reward is difficult to define in a way that avoids reward hacking \citep{skalse_defining_2022}. Alternatively, what if the test time environment or system goals change slightly? We would like a user to be able to steer a MARL system towards more desirable behaviour (human-in-the-loop). While our experiments use an LLM to simulate human interventions, the Natural Language Controller interface is designed to be directly usable by human operators, allowing seamless replacement of the LLM with human intervention. These are all key challenges that arise in real-world domains. In addition, we do not want to assume the user is a MARL expert. Ideally, the user could steer the system in an intuitive and simple way. Therefore, we consider steering a MAS using natural language. The user issues high-level strategies that an LLM then translates into actions to communicate with the MAS.
While examples of humans intervening and controlling static programs/interfaces via LLMs are pervasive \citep{hong_metagpt_2023}, we know of fewer examples controlling single-agent learning systems and no examples controlling MA learning systems.

\begin{figure}[t!]
    \centering
    \includegraphics[width=\linewidth]{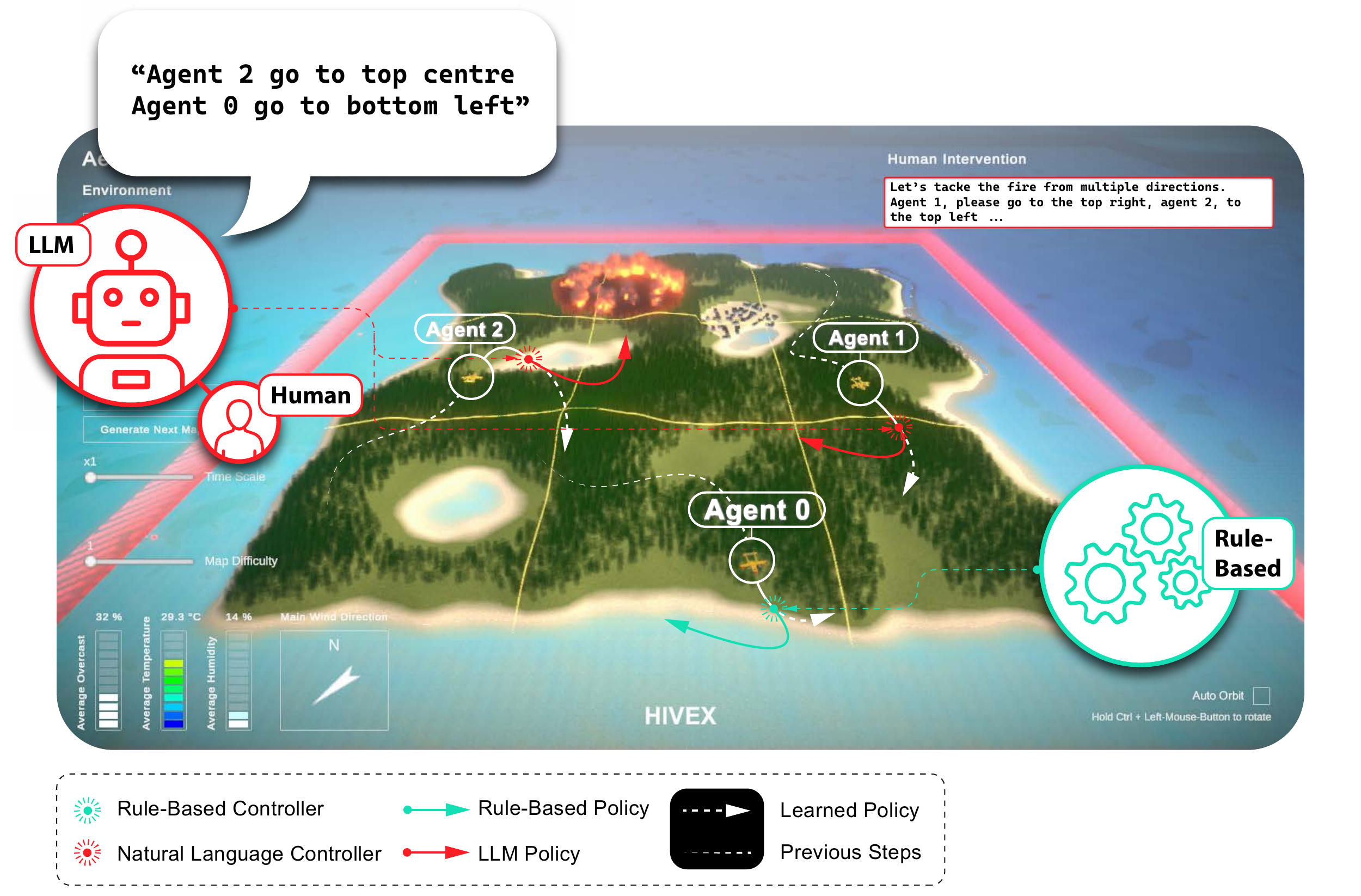}
    \caption{The Aerial Wildfire Suppression environment includes two types of controllers: Natural Language-based and Rule-Based. Controller interventions are passed to the LLM-Mediator, temporarily providing actions and overwriting the agents' learned policy actions.}
\end{figure}

Integrating LLMs with RL presents exciting opportunities for enhancing agent performance, particularly in complex MA environments. Instruction-aligned models with advanced reasoning and planning capabilities are well-suited for this task. When prompted correctly, they provide real-time, context-aware strategies that guide agents through challenges where traditional RL methods struggle. This is especially valuable in environments with large action and observation spaces or sparse rewards, particularly during early training. We envision a future where LLM-RL combinations can manage increasingly dynamic environments, with LLMs handling complex interactions and dynamically changing observation and action spaces. Our research explores this potential in MARL.
We allow users to quickly training a MARL system by guiding the agents using free-form natural language or rule-based interventions in the training process. This adaptation helps the system align more closely with the user's bespoke task requirements, ensuring that agents develop behaviours tailored to the challenges of the environment. We have specifically chosen the Aerial Wildfire Suppression (AWS) environment from the HIVEX suite \citep{siedler_hivex_2025} \footnote{Environment: \url{https://github.com/hivex-research/hivex-environments}\\Training Code: \url{https://github.com/hivex-research/llm_mediated_guidance}\\Results: \url{https://github.com/hivex-research/hivex-results}}, as it offers a relevant and intricate problem to solve.




The AWS environment presents a dynamic and high-stakes cooperative scenario, where the unpredictability of wildfire spread creates an evolving challenge. Factors such as wind direction, humidity, terrain slope, and temperature -- hidden from the agents -- add layers of complexity. Solving this environment requires seamless collaboration among agents, where strategic coordination is essential to containing fires. With AWS, users engage in a problem simulating real-world wildfire management. The combination of a physically and visually rich simulation, open-ended scenarios and environmental conditions makes AWS a demanding environment and a great challenge.

In this work, we test whether combining MARL and LLM techniques can allow users to steer and guide a MARL system towards more desirable behaviour in the challenging AWS environment. We consider two users: the simple Rule-Based (RB) Controller and a more sophisticated Natural Language (NL) Controller. The NL Controller simulates how humans might interact with the MAS, i.e., in free-form natural language. We compare these against our baseline, a setup with no interventions. Unlike prior work that uses LLMs for reward design, curriculum generation, or agent communication, our method introduces a centralized LLM-mediated intervention mechanism that temporarily overrides learned MARL policies without updating them directly, enabling sparse, human-like steering during training without modifying rewards or policies. We summarize our core contributions as follows:

\begin{itemize}[noitemsep, topsep=0pt]
    \item \textbf{Rule-Based and Natural Language Controller Generated Interventions}: We implement a novel system where rule-based and natural language-based interventions demonstrate the ability to enhance decision-making and coordination in dynamic settings like AWS.
    \item \textbf{Adaptive and Dynamic Guidance}: Our approach moves beyond static curriculum-based methods, providing real-time, adaptive interventions that respond to the evolving states of agents and environments, improving both long-term strategy and immediate decision-making.
    \item \textbf{AWS Environment}: We apply our method to the HIVEX AWS environment, simulating coordinated aerial wildfire suppression, showcasing the effectiveness of LLM-mediated interventions in managing complex and dynamic tasks in a MA environment.
    \item \textbf{Accelerated Learning and Improved Coordination}: Our results demonstrate that interventions, especially during early training, accelerate learning to reach expert-level performance more efficiently.
\end{itemize}

\section{Related Work} 


Integrating LLMs into RL has become pivotal for enhancing agent performance in complex environments. Advanced LLMs, specifically, their instruction fine-tuned versions, have demonstrated significant capabilities in providing high-level guidance, common-sense reasoning, and strategic planning, thereby possibly improving RL agents' adaptability and generalization \citep{bubeck_sparks_2023}. Recent works, such as those by \citet{wang_voyager_2023} and \citet{chiang_can_2023}, have shown that LLMs can assist RL agents by mediating natural language instructions and guiding behaviours, especially in environments where traditional reward signals are sparse or ineffective \citep{kajic_learning_2020}. However, these studies primarily focus on single-agent scenarios or environments with relatively straightforward dynamics. In contrast, our work emphasizes MA environments with complex, interdependent dynamics, demonstrating that LLM-driven interventions can significantly accelerate learning in such settings.

Historically, human-in-the-loop RL has relied on human feedback to guide the learning process \citep{kamalaruban_interactive_2019}. LLMs have emerged as scalable, real-time alternatives, providing domain-specific knowledge and policy suggestions to correct suboptimal behaviours \citep{chiang_lee_2023}. While prior work explores dynamic curriculum approaches, where models generate instructions based on agent progress \citep{narvekar_curriculum_2020}, our method instead uses LLMs for real-time interventions designed specifically to support coordination in multi-agent systems. This key distinction significantly impacts the effectiveness of the learning process in more complex environments.
LLMs also address challenges in long-term planning and common-sense reasoning \citep{hao_reasoning_2023} by offering early and intermediate guidance that traditional RL methods often lack. Previous studies in robotics have similarly leveraged LLMs as high-level strategic planners, enabling more effective decision-making in tasks that require long-term coordination and planning \citep{tang_saytap_2023, ahn_as_2022}. While these works illustrate the potential of LLMs in improving decision-making in tasks requiring extended sequences of actions, our work expands this concept by integrating LLM-driven interventions at critical points in the learning process, specifically in MA scenarios where coordinated action over long horizons is crucial.

In MA systems, LLMs show promise in improving coordination and strategic planning. Traditional MARL approaches, like MADDPG and QMIX, face limitations due to the complexity of joint action spaces and sparse rewards \citep{lowe_multi-agent_2017, rashid_qmix_2018}. Other work specifies a mediator to steer an MA system towards a desirable equilibrium without incorporating any LLM \citep{zhang_steering_2024}. While recent works, such as \citet{kwon_reward_2023}, have demonstrated that a global reward can control an MA system with a single intervention at the beginning -- showing how to cheaply design a reward model in natural language using an LLM -- these approaches do not fully address the dynamic nature of MA environments where frequent adaptations are necessary \citep{wang_large_2024}. Our research builds on these insights by demonstrating that periodic LLM interventions significantly enhance cooperation and learning efficiency, especially in dynamic and unpredictable environments such as AWS. This adaptive intervention strategy addresses the shortcomings of static coordination approaches by providing real-time guidance that aligns with the evolving state of the environment and agent interactions.

LLM interventions offer adaptive guidance that complements traditional policy shaping \citep{griffith_policy_2013}, evolving with the learning process. Our method does not fit neatly into open-loop or closed-loop categories \citep{sun_llm-based_2024}, as it temporarily replaces RL agent actions with LLM-guided interventions in both NL and RB setups. Unlike prior work using LLMs primarily for agent communication or capability building, our approach employs a centralized LLM to generate high-level strategies for coordinating multiple agents. This aligns with emerging research directions in language-enabled human-in/on-the-loop frameworks \citep{sun_llm-based_2024}, while focusing on direct intervention rather than agent-side reasoning.

While prior work leverages LLMs for curriculum generation \citep{sun_llm-based_2024}, reward design via language \citep{kwon_reward_2023}, or improving agent autonomy and exploration \citep{wang_voyager_2023}, these approaches do not intervene directly at the action level of trained MARL policies. In contrast, our method introduces a centralized LLM-mediated intervention mechanism that temporarily overrides learned agent actions during training, without modifying reward functions or policy parameters, enabling sparse, human-like steering in complex multi-agent environments.

\begin{figure*}[t]
    \centering
    \includegraphics[width=0.8\linewidth]{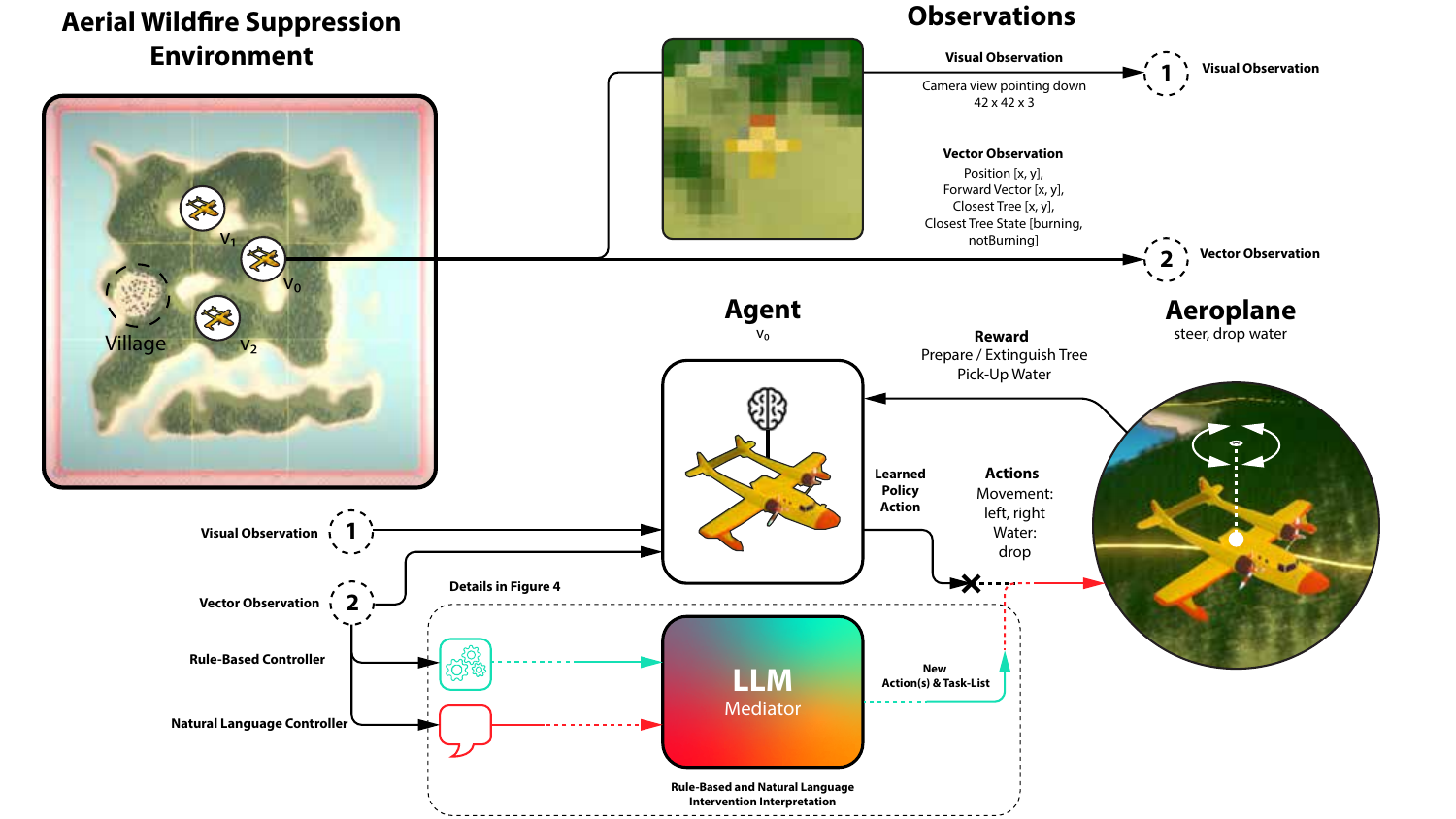}
    \caption{AWS Process Diagram: The default setup consists of three agents controlling individual aeroplanes. Each agent receives both feature vector and visual observations. Agents' actions include steering left, right, or releasing water. Rewards are given for extinguishing burning trees; smaller rewards are given for wetting living trees and picking up water. A negative reward is given for crossing the environment boundary. The LLM-Mediator interprets RB and NL Controller interventions, assigning tasks to any agent for the next 300 steps and overwriting its policy actions.}
    \label{fig:aerial_wildfire_suppression_process}
\end{figure*}

\section{The Aerial Wildfire Suppression Environment} 

\begin{figure}[H]
    \centering
    \includegraphics[width=\linewidth]{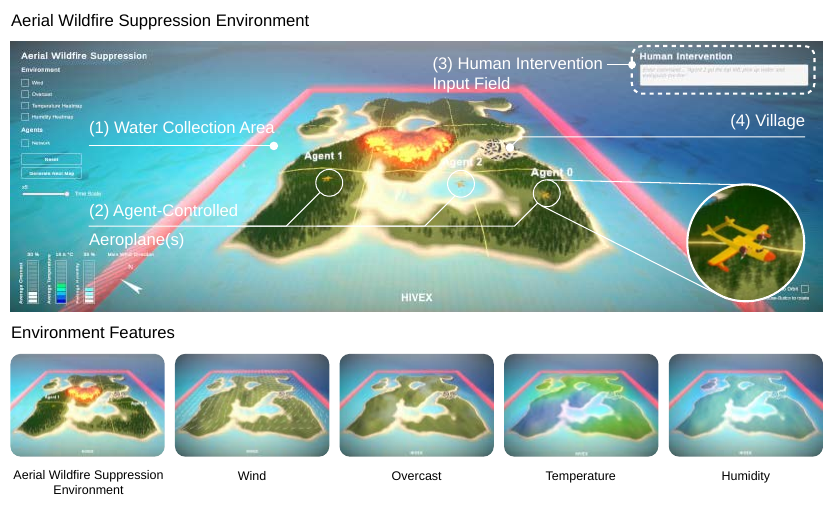}
    \vspace{-0.1cm}
    \caption{AWS Environment: (1) Water Collection Area, (2) Agent-controlled Wildfire Suppression Aeroplanes, (3) Human Natural Language Controller Input Field, (4) Village. Environment Features: Wind, overcast, temperature and humidity map sample.}
    \label{fig:aerial_wildfire_suppression_large}
\end{figure}

The Aerial Wildfire Suppression (AWS) environment presents a rich and challenging testbed for AI agents, substantially exceeding the simplicity of traditional grid-based worlds. Unlike grid environments with limited spatial structure, AWS features a continuous, three-dimensional, and dynamic landscape in which agents must adapt to stochastic and often unpredictable fire-spread dynamics. Built in Unity \citep{juliani_unity_2020}, the environment offers a semi-realistic visual presentation that contrasts with Atari-style benchmarks \citep{mnih_playing_2013}, resulting in a high-dimensional observation space composed of both vector-based features and raw visual inputs. This combination of visual complexity, real-time decision-making, and multi-agent coordination makes AWS a demanding platform for evaluating advanced AI methods in non-deterministic, safety-critical scenarios.

The environment simulates a large, bounded island on which agents are tasked with managing and mitigating an active wildfire. The primary objective is to minimize the total burning duration by extinguishing burning trees and proactively preparing surrounding forest areas to slow or redirect fire spread. Achieving this goal requires a combination of individual decision-making and coordinated teamwork across agents, as they must balance fire suppression, navigation, and resource collection while protecting key assets such as a nearby village.

AWS consists of three agents, each controlling an aerial firefighting aeroplane. Agents receive a feature vector observation space ($\mathbb{R}^8$) and a visual observation space consisting of a $42 \times 42$ RGB image captured by a downward-facing camera. Vector observations include the agent’s two-dimensional position and direction, an indicator of whether the agent is carrying water, the location of the nearest tree, and the burning state of that tree. Agents move at a constant velocity and can steer continuously while executing a discrete action to drop water when available. The operational area is bounded; crossing the boundary incurs a negative reward.

Water surrounds the island and can be collected by navigating toward it, yielding a positive reward. Additional positive rewards are granted for extinguishing burning trees, wetting unburned trees to reduce future fire spread, and fully suppressing the wildfire. Together, these reward signals encourage agents to balance immediate suppression with longer-term containment strategies. Detailed environment specifications are provided in Section~\ref{sec:env-specs}, and a full task list and reward definitions in the Appendix Section~\ref{a:reward_desc_calc}.

\section{Intervention Controllers and LLM-Mediator} 

\begin{figure}[H]
    \centering
    \includegraphics[width=\linewidth]{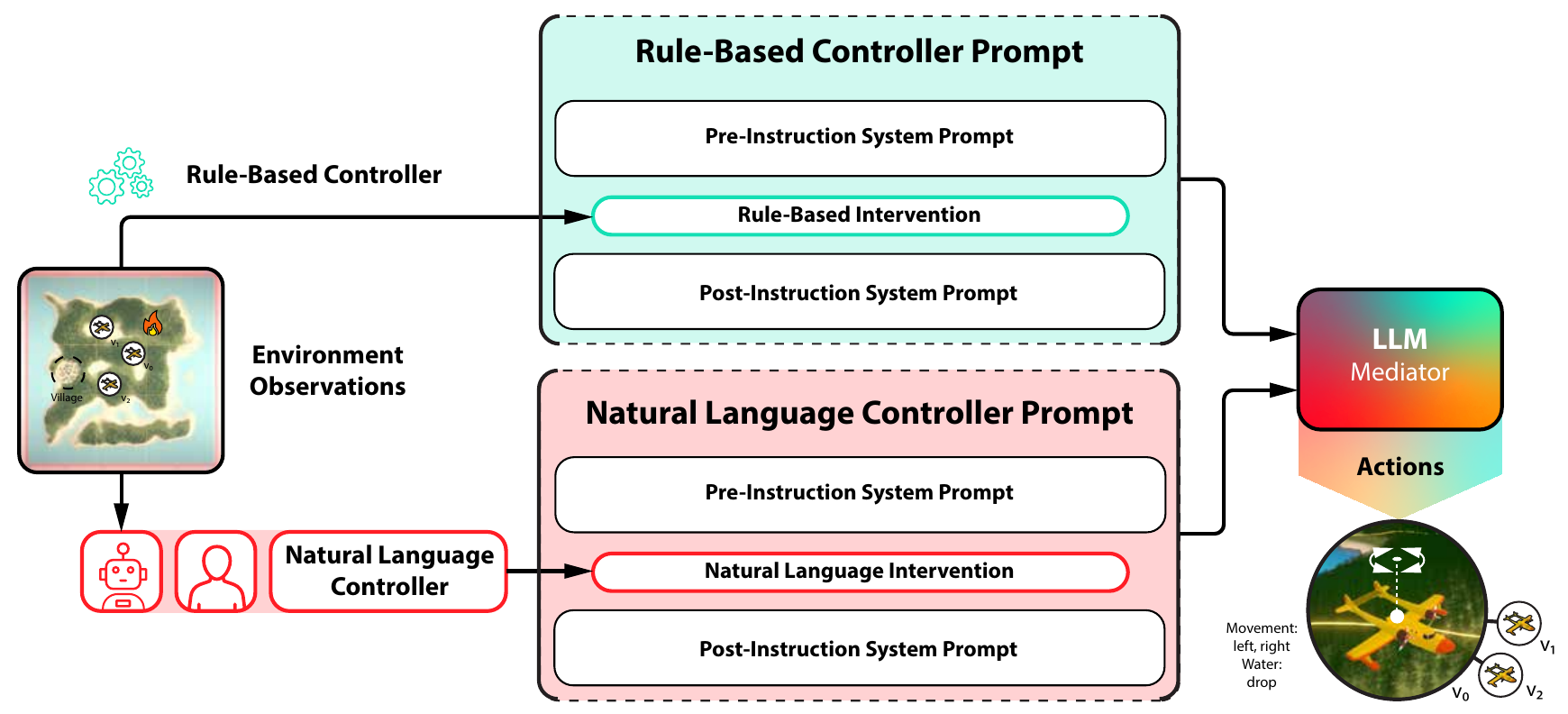}
    \caption{Overview of simplified RB and NL Controller intervention prompts sent to the LLM-Mediator, overwriting the agents' learned policy actions.} 
    \label{fig:aws_intervention_prompts}
\end{figure}

Our system supports interventions from two types of controllers: the Rule-Based (RB) and Natural Language (NL) Controller, which differ in their level of sophistication for generating interventions. The RB Controller uses predefined rules and a prompt template, producing rudimentary agent instructions. In contrast, the NL Controller communicates in free-form natural language, mimicking human behaviour. This allows it to generate more complex strategies and contextually relevant guidance. The LLM-Mediator processes both types of interventions, translating them and temporarily overwriting the agents' learned policy actions, guiding them to complete specific tasks (Figure \ref{fig:auto_aws_hi_intervention}). This framework enables adaptive guidance and control in dynamic environments (Figure \ref{fig:aws_intervention_prompts}). Importantly, LLM-issued actions are not treated as expert demonstrations, but rather as regular transitions, and policy updates remain fully on-policy PPO updates.

\subsection{Rule-Based (RB) Controller} 

\begin{figure}[h]
    \centering
    \includegraphics[width=\linewidth]{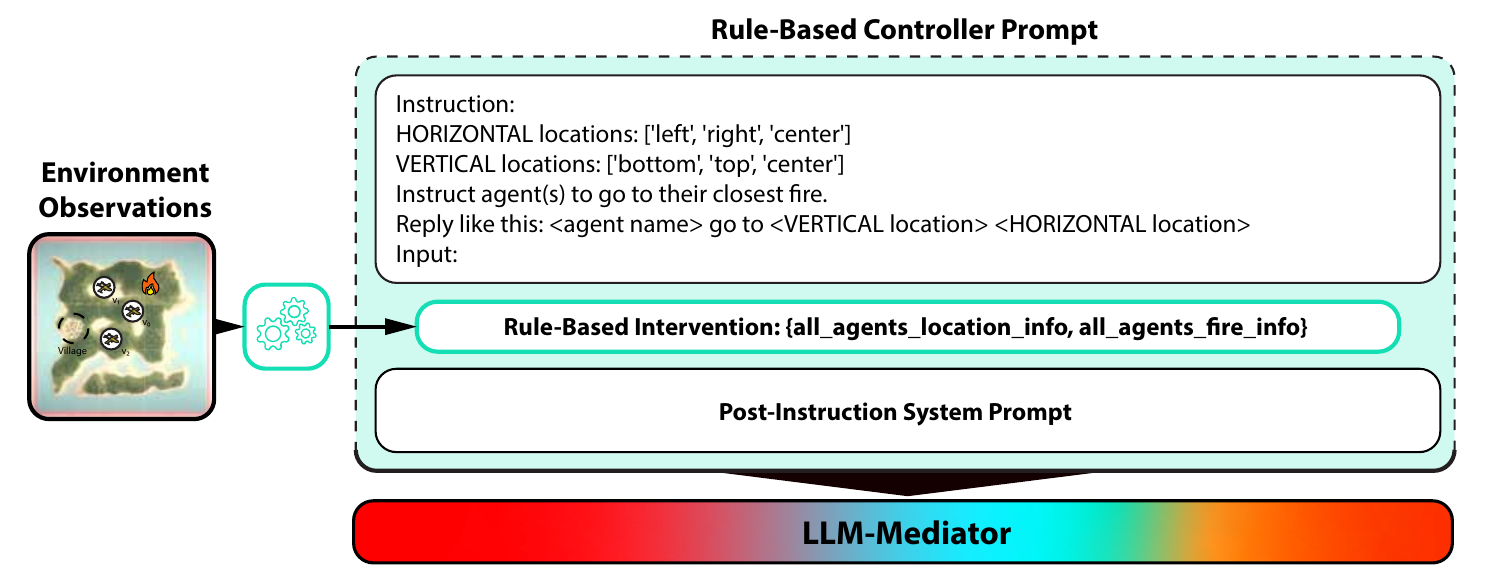}
    \caption{Abbreviated Rule-Based Controller intervention prompt template. A complete version can be found in the Appendix \ref{appendix:RB_prompt_template}.}
    \label{fig:auto_aws_hi_intervention}
\end{figure}

The RB Controller uses a prompt template that includes a subset of the agents' feature vector observations. This subset contains the agent's position and detected fire locations, which are preprocessed to natural language and integrated into the prompt template before being passed to the LLM-Mediator. The RB Controller's directive is to ``\emph{Instruct agent(s) to go to their closest fire}'', and so is considered a soft-coded intervention, as the agent and fire locations remain dynamic. Figure \ref{fig:auto_aws_hi_intervention} shows an abbreviation of the prompt template.

\subsection{Natural Language Controller} 

\begin{figure}[h]
    \centering
    \includegraphics[width=\linewidth]{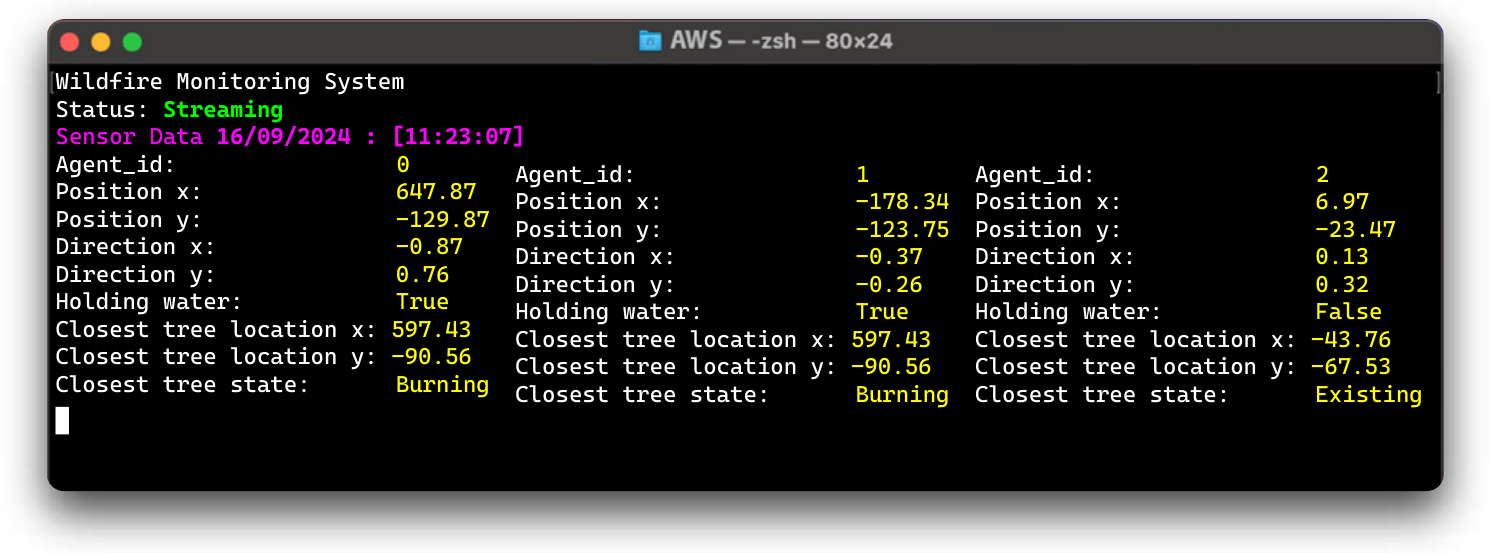}
    \vspace{-0.3cm}
    \caption{Possible AWS terminal as part of a fire-fighter dashboard. Info in this terminal is partially included in the NL strategy prompt template.}
    \label{fig:llm_aws_hi_terminal}
\end{figure}

\begin{figure}[h]
    \centering
    \includegraphics[width=0.8\linewidth]{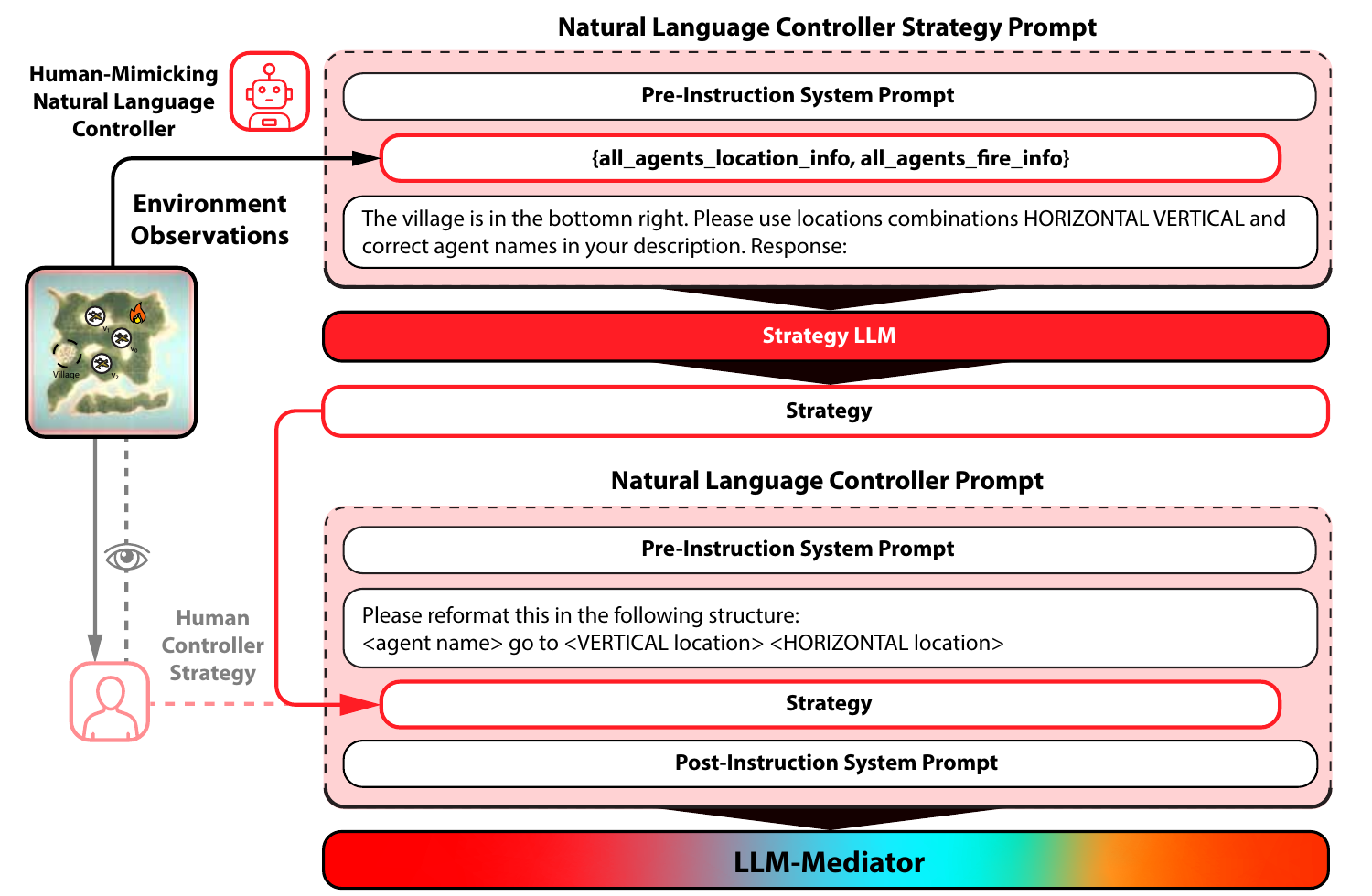}
    \caption{Abbreviated Natural Language Controller intervention prompts: 1. Human and Human-Mimicking LLM strategy prompt template generating strategies 2. A strategy as part of the prompt template is sent to the LLM-Mediator. A complete version can be found in the Appendix \ref{appendix:NL_prompt_template}.}
    \label{fig:llm_aws_hi_intervention}
\end{figure}

The NL Controller uses a prompt template with partial feature vector observation data (Figure \ref{fig:llm_aws_hi_terminal}). This information is provided as a list of all agents' observations and descriptions in natural language (Figure \ref{fig:llm_aws_hi_intervention}). The observation information formatted prompt is provided to an LLM, mimicking human behaviour, which generates a strategy directing agents to specific map locations. The NL Controller's high level directive is to ``\emph{Develop a strategy to extinguish all fires}''. The resulting strategy is then passed to the LLM-Mediator. Matching with the Rule-Based Controller, the LLM-Mediator processes this more sophisticated strategy and returns agent-readable actions. Although we use an LLM to generate natural language strategies in our experiments, the NL Controller mirrors a human-facing interface and can be directly operated by a human issuing free-form instructions.

\subsection{Mediator} 

\begin{figure}[h!]
    \centering
    \includegraphics[width=0.8\linewidth]{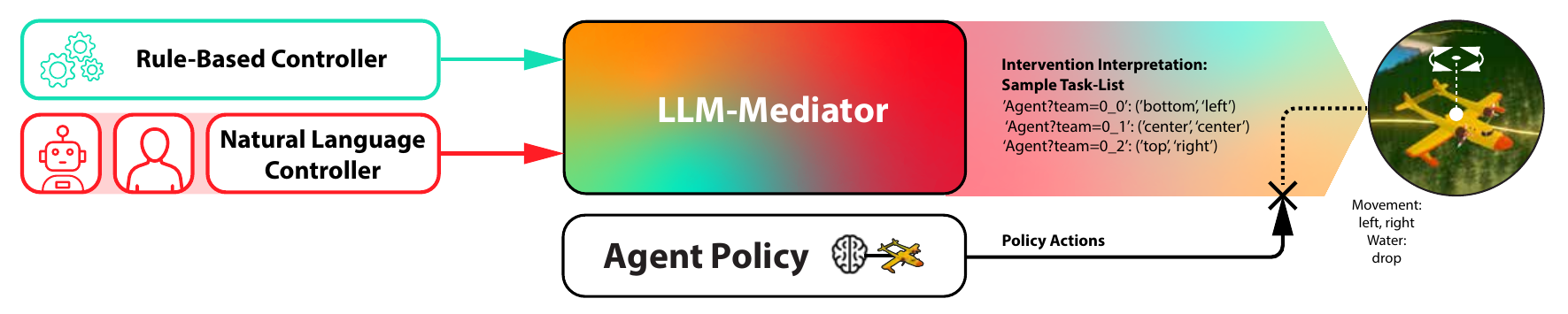}
    \caption{Rule-Based or Natural Language Controller interventions sent to LLM-Mediator, overwriting the agents' policy actions.}
    \label{fig:aws_hi_intervention}
\end{figure}

At the core, controllers act as prompt crafters. When a controller intervention prompt is issued, it is sent to the LLM-Mediator. Once the LLM-Mediator processes the intervention, a task list is generated for each agent, and a 200-time-step cooldown period begins. We empirically found this cooldown length to balance intervention effectiveness and policy autonomy; shorter cooldowns led to unstable behavior due to frequent overrides, while longer cooldowns reduced responsiveness to changing environmental conditions. During this period, agents are assigned their first task, and actions are generated to guide them toward task completion. These actions overwrite the agents’ policy actions, such as steering left or right. If the agent holds water during the intervention period, the LLM-Mediator ensures it is retained by default. Each task includes a key to identify the agent and specify a target location (Figure \ref{fig:aws_hi_intervention}). As long as the target location is not reached, actions continue to be auto-generated and issued to the agent. If the task is not completed within 200 time steps, a new intervention can be triggered. Figure \ref{fig:llm_aws_hi_terminal} illustrates a basic terminal interface, as we imagine a human controller or firefighter using it to review observations, in combination with camera feed and radar data, etc., to determine whether an intervention should be issued.

\subsection{Pseudocode: MARL with LLM Interventions}

\begin{algorithm}
    \caption{Multi-Agent PPO with LLM interventions and cooldown scheduling}
    \label{alg:marl_llm_mediation}
    \small
    \begin{algorithmic}
        \STATE Environment $\text{env}$; PPO policy $\pi_\theta$; LLM mediator $M$; intervention frequency $f$; agents $i=1..N$
        \STATE Initialize policy parameters $\theta \leftarrow \theta_0$
        \FOR{each training episode}
          \STATE Reset environment
          \STATE Initialize cooldown timers $c^i \leftarrow f$ for all agents
          \WHILE{episode not finished}
            \STATE Observe states $s_t^i$ and sample actions $a_t^i \sim \pi_\theta(\cdot \mid s_t^i)$
            \FOR{each agent $i$}
              \IF{cooldown expired ($c^i = f$)}
                \STATE Override action with LLM guidance: $a_t^i \leftarrow M(s_t^i)$
                \STATE Reset cooldown timer $c^i \leftarrow f$
              \ELSIF{agent is executing an LLM-issued task}
                \STATE Decrease cooldown timer $c^i \leftarrow c^i - 1$
                \IF{cooldown underflows}
                  \STATE Reset cooldown timer $c^i \leftarrow f$
                \ENDIF
              \ENDIF
            \ENDFOR
            \STATE Step environment with actions $\{a_t^i\}_{i=1}^N$ and observe rewards and next states
            \STATE Store transitions $(s_t^i, a_t^i, r_t^i, s_{t+1}^i)$
          \ENDWHILE
          \STATE Update PPO policy using shared experience buffer
        \ENDFOR
    \end{algorithmic}
\end{algorithm}

Our algorithm leverages a shared policy (\(\pi_\theta\)) for all agents, enabling simultaneous learning through centralized training. Experiences from all agents update the shared parameters. The LLM-Mediator selectively overrides agent actions based on cooldowns, while all collected experiences contribute to a single policy update, ensuring coordinated learning across agents. More details can be found in the code provided as well as the pseudocode in Algorithm \ref{alg:marl_llm_mediation}.

\section{Experiments} 

To evaluate the effectiveness of RB and NL Controller interventions in our MARL framework, we conducted experiments within a custom AWS environment, part of the HIVEX suite. The experiments were designed to compare agents' performance under three different intervention setups: No Controller, RB and NL Controller. For LLMs, we used Pharia-1-LLM-7B-control-aligned \citep{alephalpha_introducing_2024} or Llama-3.1-8B Instruct \citep{meta_llama_2023}. Experiments assess how well intervention and non-intervention-supported agents can learn and perform. All experiment setups utilize Proximal Policy Optimization (PPO) as the MARL algorithm \citep{schulman_proximal_2017} and are trained on $3 \cdot 10^5$ time-steps.
We use the default task ($0$) and terrain elevation level ($1$) of the AWS environment, but re-shaped rewards to focus on maximizing extinguishing tree rewards. We re-shaped the pick-up water reward from $1$ to $0.1$, the max preparing trees reward from $1$ to $0.1$ per tree, fire out reward from $10$ to $0$, too close to village reward from $-50$ to $0$, and the max extinguishing trees reward from $5$ to $1000$ per tree. Although the extinguishing reward is aggressively rescaled to emphasize the primary task objective, all controller and baseline experiments use identical reward functions, ensuring that observed performance differences are attributable to intervention strategies rather than reward design.

\section{Results} 

\begin{table}[H]
\caption{No controller, RB and NL Controller performance on \emph{Episode Reward Mean\textsuperscript{1}} and \emph{Extinguishing Trees Reward\textsuperscript{2}} for Llama-3.1-8B Instruct and Pharia-1-LLM-control-aligned. \emph{Average Wall-Time} per training run is in hour(s)\textsuperscript{3}.}
\label{table:interpretation_performance}
\centering
\small
\setlength{\tabcolsep}{3pt}
\begin{adjustbox}{max width=\columnwidth}
\begin{tabular}{@{}llllll@{}}
\multicolumn{1}{c}{Mediator} &
\multicolumn{1}{c}{Size} &
\multicolumn{1}{c}{Controller} &
\multicolumn{1}{c}{Episode R.\ Mean$^{1}$} &
\multicolumn{1}{c}{Ext.\ Trees R.$^{2}$} &
\multicolumn{1}{c}{Wall-Time$^{3}$} \\
\hline\hline
 &  & None               & 238.34 (±14.34)            & 1.18 (±0.16)              & \textbf{2.65} \\
Pharia-1-LLM & 7B & Rule-Based         & \textbf{437.65} (±43.28)   & 13.75 (±1.38)             & 2.96 \\
 &  & Natural Language   & 372.05 (±24.45)            & 5.89 (±0.79)              & 3.99 \\
Llama-3.1    & 8B & Rule-Based         & 376.18 (±21.98)            & \textbf{15.76} (±1.76)    & 3.13 \\
 &  & Natural Language   & 331.22 (±39.88)            & 6.73 (±0.81)              & 5.72 \\
\end{tabular}
\end{adjustbox}
\end{table}

\begin{figure}[h!]
    \centering
    \label{results:train_episode_mean_reward}
    \begin{minipage}{0.5\textwidth}
        \centering
        \includegraphics[width=\linewidth]{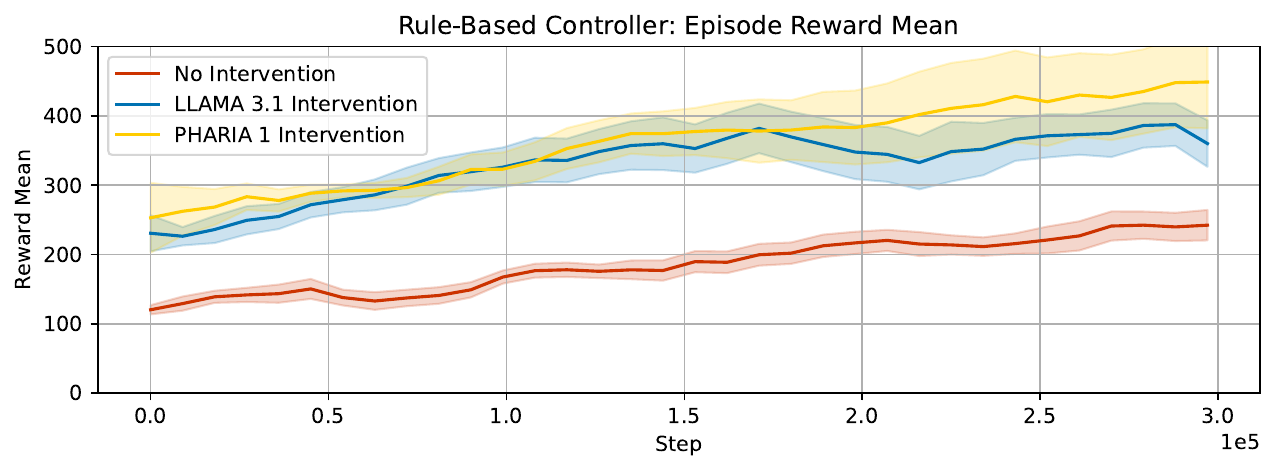}
    \end{minipage}\hfill
    \begin{minipage}{0.5\textwidth}
        \centering
        \includegraphics[width=\linewidth]{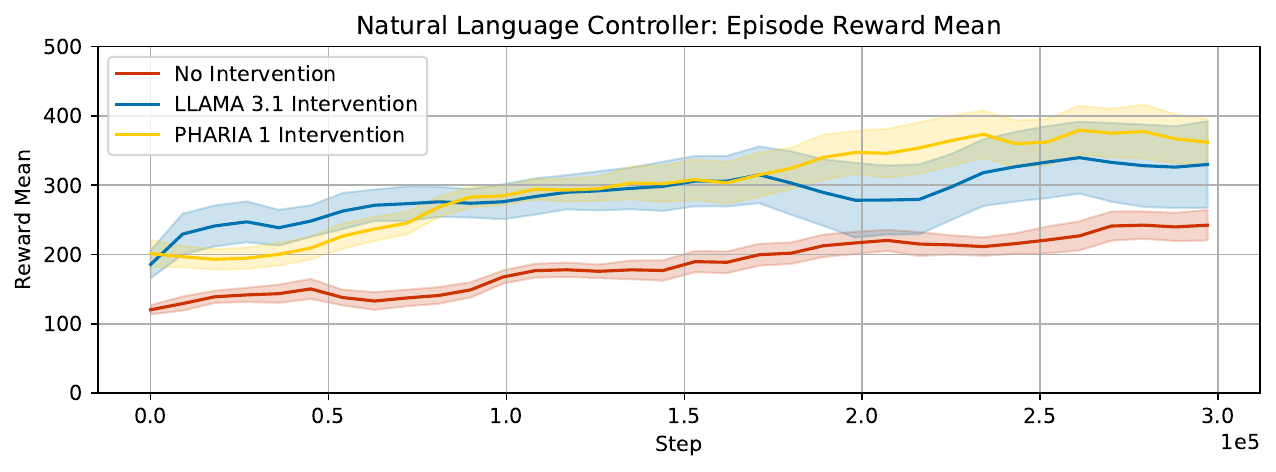}
    \end{minipage}
    \caption{Episode Reward Mean: Left: No controller baseline VS Rule-Based Controller with Llama-3.1-8B Instruct and Pharia-1-LLM-control-aligned-Mediator. Right:  No controller baseline VS Natural Language Controller with Llama-3.1-8B Instruct and Pharia-1-LLM-7B-control-aligned-Mediator.}
    \label{results:train_rb_nl_episode_mean_reward}
\end{figure}

\begin{figure}[h!]
    \centering
    \label{results:train_ext_tree}
    \begin{minipage}{0.5\textwidth}
        \centering
        \includegraphics[width=\linewidth]{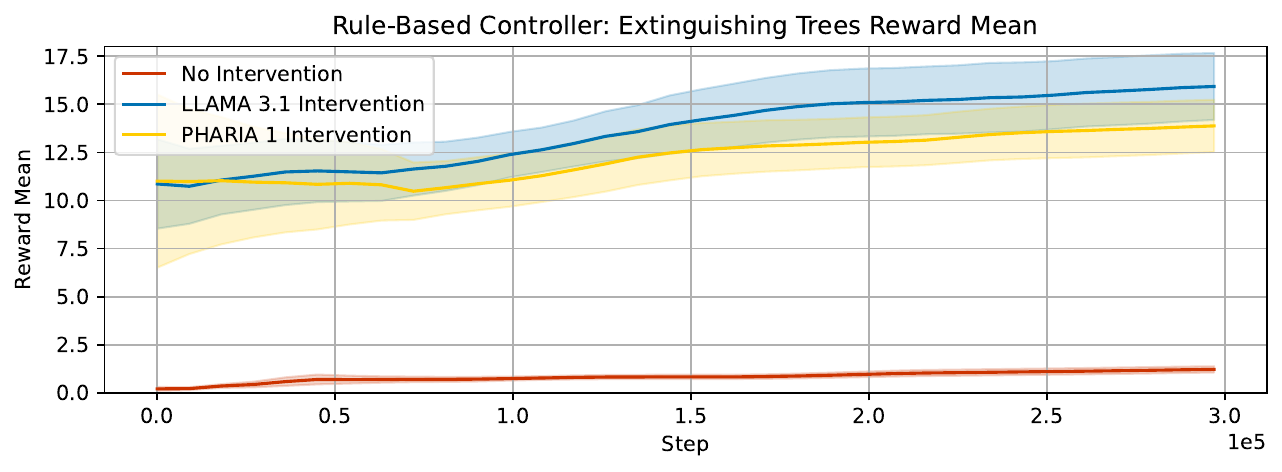}
    \end{minipage}\hfill
    \begin{minipage}{0.5\textwidth}
        \centering
        \includegraphics[width=\linewidth]{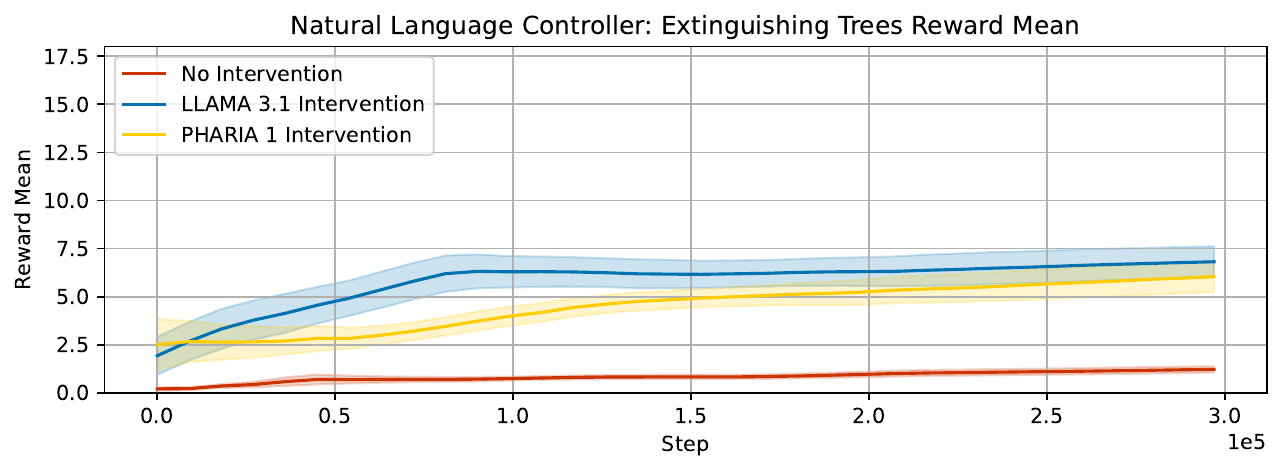}
    \end{minipage}
    \caption{Extinguishing Trees Reward Mean: Left: No controller baseline VS Rule-Based Controller with Llama-3.1-8B Instruct and Pharia-1-LLM-control-aligned-Mediator. Right:  No controller baseline VS Natural Language Controller with Llama-3.1-8B Instruct and Pharia-1-LLM-control-aligned-Mediator.}
    \label{results:train_rb_nl_ext_tree}
\end{figure}

Our results show that the RB and NL Controller interventions outperform the baseline without interventions, highlighting the potential of LLM-mediated guidance to accelerate training and enhance MARL performance in challenging environments. Generally, we can say that intervention is better than none, even with sparse supervision. In addition, both intervention controllers achieve a high-performance level and adapt to the demands of the new environment directive. Table \ref{table:interpretation_performance} shows performance on extinguishing trees reward and episode mean reward for three controller setups: None, RB and NL for Pharia-1-7B-control-aligned and LLama-3.1-8B Instruct. In Figure \ref{results:train_rb_nl_ext_tree}, we show mean \emph{Extinguishing Trees Reward} and in Figure \ref{results:train_rb_nl_episode_mean_reward} \emph{Episode Reward Mean} over $10$ trials for each controller experiment, RB and NL versus the baseline without interventions. Please see Appendix \ref{sec:additional_results} for additional results. All reported results are averaged over 10 independent training runs with different random seeds; observed trends were consistent across runs.

\begin{figure}[h!]
    \centering
    \begin{minipage}{0.5\textwidth}
        \centering
        \includegraphics[width=\linewidth]{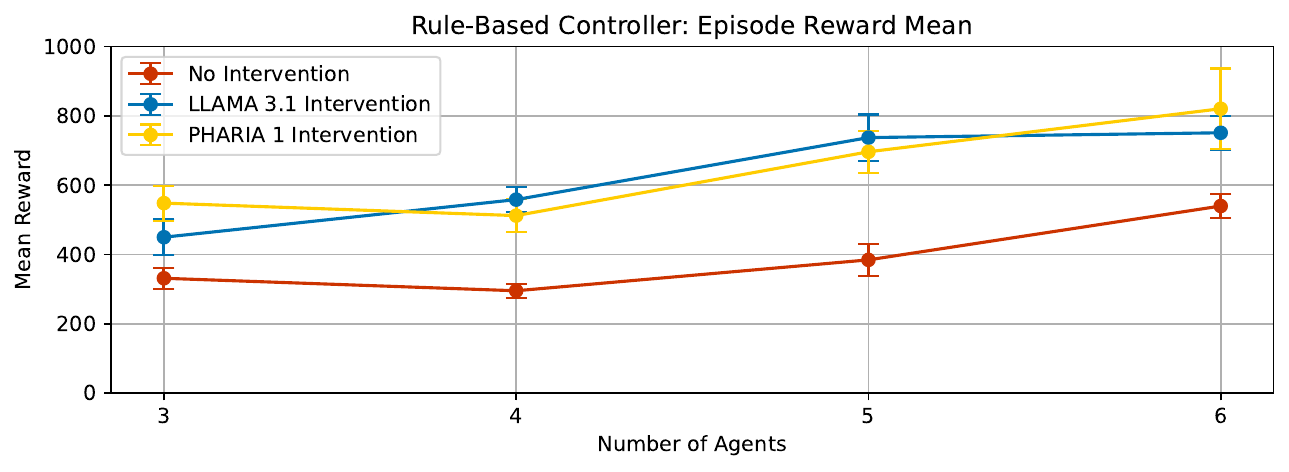}
    \end{minipage}\hfill
    \begin{minipage}{0.5\textwidth}
        \centering
        \includegraphics[width=\linewidth]{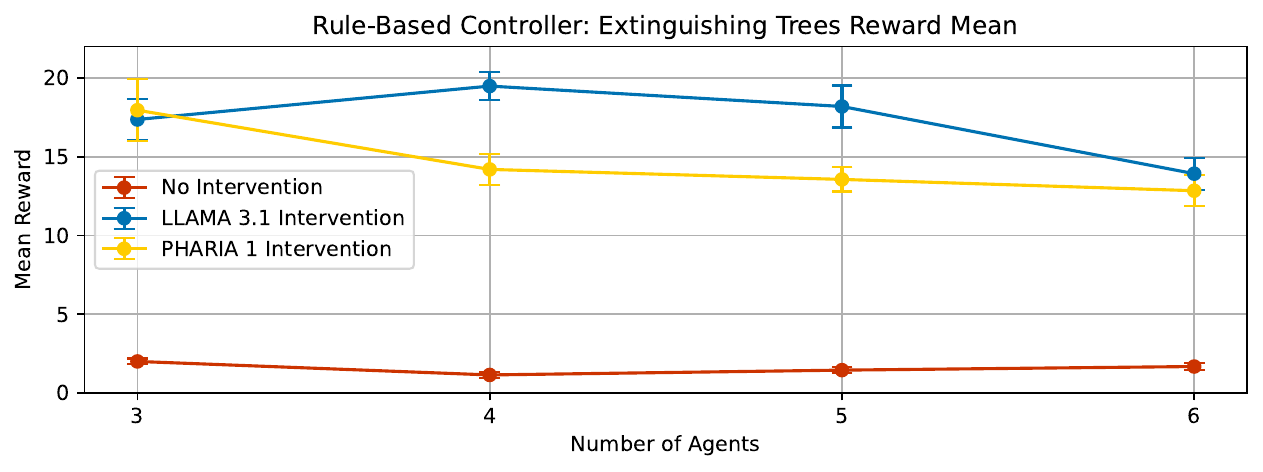}
    \end{minipage}
    \caption{Scalability Experiment with 3 (default), 4, 5 and 6 agents: No controller baseline VS Rule-Based Controller with Llama-3.1-8B Instruct and Pharia-1-LLM-control-aligned-Mediator: Episode Reward Mean (left), Extinguishing Trees Reward Mean (right).}
    \label{results:scalability}
\end{figure}

We also investigated the scalability of our method by extending the default three-agent setup to configurations with four, five, and six agents. Performance was compared between RB interventions and the no-intervention baseline using \emph{Episode Reward Mean} and \emph{Extinguishing Trees Reward Mean} for Pharia-1-7B-control-aligned and LLama-3.1-8B Instruct (Figure \ref{results:scalability}). As the number of agents increases, overall performance gradually decreases, indicating that coordination becomes more challenging in larger teams. However, this degradation is graceful, and LLM-mediated interventions consistently outperform the no-intervention baseline. Notably, the relative performance gap between intervention-based and non-intervention-based training widens with more agents, suggesting that centralized LLM guidance becomes increasingly beneficial as coordination complexity grows.

\section{Discussion}

The results of our experiments provide valuable insights into the effectiveness of LLM-based interventions in MARL. Our findings show that periodic interventions, mimicking human behaviour, can significantly enhance agents' performance in complex environments like AWS, where coordinated actions across multiple agents are crucial.

A key observation is the comparative advantage of NL Controller interventions over non-intervention baselines. Pharia-1-LLM-7B-control-aligned outperformed in the Rule-Based Environment Mean Rewards, while Llama-3.1-8B Instruct excelled in the Extinguishing Trees Reward category. This suggests that Pharia-1-LLM-7B-control-aligned handles structured interventions better, while Llama-3.1-8B Instruct is more adept at free-form natural language interventions. The $200$-step intervention cooldown allowed agents to consolidate learning by operating independently between interventions. The adaptability of LLMs in real-time, context-sensitive guidance is evident. However, each model excels in different dimensions. Both would benefit from memory of past tasks to refine strategies and enhance their adaptability in rapidly changing environments. The scalability experiments show that RB interventions consistently outperform the no-intervention baseline as agent numbers increase. Pharia-1 slightly outperforms LLama-3.1 in Episode Reward Mean, while both show a small decline in Extinguishing Trees Reward Mean with more agents, indicating coordination challenges in larger teams.

These findings suggest that LLM-based NL Controller interventions offer a promising approach for improving MARL systems. This is particularly relevant in settings where traditional RL methods face fundamental limitations. The distinct strengths of Pharia-1-LLM-7B-control-aligned and Llama-3.1-8B Instruct underscore the need for continued research to enhance LLM reasoning and planning capabilities. Further studies in more realistic environments are needed to validate these results across different domains.

\section{Limitations and Potential Impacts} 

While our research demonstrates the significant potential of integrating LLMs into MA systems, several limitations and considerations must be acknowledged, particularly concerning bias, safety, the realism of the environment, and the transferability of our findings to other domains.\\
\textbf{Realism of the Environment}: One limitation is the realism of the experimental environment. Although the AWS environment simulates real-world challenges, discrepancies remain between the simulation, actual wildfire scenarios, and the control mechanisms of autonomous aeroplanes. These differences may affect the generalizability of our findings, as agents trained in a simulated setting may underperform in real-world conditions. Moreover, fine-tuning the models using real-world data could be costly. Enhancing the simulation to mirror real-world conditions and incorporating additional realistic variables more closely would help mitigate this limitation.\\
\textbf{Transferability to Other Domains}: Our LLM-Mediator approach's success in the AWS environment context raises questions about its transferability to other domains. While the adaptive and context-sensitive nature of LLM-mimicked human interventions shows promise, different tasks and environments may require tailored adjustments to achieve similar levels of effectiveness. The complexity of the task, the nature of agent interactions, and the specific challenges of the domain in question all influence how well this approach can be applied elsewhere. Future research should explore the adaptability of intervention and LLM-driven mediation across various MARL applications to investigate its broader applicability.\\
\textbf{Bias and Safety Concerns}: A key limitation of using LLMs is the risk of bias in their human-mimicked interventions, stemming from the potentially biased datasets they are trained on. Such biases could result in suboptimal or harmful behaviours, particularly in critical tasks like wildfire suppression. Additionally, deploying LLMs in real-world environments raises safety concerns due to unpredictable outcomes. Rigorous testing and validation in controlled settings are essential to mitigate these risks.\\
\textbf{Resources and Inference Cost}: Another important consideration is the inference cost associated with the human LLM-mimicked interventions and LLM-Mediator. Out of the $3000$ total steps per agent per episode, the inference cost is only a fraction, as interventions are introduced every $200$ steps and typically influence agent behaviour for approximately $\sim120$ steps. This periodic intervention minimizes the computational overhead, allowing agents to continue operating efficiently under the learned policy for the remaining $\sim80$ steps. By balancing intervention frequency and task completion duration, we ensure that the computational load is manageable while still leveraging the benefits of real-time guidance from LLMs. Future work could further explore optimising this balance, reducing the task completion duration or intervention frequency while maintaining or improving agent performance. The training and testing of our experiment have been conducted on accessible, end-user hardware featuring an NVIDIA GeForce RTX $3090$ GPU, an AMD Ryzen $9$ $7950$X $16$-Core Processor, and $64$ GB of RAM. While these specifications align with high-end gaming laptops and desktop computers, the configuration could still be adapted to low-budget and non-GPU environments. This eliminates the need for specialized computational clusters, ensuring that researchers and practitioners with mid-range to high-end hardware can readily replicate our results using only consumer-grade equipment and an API for the LLM-Mediator.
\textbf{Potential Impacts}: Despite these limitations, the potential impacts of our research are substantial. By demonstrating the effectiveness of intervention and LLM-driven mediation in accelerating learning and improving coordination among agents, our approach offers a scalable solution for enhancing MARL systems in complex, dynamic environments. The findings suggest that human-like reasoning can lead to more efficient and effective learning processes, potentially reducing the computational resources required to train agents in complex environments. As these methods are refined and adapted to other domains, they could significantly advance the field of RL, contributing to more resilient and intelligent MA systems capable of tackling a wide range of real-world challenges.\\

\section{Conclusion} 

This paper demonstrates the potential of integrating LLMs into MARL environments, particularly in interpreting complex environmental observations and mediating real-time, context-sensitive interventions. Our experiments within the MA Aerial Wildfire Suppression environment part of the HIVEX suite show that periodic LLM guidance significantly improves agent performance, surpassing rule-based and non-guided baselines. Pharia-1-LLM-7B-control-aligned excelled in structured, rule-based tasks, while Llama-3.1-8B Instruct performed better in dynamic, situational challenges, highlighting the complementary strengths of different LLMs as mediators. This work underscores the scalability and efficiency of LLMs, particularly when mimicking human expertise, as a promising alternative to direct human guidance.\\
In conclusion, our findings suggest that LLMs and MARL techniques have matured to a point where they can effectively adapt systems to complex, dynamic environments -- an essential capability for tackling real-world challenges. The versatility of LLM-mediated interventions allows for easy adaptation to other domains, enabling efficient training of MARL systems for specific tasks. While fully automating curriculum design remains challenging, minimal real-time human supervision can provide cost-effective, sparse guidance, helping agents develop more efficient policies and address increasingly complex tasks.

\clearpage

\bibliography{references, additional_references}
\bibliographystyle{icml2026}

\newpage
\appendix
\onecolumn

\section{Appendix}

\subsection{Hyperparameters}

\subsubsection{No Intervention}

\begin{verbatim}
name: "NO_INTERVENTION"
env_parameters:
  training: 1
  human_intervention: 0
  task: 0
  ext_fire_reward: 1000
  prep_tree_reward: 0.1
  water_pickup_reward: 0.1
  fire_out_reward: 0
  crash_reward: -100
  fire_close_to_village_reward: 0
no_graphics: True
intervention_type: "none"
lr: 0.005
lambda_: 0.95
gamma: 0.99
sgd_minibatch_size: 900
train_batch_size: 9000
num_sgd_iter: 3
clip_param: 0.2
\end{verbatim}

\subsubsection{Rule-Based Llama-3.1-8B Instruct}

\begin{verbatim}
name: "RB_LLAMA_3.1"
env_parameters:
  training: 1
  human_intervention: 0
  task: 0
  ext_fire_reward: 1000
  prep_tree_reward: 0.1
  water_pickup_reward: 0.1
  fire_out_reward: 0
  crash_reward: -100
  fire_close_to_village_reward: 0
no_graphics: True
intervention_type: "auto"
model: "llama-3.1-8b-instruct"
shot: "few"
lr: 0.005
lambda_: 0.95
gamma: 0.99
sgd_minibatch_size: 900
train_batch_size: 9000
num_sgd_iter: 3
clip_param: 0.2
\end{verbatim}

\newpage

\subsubsection{Rule-Based Pharia-1-LLM-7B-control-aligned}

\begin{verbatim}
name: "RB_PHARIA_1"
env_parameters:
  training: 1
  human_intervention: 0
  task: 0
  ext_fire_reward: 1000
  prep_tree_reward: 0.1
  water_pickup_reward: 0.1
  fire_out_reward: 0
  crash_reward: -100
  fire_close_to_village_reward: 0
no_graphics: True
intervention_type: "auto"
model: "Pharia-1-LLM-7B-control-aligned"
shot: "few"
lr: 0.005
lambda_: 0.95
gamma: 0.99
sgd_minibatch_size: 900
train_batch_size: 9000
num_sgd_iter: 3
clip_param: 0.2
\end{verbatim}

\subsubsection{Natural Language Llama-3.1-8B Instruct}

\begin{verbatim}
name: "NL_LLAMA_3.1"
env_parameters:
  training: 1
  human_intervention: 0
  task: 0
  ext_fire_reward: 1000
  prep_tree_reward: 0.1
  water_pickup_reward: 0.1
  fire_out_reward: 0
  crash_reward: -100
  fire_close_to_village_reward: 0
no_graphics: True
intervention_type: "llm"
model: "llama-3.1-8b-instruct"
shot: few
lr: 0.005
lambda_: 0.95
gamma: 0.99
sgd_minibatch_size: 900
train_batch_size: 9000
num_sgd_iter: 3
clip_param: 0.2
\end{verbatim}

\newpage

\subsubsection{Natural Language Pharia-1-LLM-7B-control-aligned}

\begin{verbatim}
name: "NL_PHARIA_1"
env_parameters:
  training: 1
  human_intervention: 0
  task: 0
  ext_fire_reward: 1000
  prep_tree_reward: 0.1
  water_pickup_reward: 0.1
  fire_out_reward: 0
  crash_reward: -100
  fire_close_to_village_reward: 0
no_graphics: True
intervention_type: "llm"
model: "Pharia-1-LLM-7B-control-aligned"
shot: few
lr: 0.005
lambda_: 0.95
gamma: 0.99
sgd_minibatch_size: 900
train_batch_size: 9000
num_sgd_iter: 3
clip_param: 0.2
\end{verbatim}

\newpage

\subsection{Environment Specification}\label{sec:env-specs}

\begin{itemize}[noitemsep,nolistsep]
    \item Episode Length: $3000$
    \item Agent Count: $3$
    \item Neighbour Count: $0$
\end{itemize}

\textbf{Feature Vector Observations (8)} - Stacks: 1 - Normalized: True
    \begin{itemize}[noitemsep,nolistsep]
        \item Local Position (2): $\Vec{p}(x, y)$
        \item Direction (2): $\Vec{dir}(x, y)$
        \item Holding Water (1): $hw = [0, 1]$
        \item Closest Tree Location (2): $\Vec{ct}(x, y)$
        \item Closest Tree Burning (1): $ctb = [0, 1]$
    \end{itemize}

\textbf{Visual Observations (42, 42, 3)} - Stacks: 1 - Normalized: True
    \begin{itemize}[noitemsep,nolistsep]
        \item Downward Pointing Camera in RGB (1764): $[r, g, b] = [[0, 1], [0, 1], [0, 1]]$
    \end{itemize}

\textbf{Continous Actions (1)}:
    \begin{itemize}[noitemsep,nolistsep]
        \item Steer Left / Right (1): $[-1, 1]$
    \end{itemize}
\textbf{Discrete Actions (1)}:
    \begin{itemize}[noitemsep,nolistsep]
        \item Branch 0 - Drop Water (2): { 0: Do Nothing, 1: Drop Water }
    \end{itemize}

\newpage

\subsection{Un-shaped Reward Description and Calculation}\label{a:reward_desc_calc}

\textbf{Reward Description}

\begin{enumerate}[noitemsep,nolistsep]
    \item \textbf{Crossed Border} - This is a negative reward of $-100$ given when the border of the environment is crossed. The border is a square around the island in the size of $1500$ by $1500$. The island is $1200$ by $1200$.
    
    \item \textbf{Pick-up Water} - This is a positive reward of $1$ given when the agent steers the aeroplane towards the water. The island is $1200$ by $1200$ and there is a surrounding band of water around the island with a width of $300$.
    
    \item \textbf{Fire Out} - This is a positive reward of $10$ given when the fire on the whole island dies out, with or without the active assistance of the agent.
    
    \item \textbf{Too Close to Village} - This is a negative reward of $-50$ given when the fire is closer than $150$ to the centre of the village.
    
    \item \textbf{Time Step Burning} - This is a negative reward of $-0.01$ given at each time-step, while the fire is burning.
        
    \item \textbf{Extinguishing Tree} - This is a positive reward in the range of $[0, 5]$ given for each tree that has been in the state burning in time-step $t_{-1}$ and is now extinguished by dropping water at its location.
    
    \item \textbf{Preparing Tree} - This is a positive reward in the range of $[0, 1]$ given for each tree that has been in the state not burning in time-step $t_{-1}$ and is now wet by dropping water at its location.
\end{enumerate}
\vspace{1cm}
\textbf{Reward Calculation}

\textbf{1. Crossed Border} - To calculate the Crossed Border reward, let us define the following:
\begin{itemize}[noitemsep,nolistsep]
    \item $eh = 750$ --- The environment half extend.
    \item $\vec{p}$ --- The drone position.
    \item $r_{cb}$ --- Crossed boundary reward.
\end{itemize}

Calculation steps:

\begin{enumerate}[noitemsep,nolistsep]
    \item We can now calculate the Crossed Border reward:
    \begin{equation} \label{AWS:eq:1}
    r_{cb} =
    \begin{cases}
    -100 & \text{if } \left( p_x > eh \text{ or } p_x < -eh \text{ or } p_y > eh \text{ or } p_y < -eh \right) \\
    0 & \text{otherwise}
    \end{cases}
    \end{equation}
\end{enumerate}

\textbf{2. Pick-up Water} - To calculate the Pick-up Water reward, let us define the following:

\begin{itemize}[noitemsep,nolistsep]
    \item $eh = 750$ --- The environment half extend.
    \item $ih = 600$ --- Island half extend.
    \item $\vec{p}$ --- The drone position.
    \item $r_{pw}$ --- Pick-up Water reward.
\end{itemize}

Calculation steps:

\begin{enumerate}[noitemsep,nolistsep]
    \item We can now calculate the Pick-up Water reward:
    \begin{equation} \label{DBR:eq:14}
    r_{pw} = 
    \begin{cases}
    1 & \text{if } \left( p_x < eh \text{ or } p_x > -eh \text{ or } p_y < eh \text{ or } p_y > -eh \right) \\
      & \text{and } \left( p_x > ih \text{ or } p_x < -ih \text{ or } p_y > ih \text{ or } p_y < -ih \right) \\
    0 & \text{otherwise}
    \end{cases}
    \end{equation}
\end{enumerate}

\textbf{3. Fire Out} - To calculate the Fire Out reward, let us define the following:

\begin{itemize}[noitemsep,nolistsep]
    \item $T$ --- All tree states.
    \item $r_{nb}$ --- No burning tree reward.
\end{itemize}

Calculation steps:

\begin{enumerate}[noitemsep,nolistsep]
    \item We can now calculate the Fire Out reward:
    \begin{equation}
    r_{nb} =
    \begin{cases}
    10 & \text{if } \forall t \in T, \, t \neq \text{"burning"} \\ 
    0 & \text{otherwise} \\
    \end{cases}
    \end{equation}
\end{enumerate}

\textbf{4. Too Close to Village} - To calculate the Too Close to Village reward, let us define the following:

\begin{itemize}[noitemsep,nolistsep]
    \item $T_c$ --- All tree states, closer to or equal to $150$ to the village.
    \item $r_{cv}$ --- Too Close to Village reward.
\end{itemize}

Calculation steps:

\begin{enumerate}[noitemsep,nolistsep]
    \item We can now calculate the Fire Out reward:
    \begin{equation}
    r_{cc} =
    \begin{cases}
    -50 & \text{if } \exists t \in T_c, \, t = \text{"burning"} \\ 
    0 & \text{otherwise}
    \end{cases}
    \end{equation}
\end{enumerate}

\textbf{5. Time Step Burning} - To calculate the Time Step Burning reward, let us define the following:

\begin{itemize}[noitemsep,nolistsep]
    \item $T$ --- All tree states.
    \item $r_{tsb}$ --- Time Step Burning reward.
\end{itemize}

Calculation steps:

\begin{equation}
r_{tsb} =
\begin{cases}
-0.01 & \text{if } \exists t \in T \text{ such that } t = \text{``burning''} \\
0 & \text{otherwise}
\end{cases}
\end{equation}

\textbf{6. Extinguishing Tree} - To calculate the Extinguish Tree reward, let us define the following:

\begin{itemize}[noitemsep,nolistsep]
    \item $T$ --- All tree states.
    \item $r_{e}$ --- Extinguish Tree reward.
\end{itemize}

Calculation steps:

\begin{equation}
r_e =
5 \sum_{t \in T}
\mathbb{I}\!\left(
t_{\text{previous}} = \text{``burning''}
\;\land\;
t_{\text{current}} = \text{``extinguished''}
\right)
\end{equation}

\textbf{7. Preparing Tree} - To calculate the Preparing Tree reward, let us define the following:

\begin{itemize}[noitemsep,nolistsep]
    \item $T$ --- All tree states.
    \item $r_{p}$ --- Preparing Tree reward.
\end{itemize}

Calculation steps:

\begin{equation}
r_p =
\sum_{t \in T}
\mathbb{I}\!\left(
t_{\text{previous}} = \text{``not burning''}
\;\land\;
t_{\text{current}} = \text{``wet''}
\right)
\end{equation}

\newpage

\subsection{Prompt Templates \& Samples}

\subsubsection{Rule-Based Controller Prompt Template: LLM-Mediator}

\begin{figure}[h!]
    \centering
    \includegraphics[width=\linewidth]{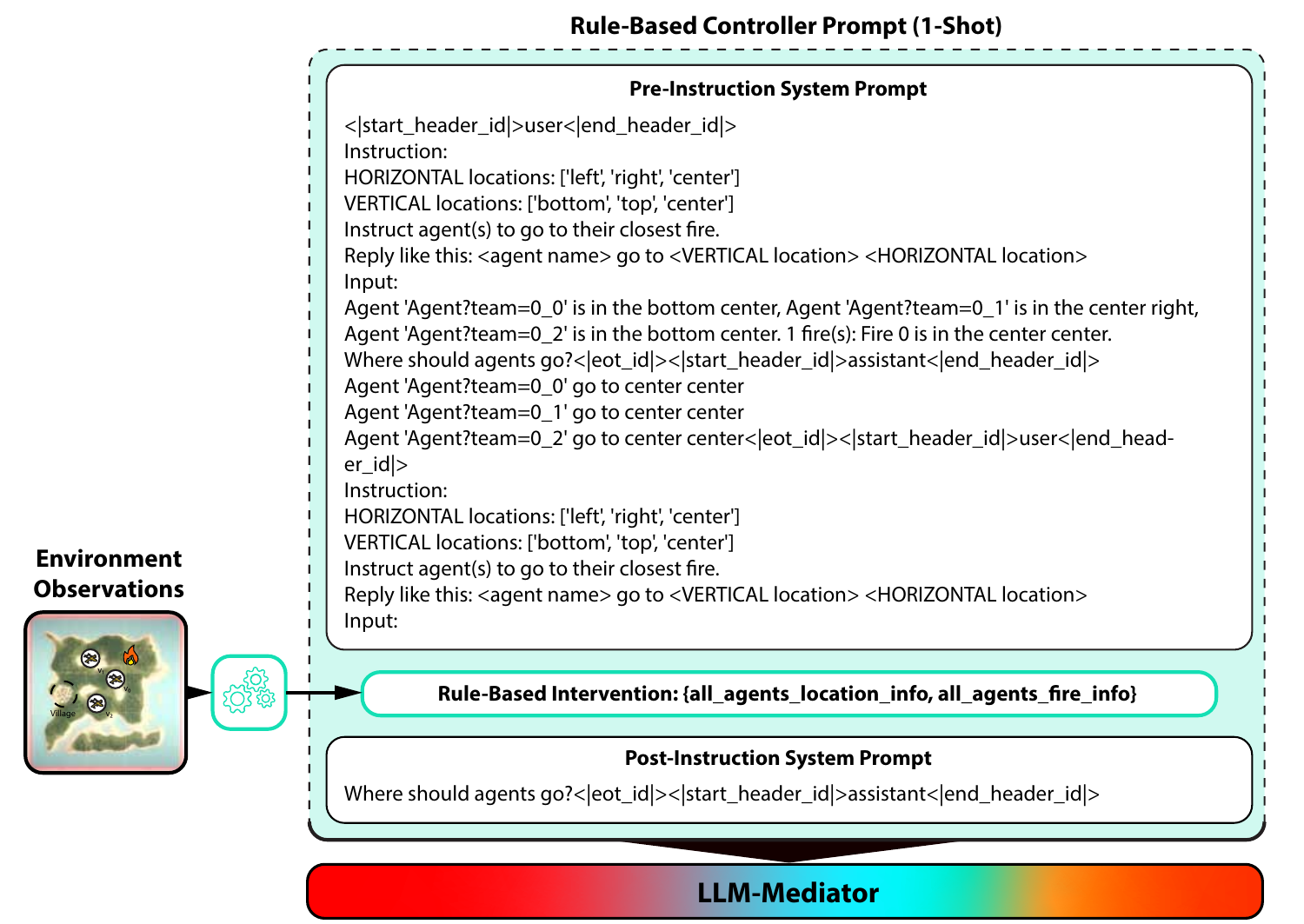}
    \caption{Complete prompt template for the Rule-Based Controller. This prompt is sent to the LLM-Mediator.}
    \label{appendix:RB_prompt_template}
\end{figure}

\clearpage

\subsubsection{Natural Language Controller Prompt Template: Strategy and LLM-Mediator}

\begin{figure}[h!]
    \centering
    \includegraphics[width=\linewidth]{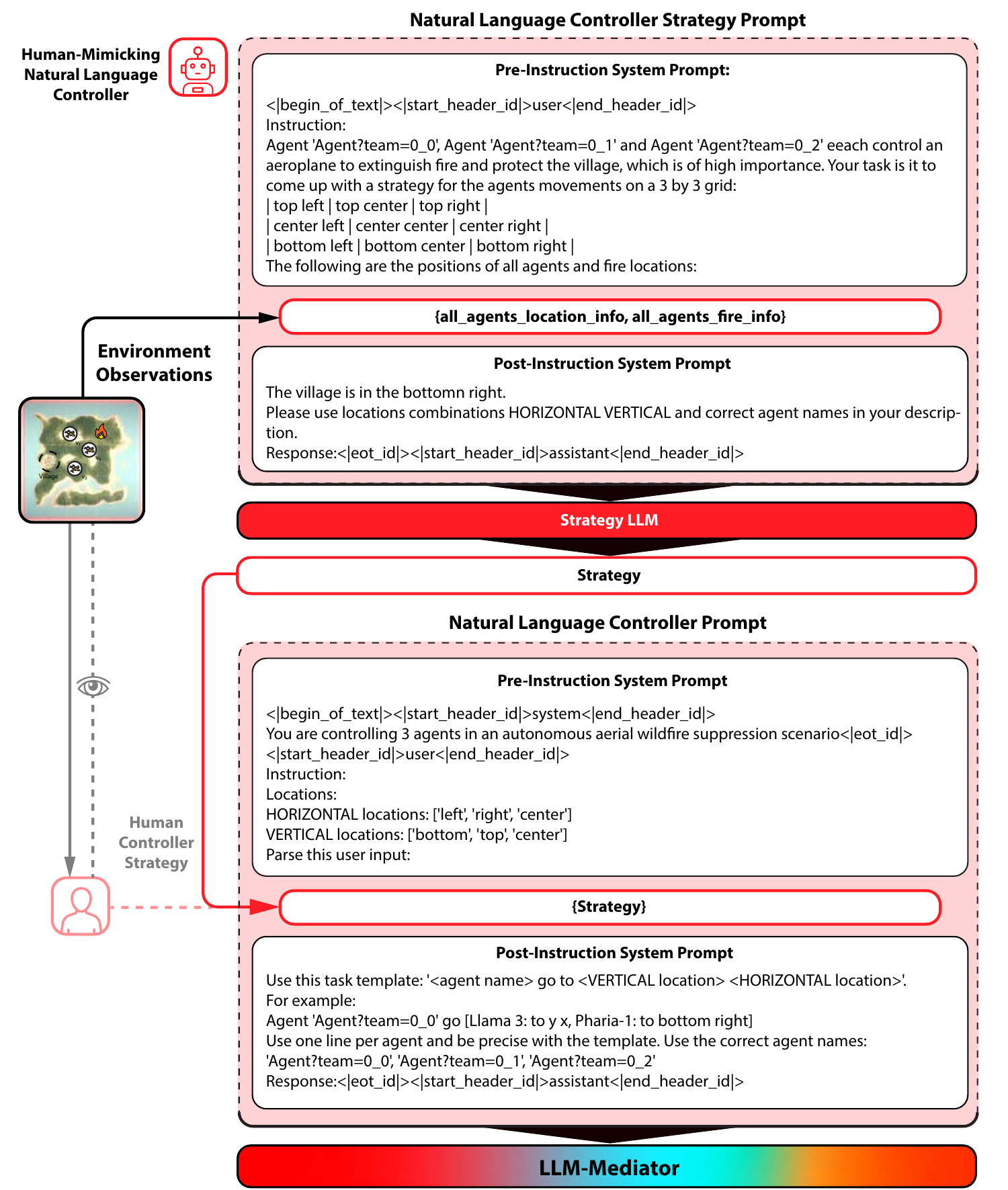}
    \caption{Complete prompt templates for the Natural Language Controller. The first prompt template is to generate a strategy, which is then integrated in the second prompt template that is sent to the LLM-Mediator.}
    \label{appendix:NL_prompt_template}
\end{figure}

\clearpage

\subsubsection{Rule-Based and Natural Language Controller Vector Observation Data Samples}

\begin{figure}[H]
    \centering
    \includegraphics[width=0.83\linewidth]{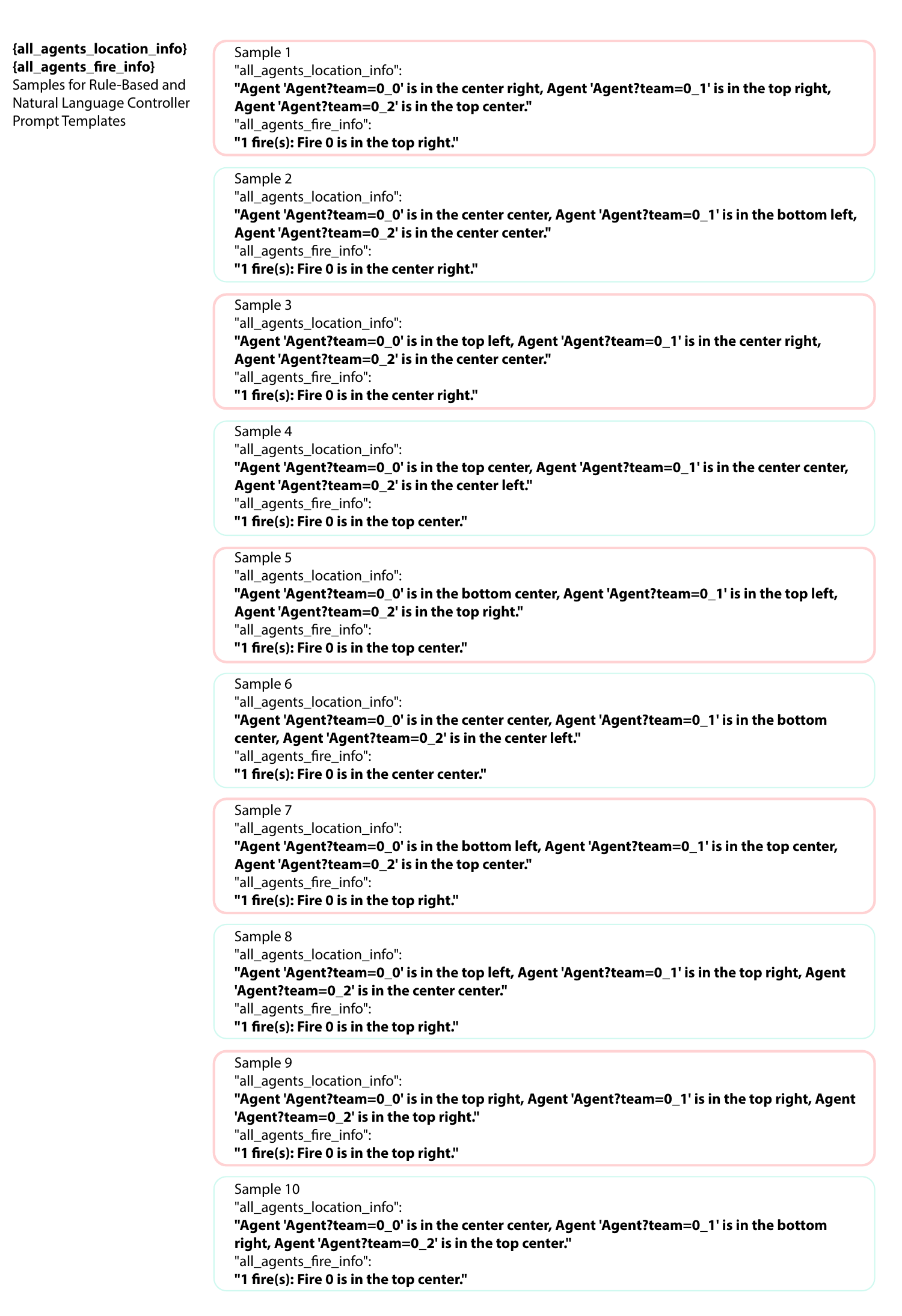}
    \caption{Feature Vector observation data samples in natural language of $ \mathtt{\{all\_agents\_location\_info\}} $ and $ \mathtt{\{all\_agents\_fire\_info\}} $, integrated in the Rule-Based Controller prompt template as well as the strategy prompt template as part of the Natural Language Controller.}
    \label{appendix:feature_nl_samples}
\end{figure}

\clearpage

\subsubsection{Natural Language Strategy Samples: Pharia-1-LLM-7B-control-aligned}

\begin{figure}[H]
    \centering
    \includegraphics[width=0.82\linewidth]{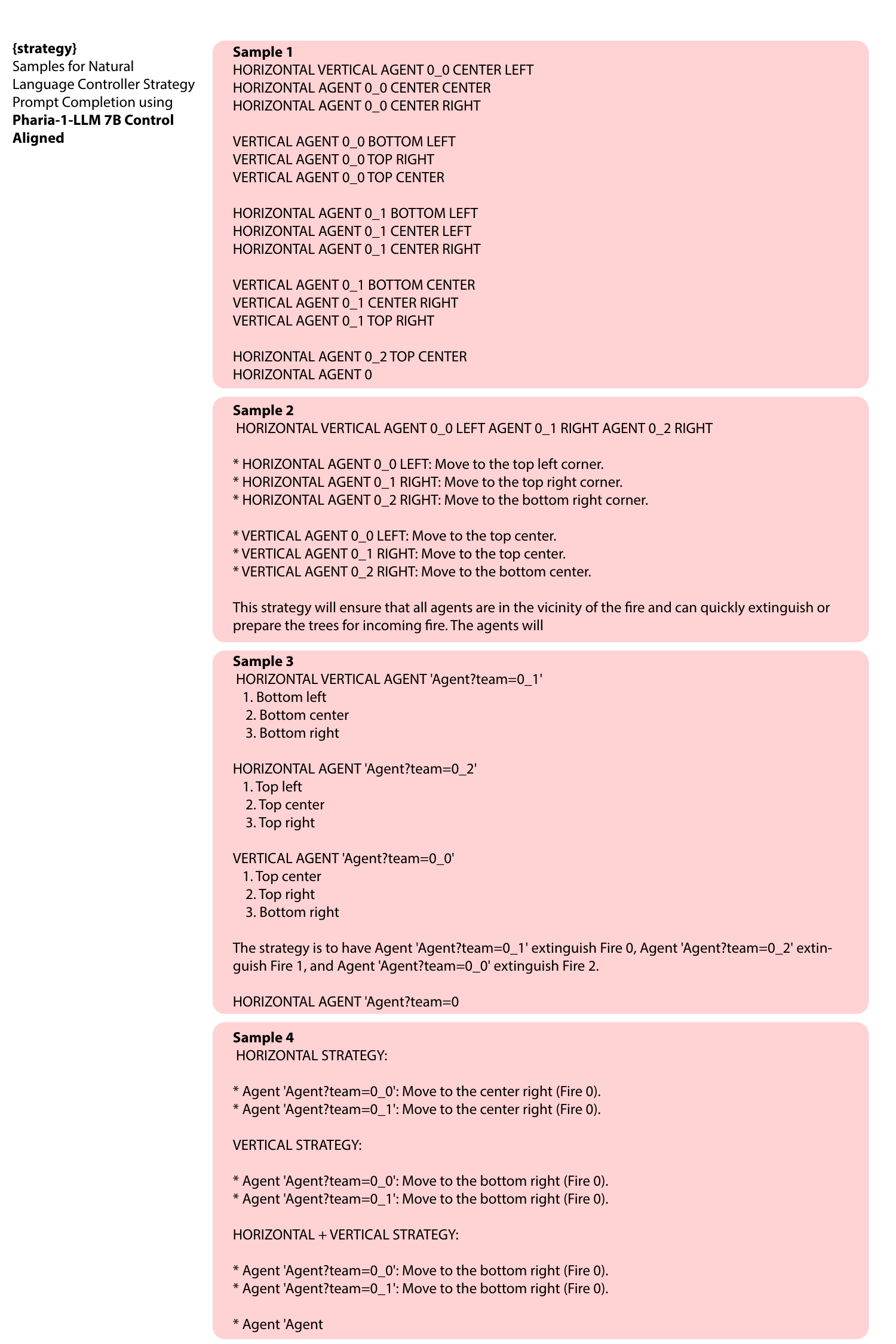}
    \caption{Pharia-1-LLM-7B-control-aligned samples for $ \mathtt{\{strategy\}} $, to be integrated in the Natural Language Controller prompt template, sent to the LLM-Mediator.}
    \label{appendix:strategy_samples_pharia}
\end{figure}

\clearpage

\subsubsection{Natural Language Strategy Samples: Llama-3.1-8B Instruct}

\begin{figure}[h!]
    \centering
    \includegraphics[width=0.82\linewidth]{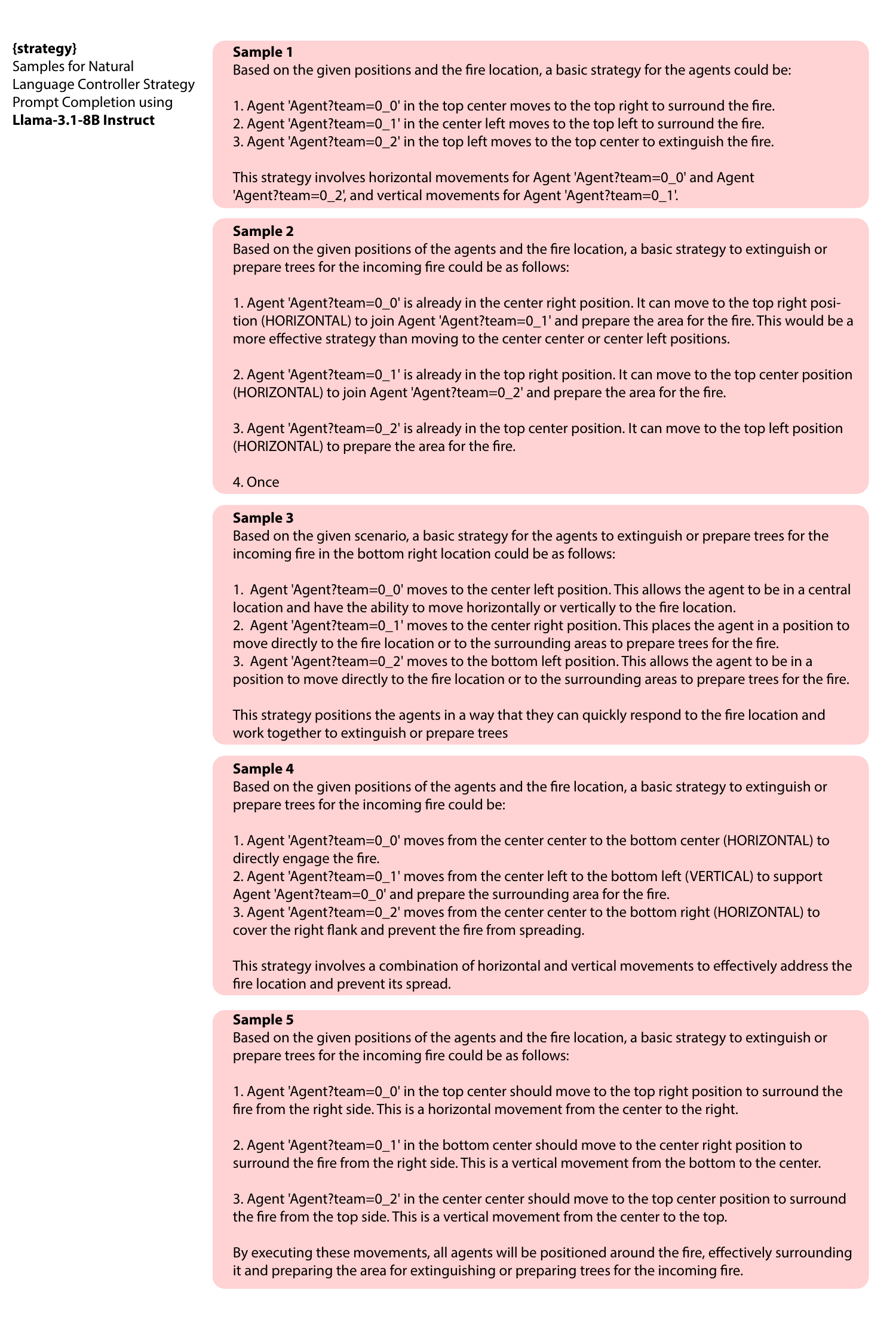}
    \caption{LLama-3.1-8B Instruct samples for $ \mathtt{\{strategy\}} $, to be integrated in the Natural Language Controller prompt template, sent to the LLM-Mediator.}
    \label{appendix:strategy_samples_llama}
\end{figure}

\clearpage

\subsection{Additional Results}\label{sec:additional_results}

\begin{figure}[h!]
    \centering
    \label{results:crash_count_RB}
    \begin{minipage}{0.33\textwidth}
        \centering
        \includegraphics[width=\linewidth]{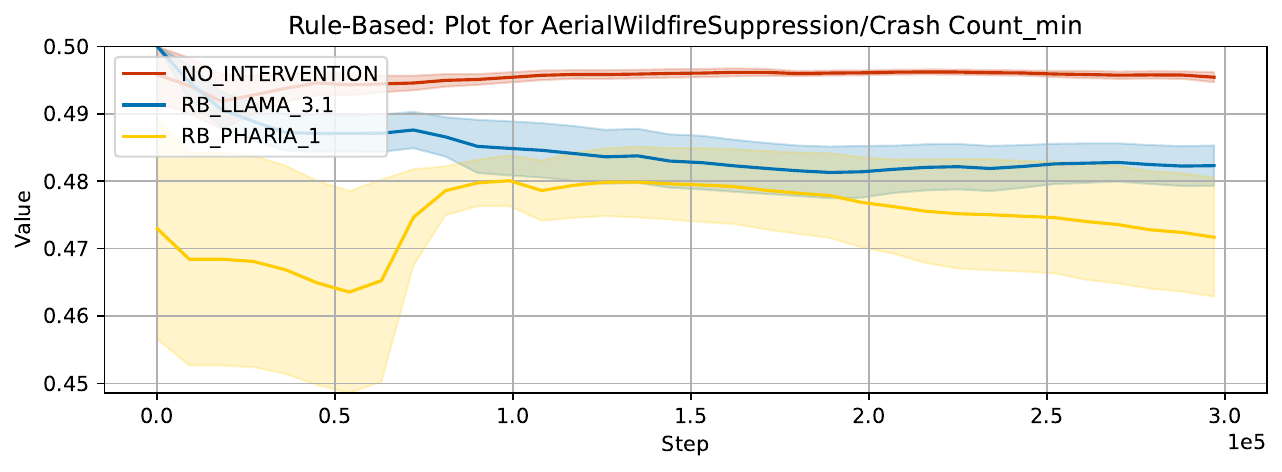}
    \end{minipage}\hfill
    \begin{minipage}{0.33\textwidth}
        \centering
        \includegraphics[width=\linewidth]{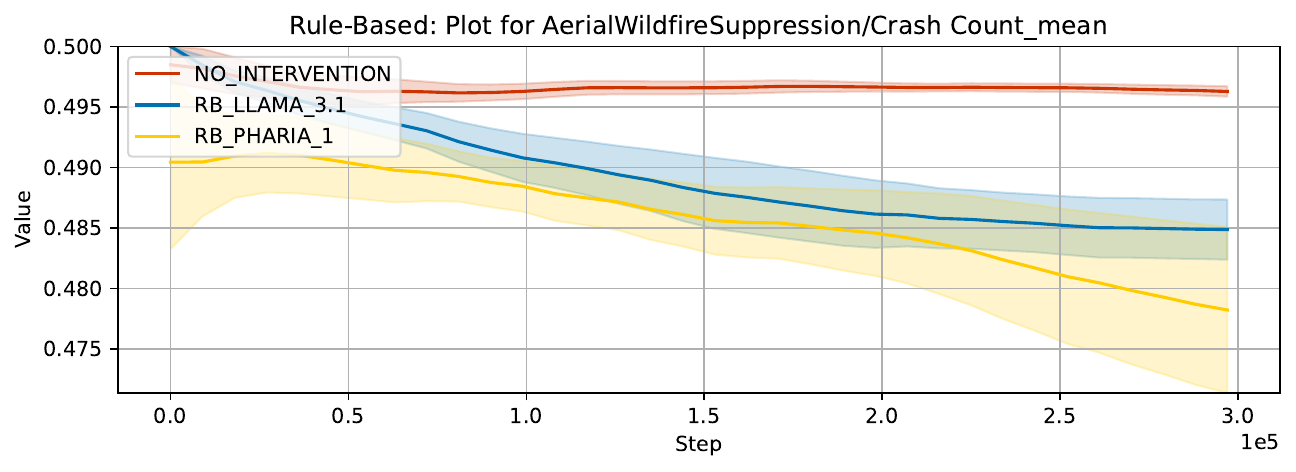}
    \end{minipage}
    \begin{minipage}{0.33\textwidth}
        \centering
        \includegraphics[width=\linewidth]{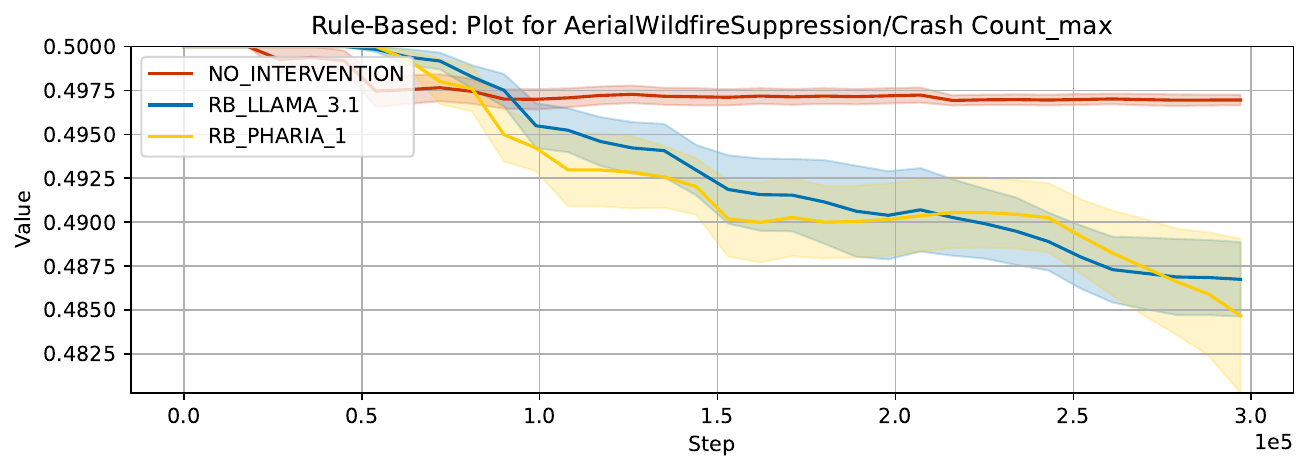}
    \end{minipage}
    \caption{Crash Count \textbf{(Rule-Based)} - No controller baseline VS Rule-Based Controller with Llama-3.1-8B Instruct: min, mean and max.}
\end{figure}

\begin{figure}[h!]
    \centering
    \label{results:crash_count_NL}
    \begin{minipage}{0.33\textwidth}
        \centering
        \includegraphics[width=\linewidth]{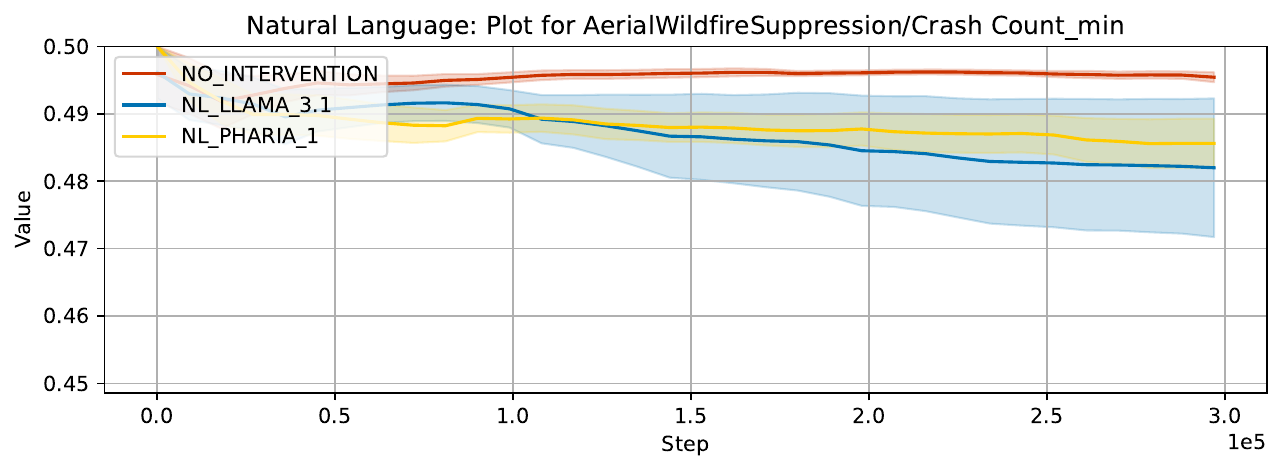}
    \end{minipage}\hfill
    \begin{minipage}{0.33\textwidth}
        \centering
        \includegraphics[width=\linewidth]{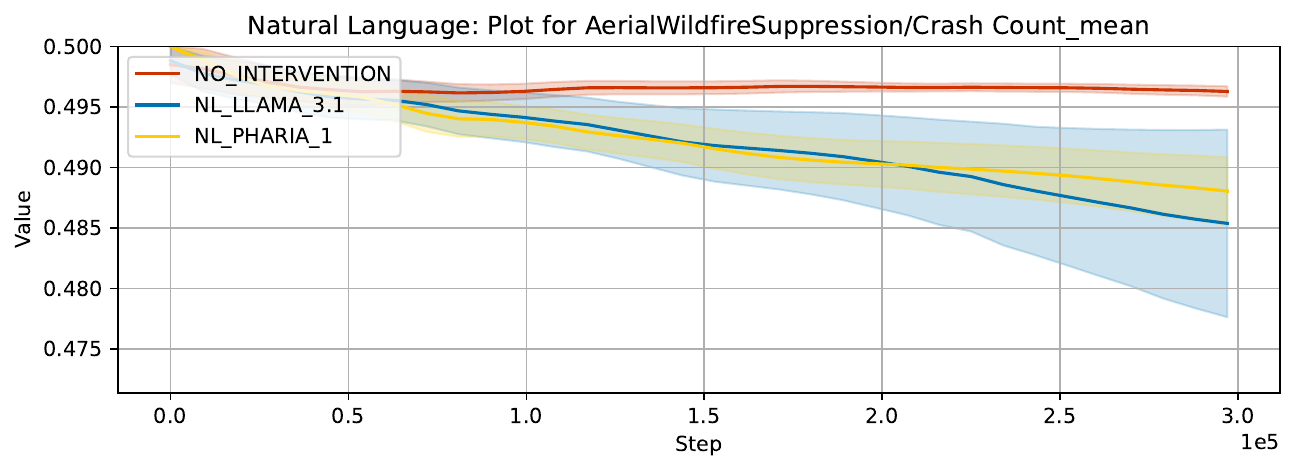}
    \end{minipage}
    \begin{minipage}{0.33\textwidth}
        \centering
        \includegraphics[width=\linewidth]{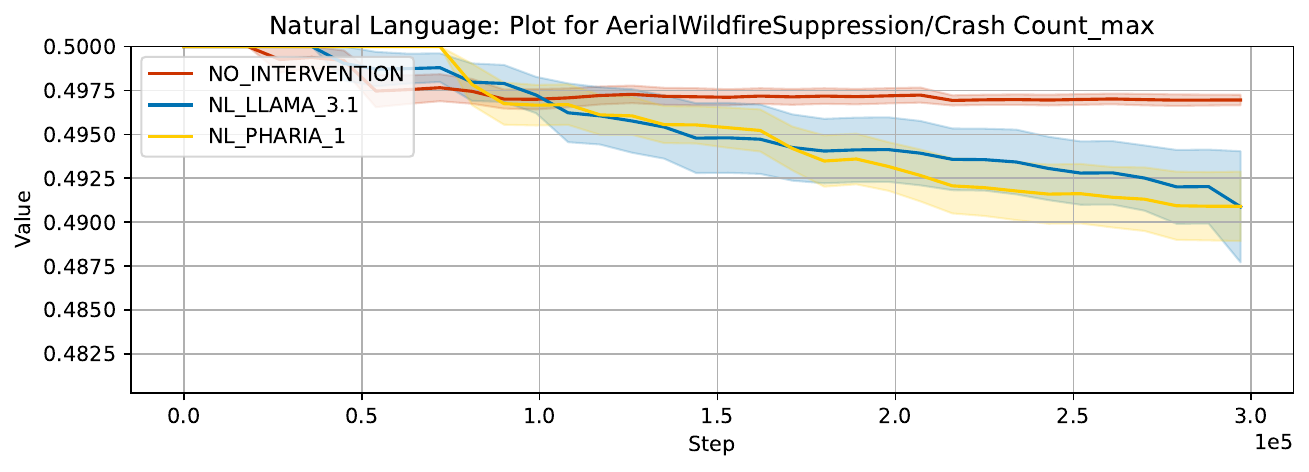}
    \end{minipage}
    \caption{Crash Count \textbf{(Natural Language)} - No controller baseline VS Natural Language Controller with Llama-3.1-8B Instruct: min, mean and max.}
\end{figure}

\begin{figure}[h!]
    \centering
    \label{results:episode_RB}
    \begin{minipage}{0.33\textwidth}
        \centering
        \includegraphics[width=\linewidth]{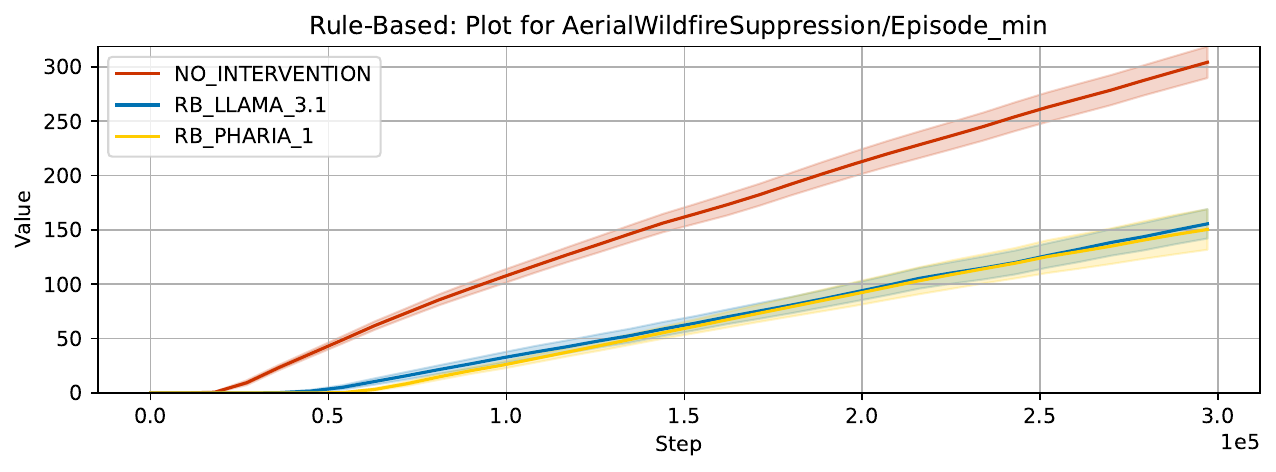}
    \end{minipage}\hfill
    \begin{minipage}{0.33\textwidth}
        \centering
        \includegraphics[width=\linewidth]{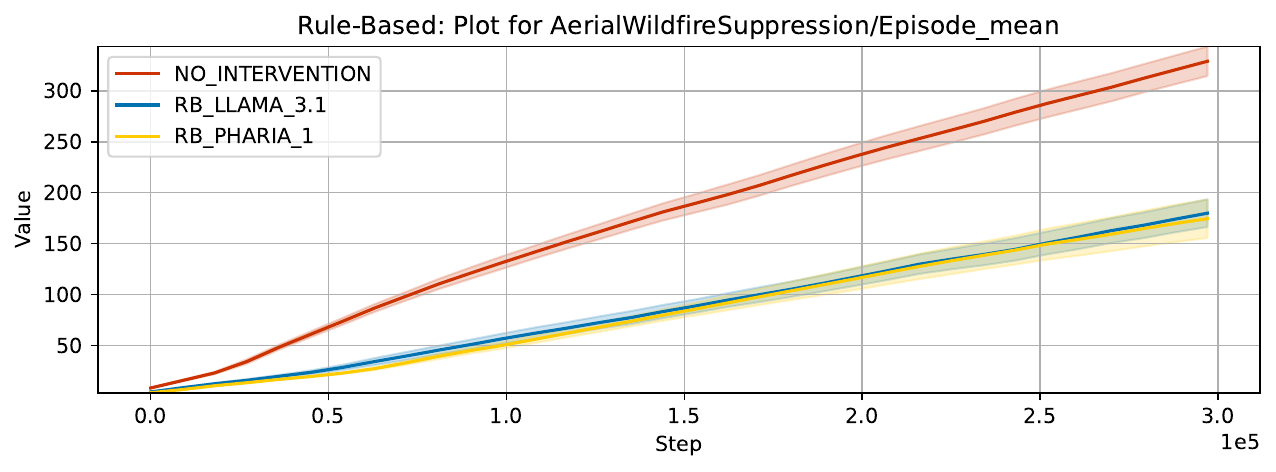}
    \end{minipage}
    \begin{minipage}{0.33\textwidth}
        \centering
        \includegraphics[width=\linewidth]{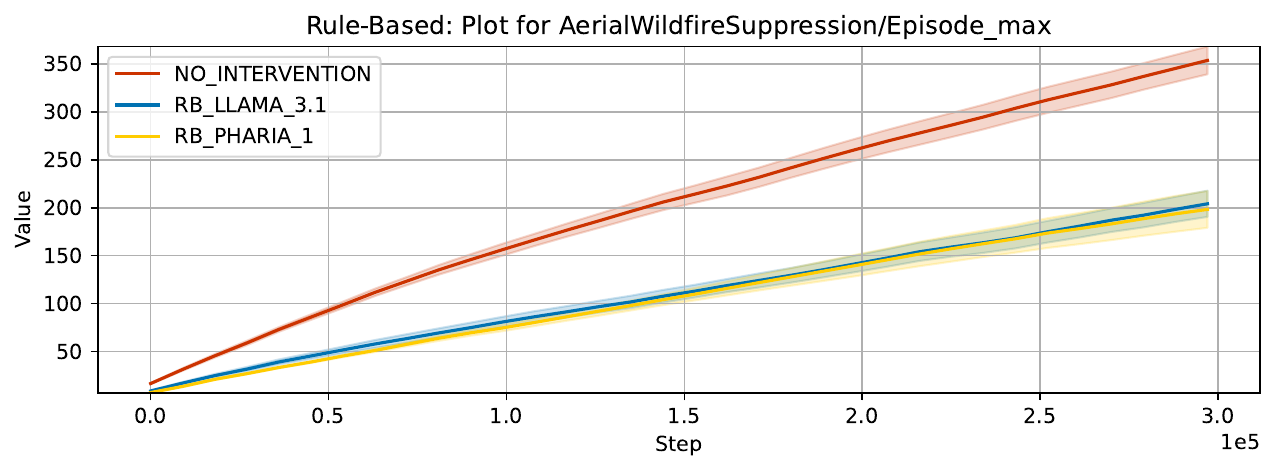}
    \end{minipage}
    \caption{Episode Count \textbf{(Rule-Based)} - No controller baseline VS Rule-Based Controller with Llama-3.1-8B Instruct: min, mean and max.}
\end{figure}

\begin{figure}[h!]
    \centering
    \label{results:episode_NL}
    \begin{minipage}{0.33\textwidth}
        \centering
        \includegraphics[width=\linewidth]{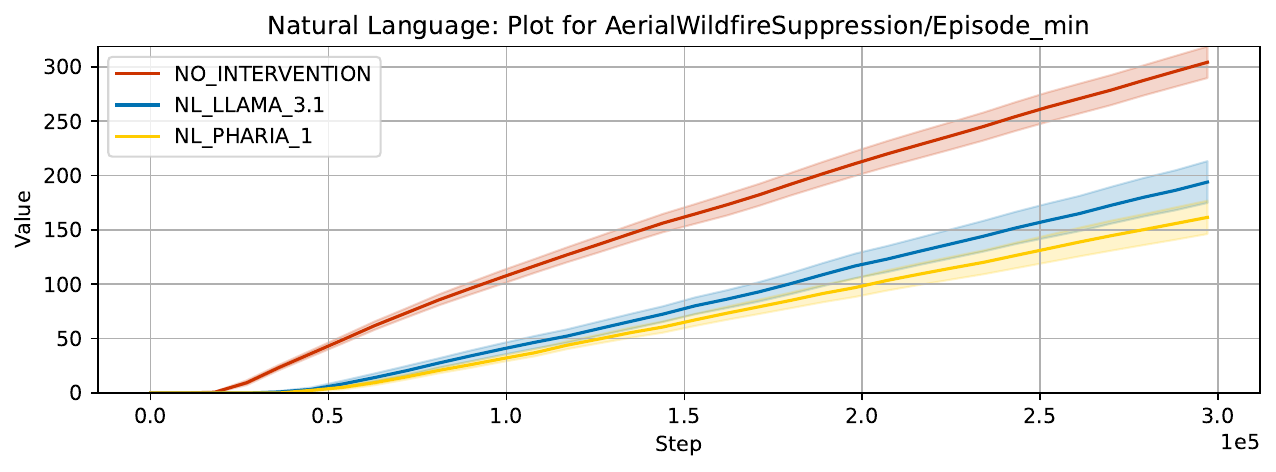}
    \end{minipage}\hfill
    \begin{minipage}{0.33\textwidth}
        \centering
        \includegraphics[width=\linewidth]{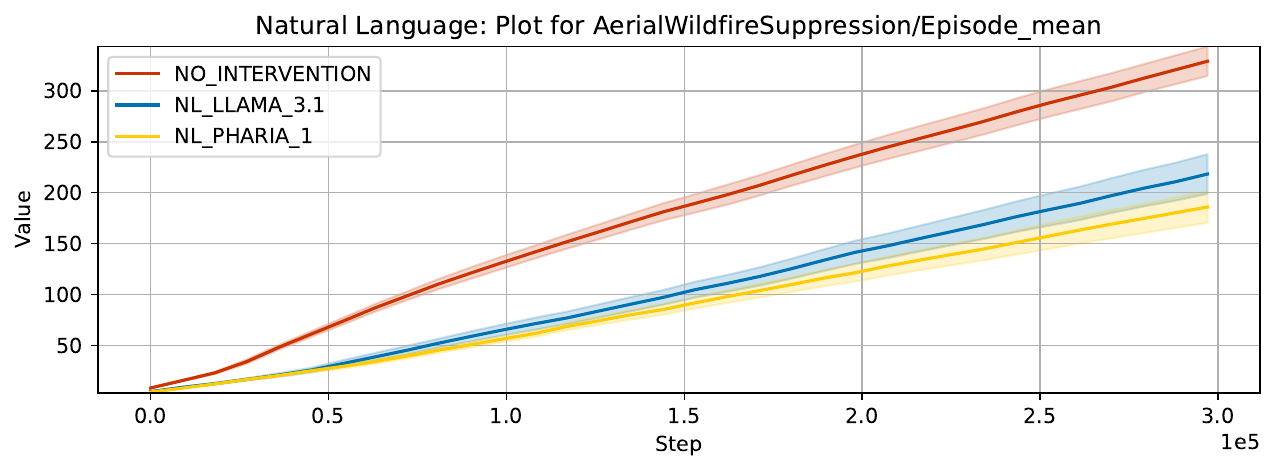}
    \end{minipage}
    \begin{minipage}{0.33\textwidth}
        \centering
        \includegraphics[width=\linewidth]{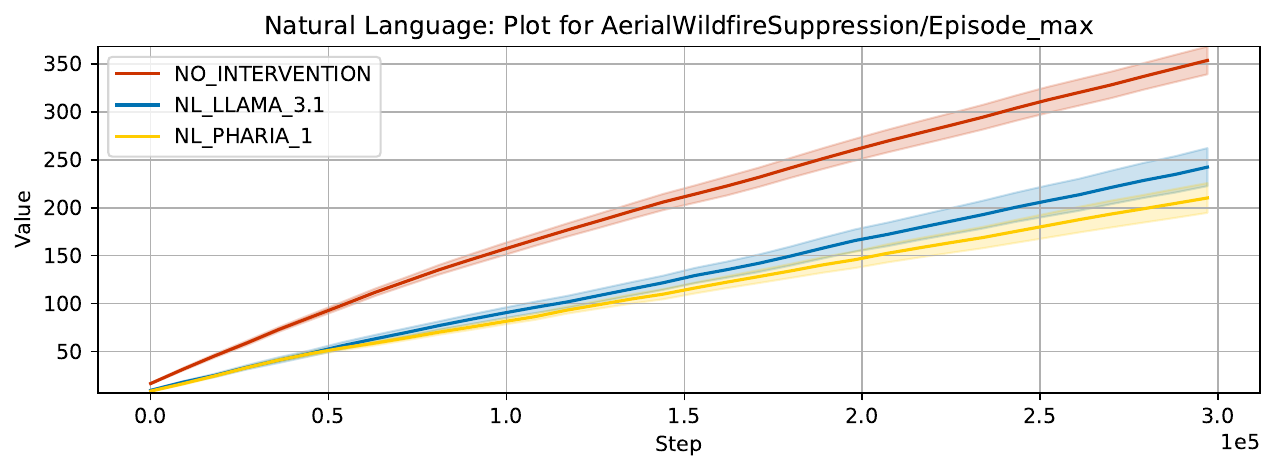}
    \end{minipage}
    \caption{Episode Count \textbf{(Natural Language)} - No controller baseline VS Natural Language Controller with Llama-3.1-8B Instruct: min, mean and max.}
\end{figure}

\begin{figure}[h!]
    \centering
    \label{results:extunguishing_trees_RB}
    \begin{minipage}{0.33\textwidth}
        \centering
        \includegraphics[width=\linewidth]{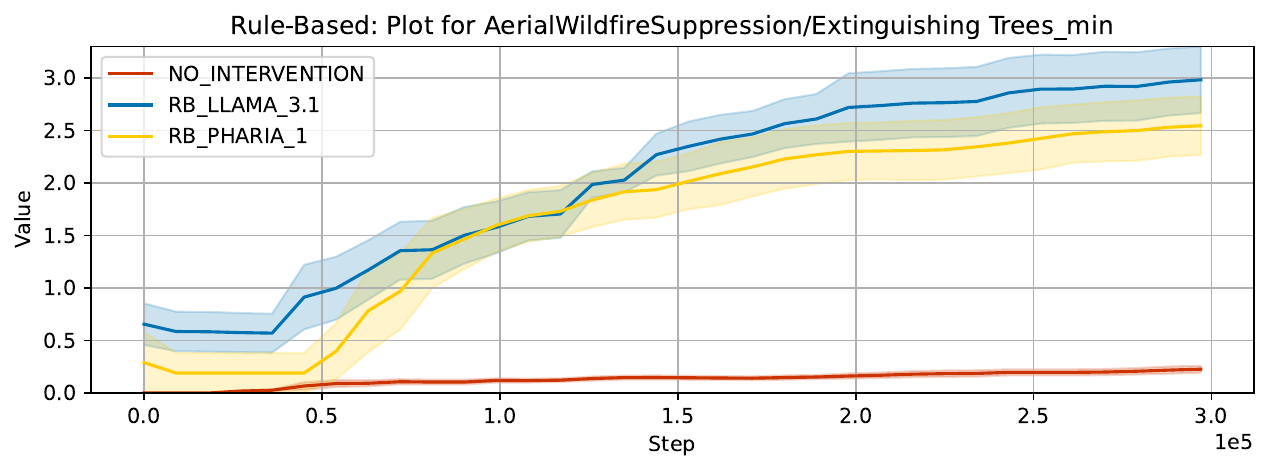}
    \end{minipage}\hfill
    \begin{minipage}{0.33\textwidth}
        \centering
        \includegraphics[width=\linewidth]{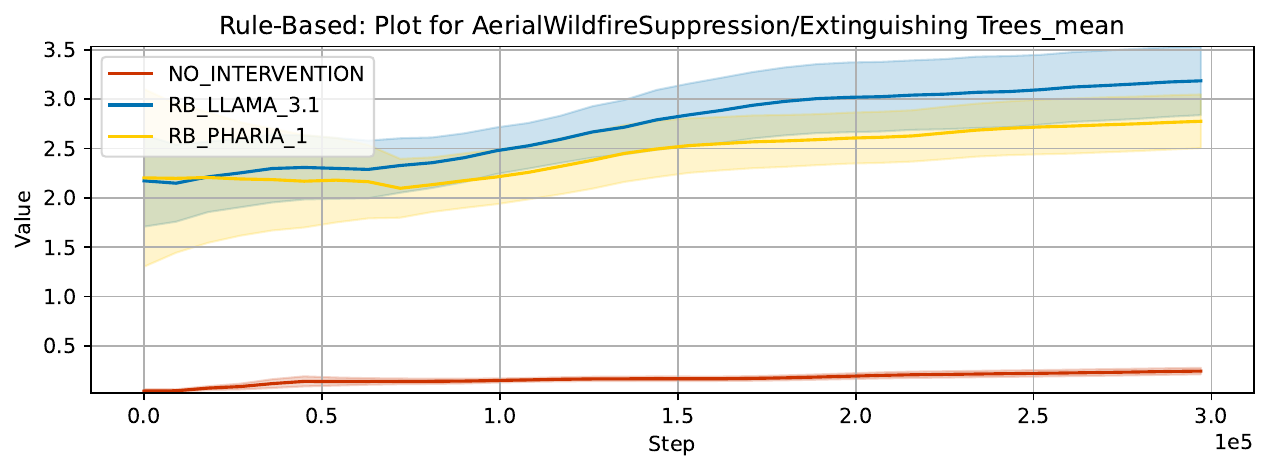}
    \end{minipage}
    \begin{minipage}{0.33\textwidth}
        \centering
        \includegraphics[width=\linewidth]{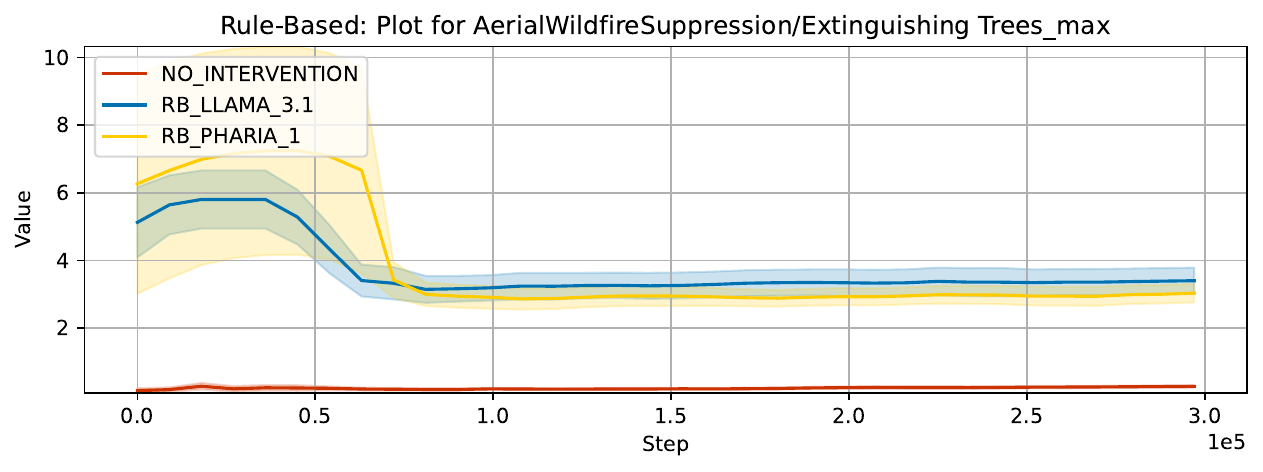}
    \end{minipage}
    \caption{Extinguishing Trees \textbf{(Rule-Based)} - No controller baseline VS Rule-Based Controller with Llama-3.1-8B Instruct: min, mean and max.}
\end{figure}

\begin{figure}[h!]
    \centering
    \label{results:extunguishing_trees_NL}
    \begin{minipage}{0.33\textwidth}
        \centering
        \includegraphics[width=\linewidth]{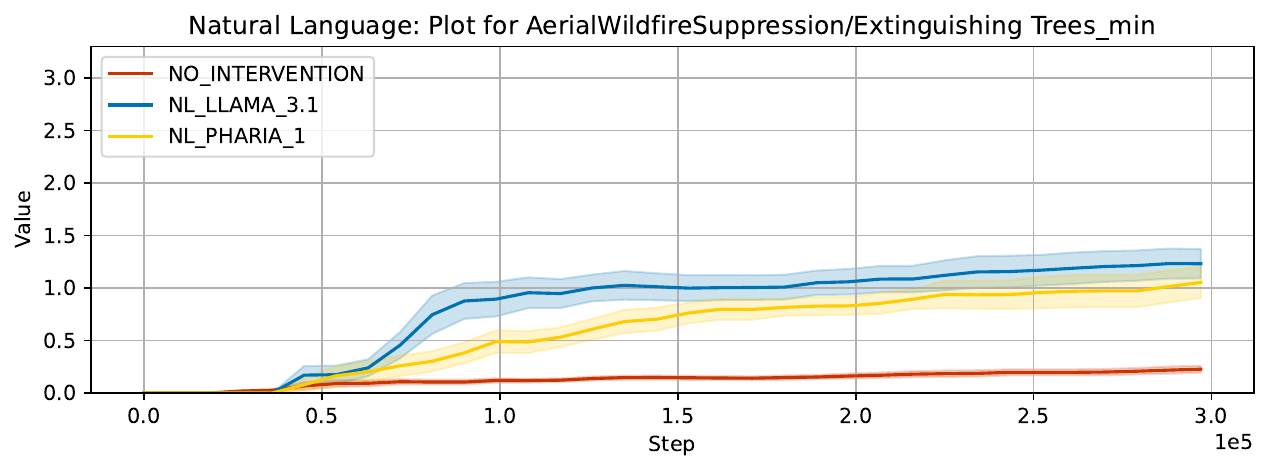}
    \end{minipage}\hfill
    \begin{minipage}{0.33\textwidth}
        \centering
        \includegraphics[width=\linewidth]{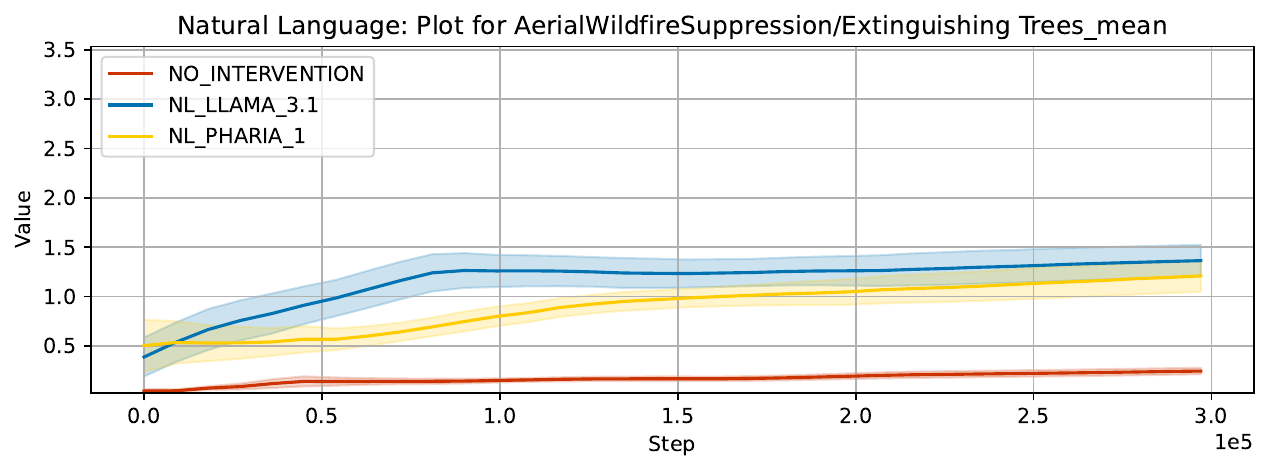}
    \end{minipage}
    \begin{minipage}{0.33\textwidth}
        \centering
        \includegraphics[width=\linewidth]{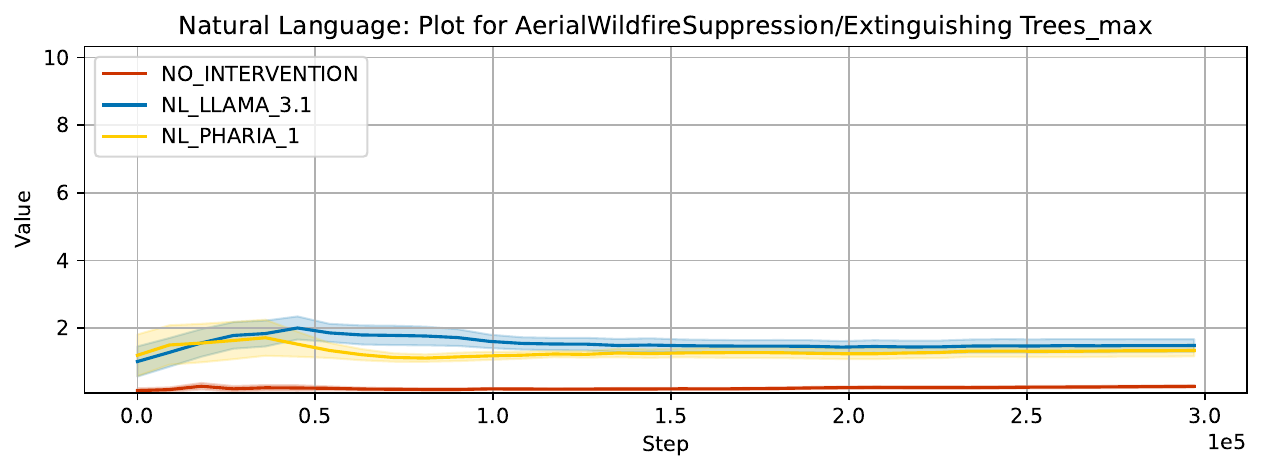}
    \end{minipage}
    \caption{Extinguishing Trees \textbf{(Natural Language)} - No controller baseline VS Natural Language Controller with Llama-3.1-8B Instruct: min, mean and max.}
\end{figure}

\begin{figure}[h!]
    \centering
    \label{results:extunguishing_trees_reward_RB}
    \begin{minipage}{0.33\textwidth}
        \centering
        \includegraphics[width=\linewidth]{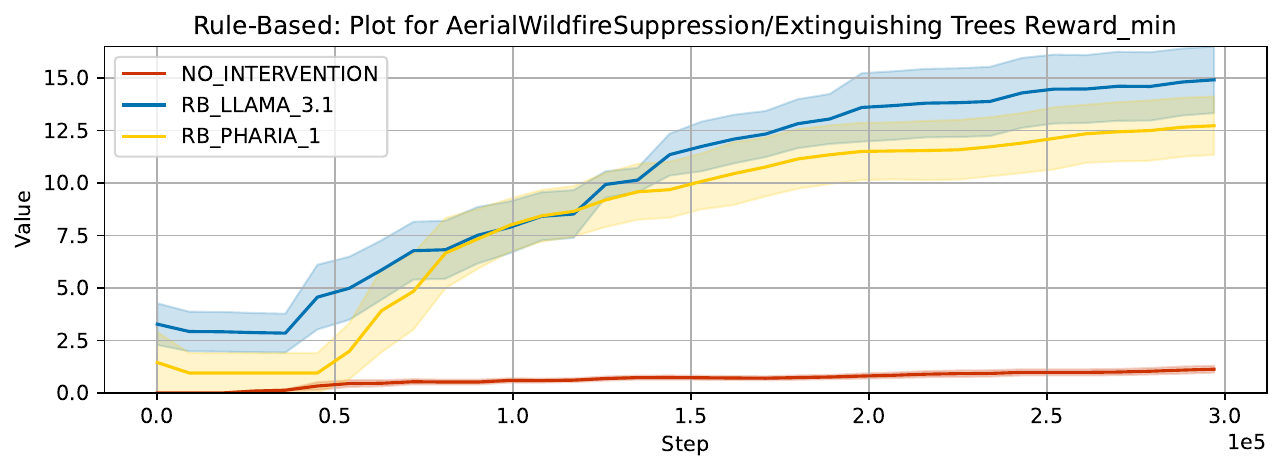}
    \end{minipage}\hfill
    \begin{minipage}{0.33\textwidth}
        \centering
        \includegraphics[width=\linewidth]{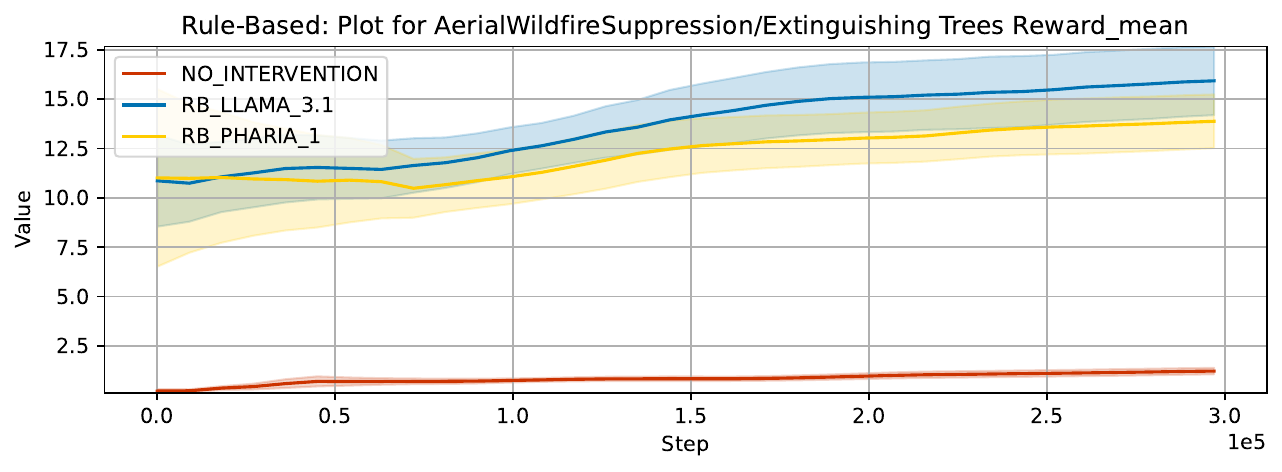}
    \end{minipage}
    \begin{minipage}{0.33\textwidth}
        \centering
        \includegraphics[width=\linewidth]{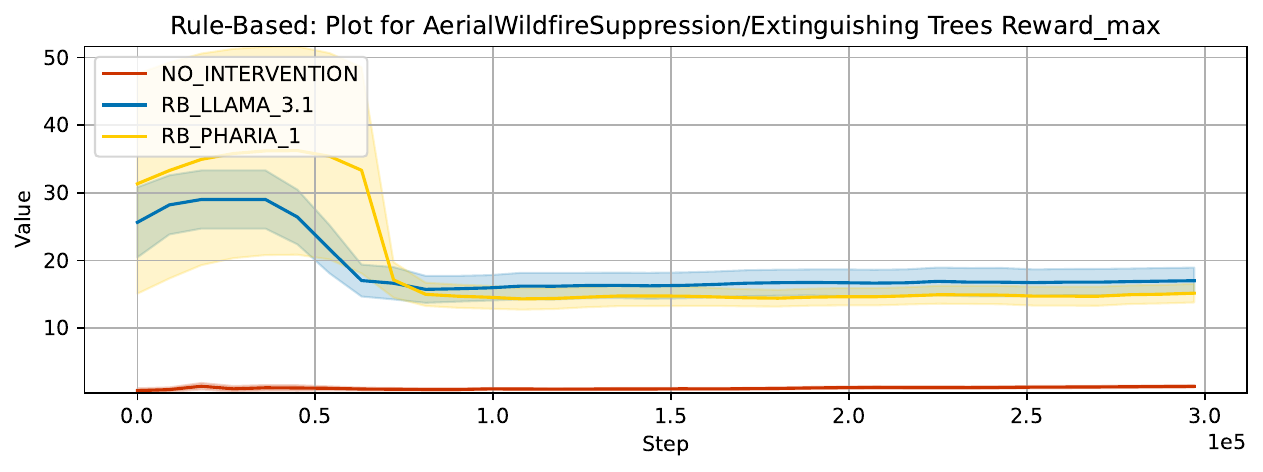}
    \end{minipage}
    \caption{Extinguishing Trees Reward \textbf{(Rule-Based)} - No controller baseline VS Rule-Based Controller with Llama-3.1-8B Instruct: min, mean and max.}
\end{figure}

\begin{figure}[h!]
    \centering
    \label{results:extunguishing_trees_reward_NL}
    \begin{minipage}{0.33\textwidth}
        \centering
        \includegraphics[width=\linewidth]{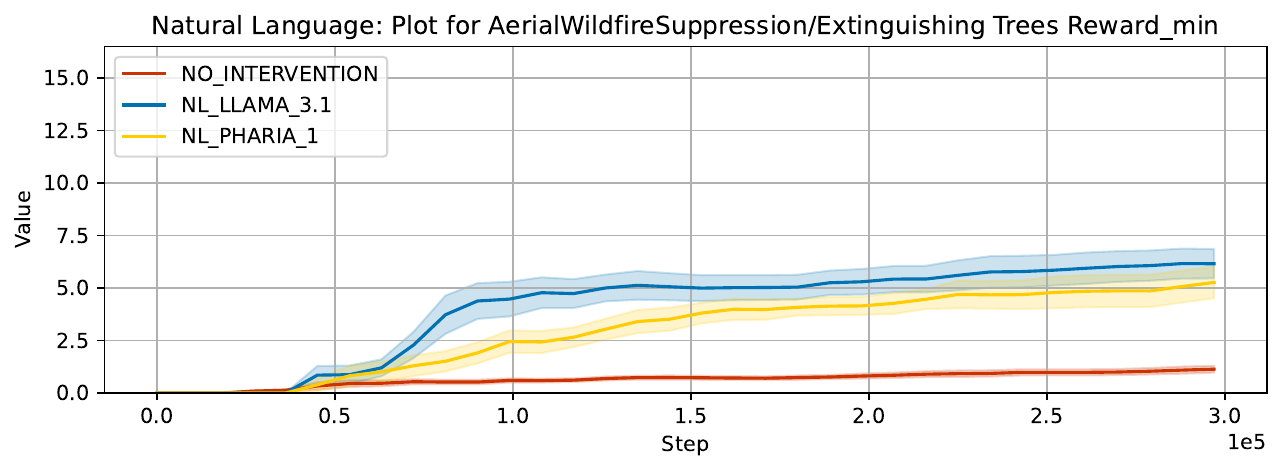}
    \end{minipage}\hfill
    \begin{minipage}{0.33\textwidth}
        \centering
        \includegraphics[width=\linewidth]{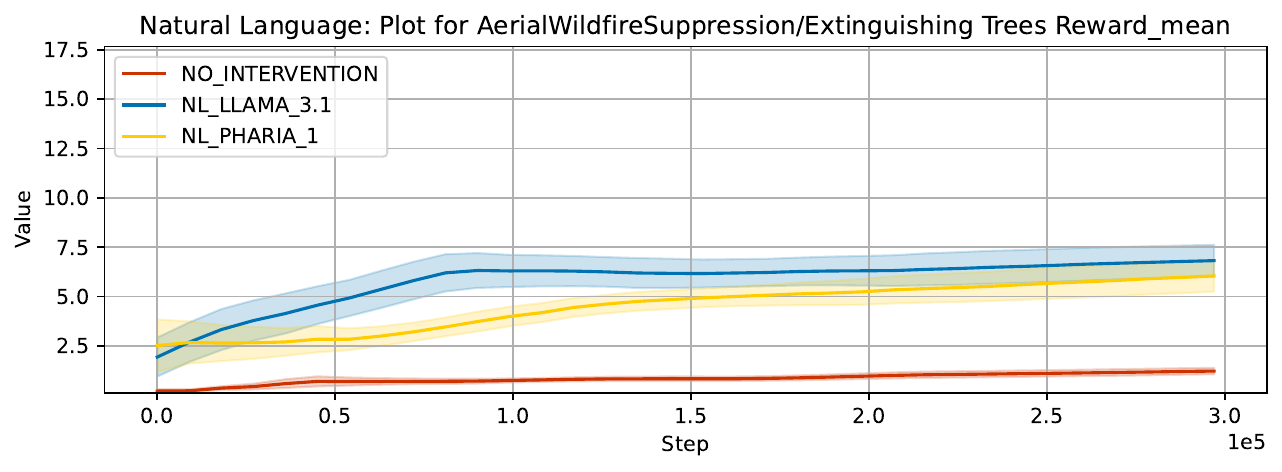}
    \end{minipage}
    \begin{minipage}{0.33\textwidth}
        \centering
        \includegraphics[width=\linewidth]{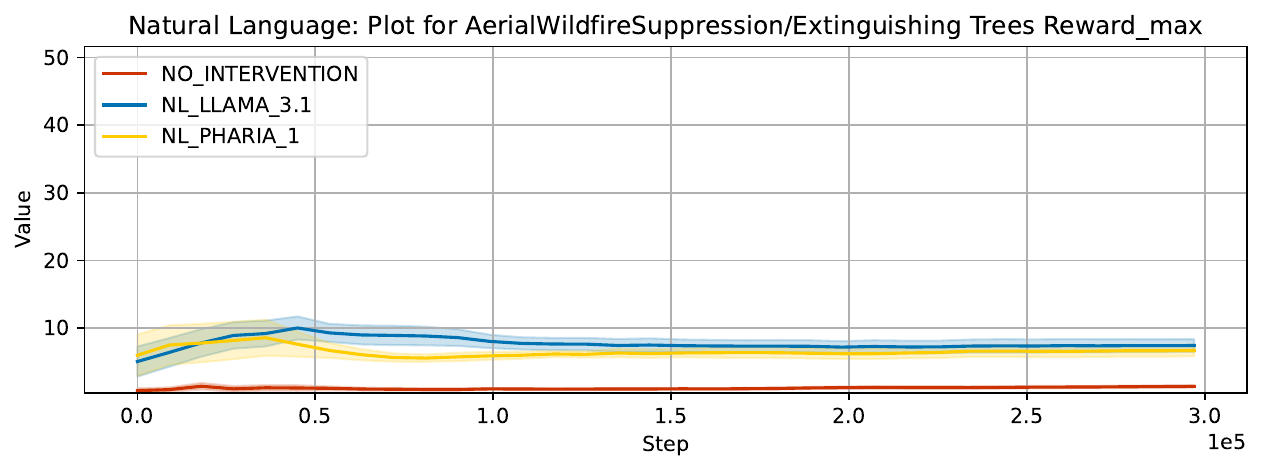}
    \end{minipage}
    \caption{Extinguishing Trees Reward \textbf{(Natural Language)} - No controller baseline VS Natural Language Controller with Llama-3.1-8B Instruct: min, mean and max.}
\end{figure}

\begin{figure}[h!]
    \centering
    \label{results:fire_out_RB}
    \begin{minipage}{0.33\textwidth}
        \centering
        \includegraphics[width=\linewidth]{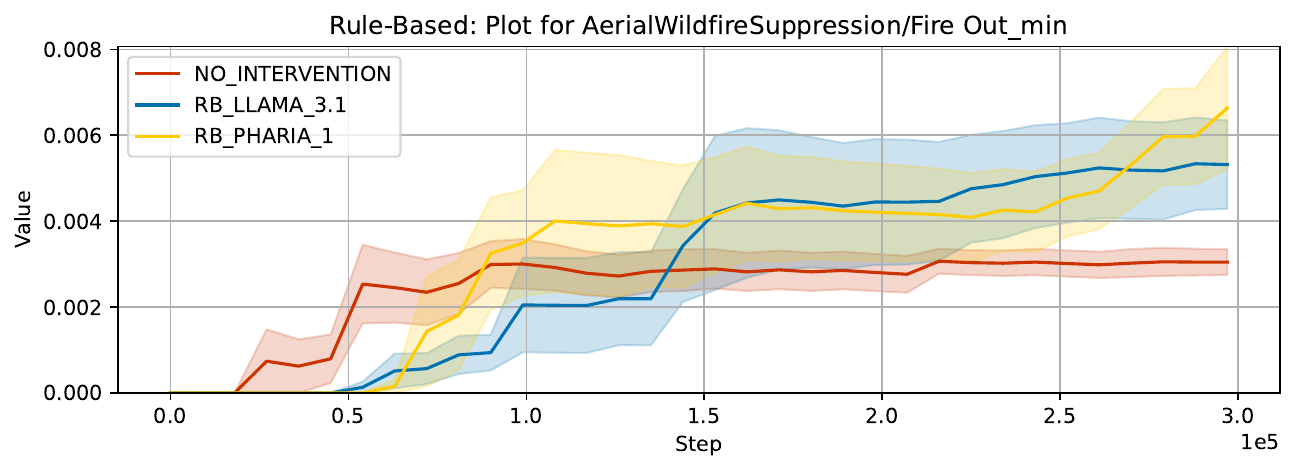}
    \end{minipage}\hfill
    \begin{minipage}{0.33\textwidth}
        \centering
        \includegraphics[width=\linewidth]{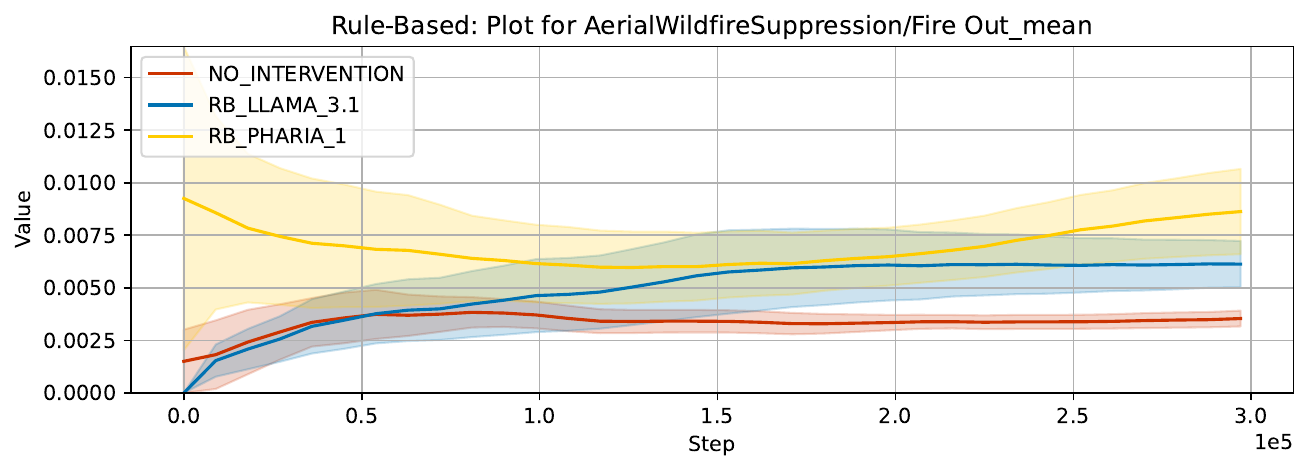}
    \end{minipage}
    \begin{minipage}{0.33\textwidth}
        \centering
        \includegraphics[width=\linewidth]{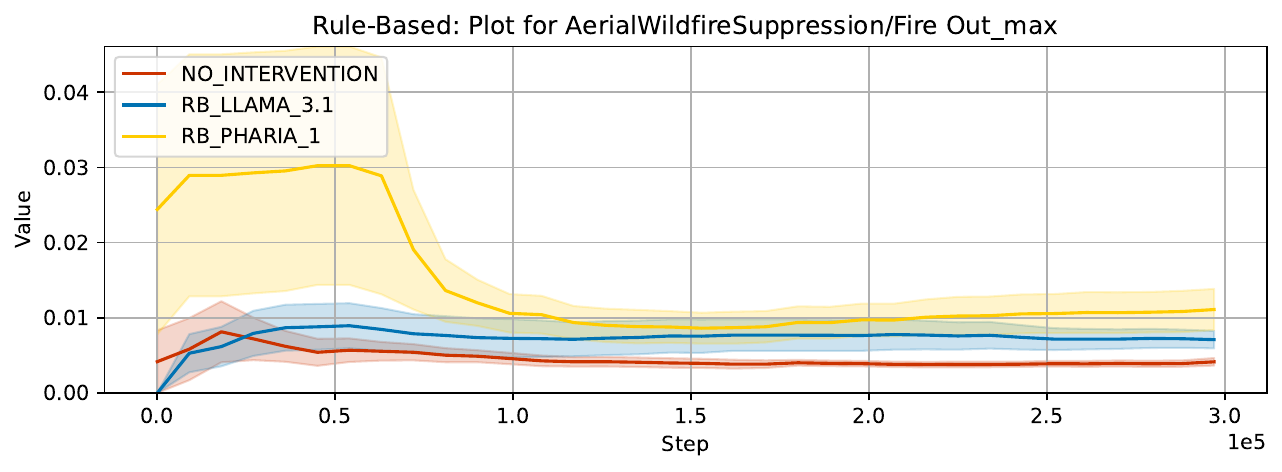}
    \end{minipage}
    \caption{Fire Out Count \textbf{(Rule-Based)} - No controller baseline VS Rule-Based Controller with Llama-3.1-8B Instruct: min, mean and max.}
\end{figure}

\begin{figure}[h!]
    \centering
    \label{results:fire_out_NL}
    \begin{minipage}{0.33\textwidth}
        \centering
        \includegraphics[width=\linewidth]{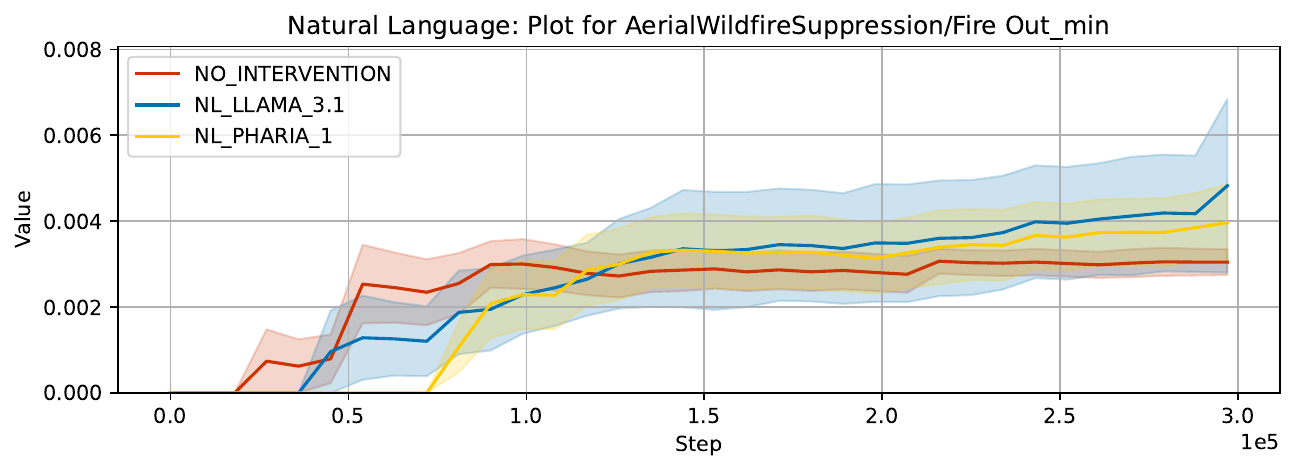}
    \end{minipage}\hfill
    \begin{minipage}{0.33\textwidth}
        \centering
        \includegraphics[width=\linewidth]{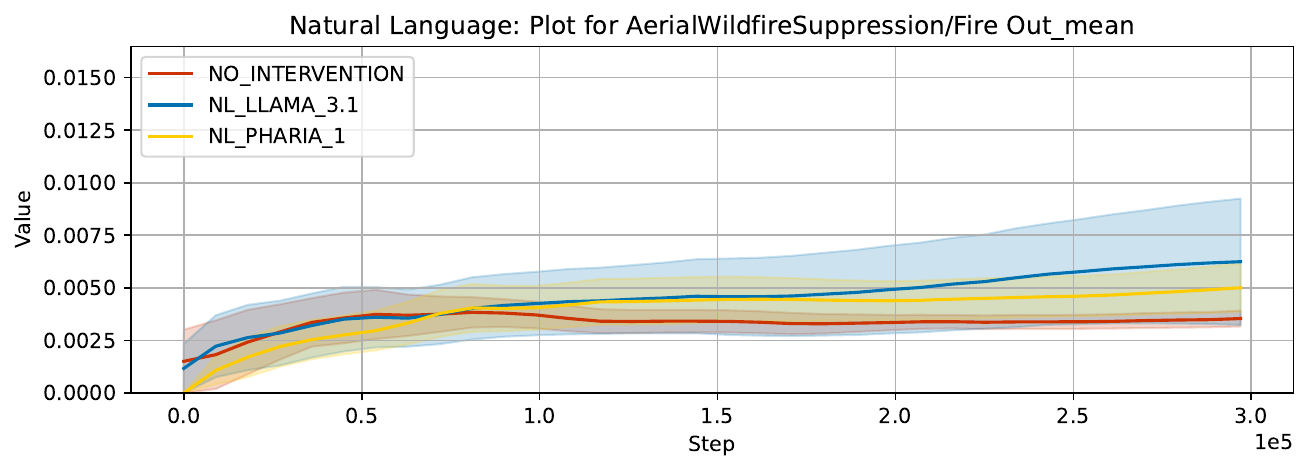}
    \end{minipage}
    \begin{minipage}{0.33\textwidth}
        \centering
        \includegraphics[width=\linewidth]{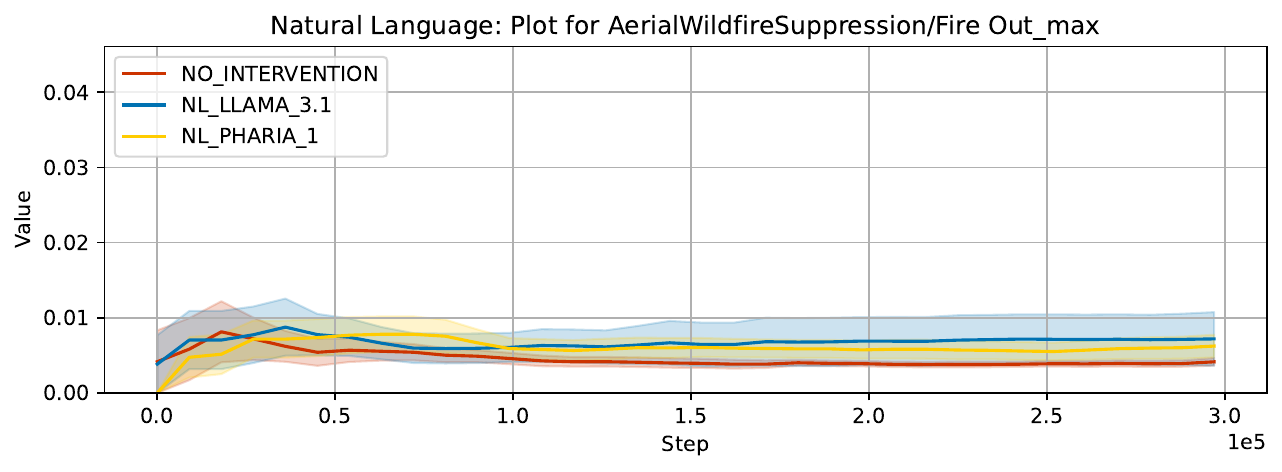}
    \end{minipage}
    \caption{Fire Out Count \textbf{(Natural Language)} - No controller baseline VS Natural Language Controller with Llama-3.1-8B Instruct: min, mean and max.}
\end{figure}

\begin{figure}[h!]
    \centering
    \label{results:fire_too_close_to_village_RB}
    \begin{minipage}{0.33\textwidth}
        \centering
        \includegraphics[width=\linewidth]{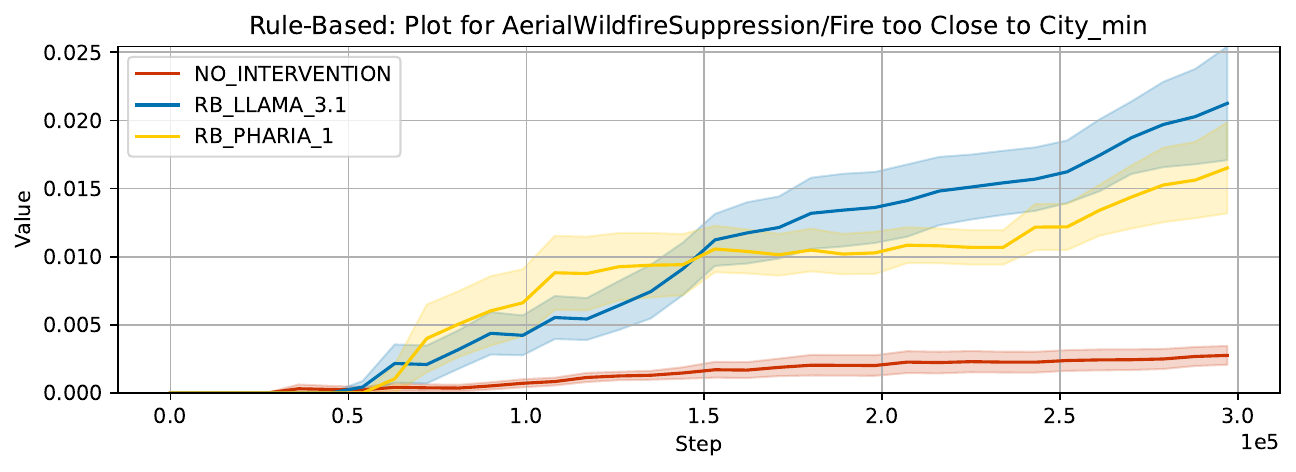}
    \end{minipage}\hfill
    \begin{minipage}{0.33\textwidth}
        \centering
        \includegraphics[width=\linewidth]{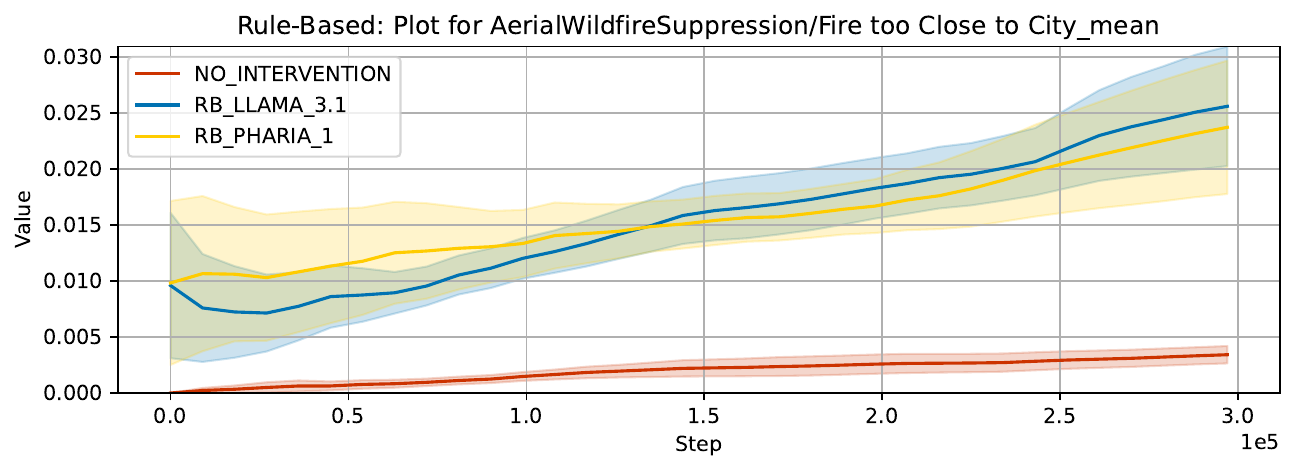}
    \end{minipage}
    \begin{minipage}{0.33\textwidth}
        \centering
        \includegraphics[width=\linewidth]{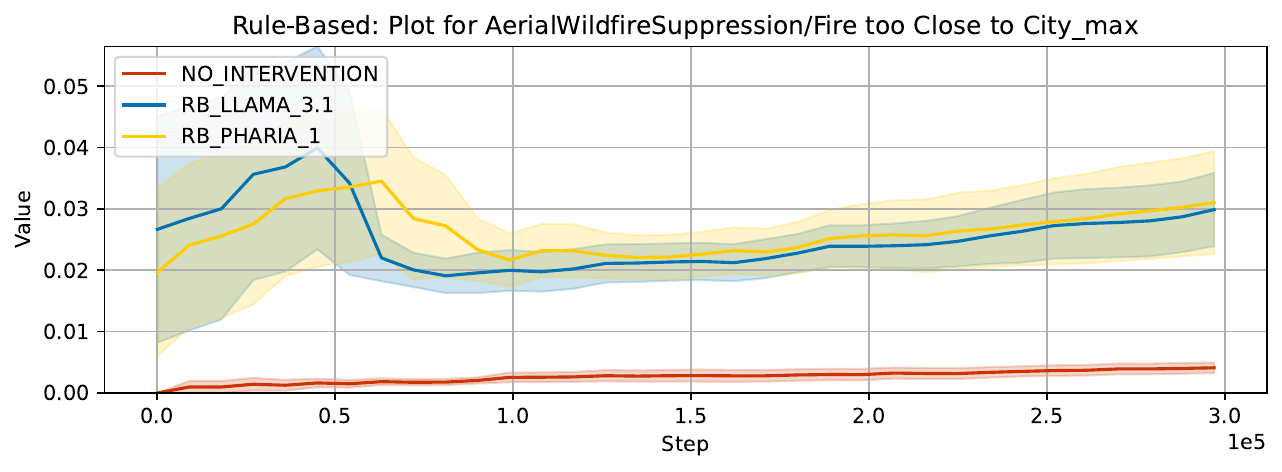}
    \end{minipage}
    \caption{Fire too Close to Village \textbf{(Rule-Based)} - No controller baseline VS Rule-Based Controller with Llama-3.1-8B Instruct: min, mean and max.}
\end{figure}

\begin{figure}[h!]
    \centering
    \label{results:fire_too_close_to_village_NL}
    \begin{minipage}{0.33\textwidth}
        \centering
        \includegraphics[width=\linewidth]{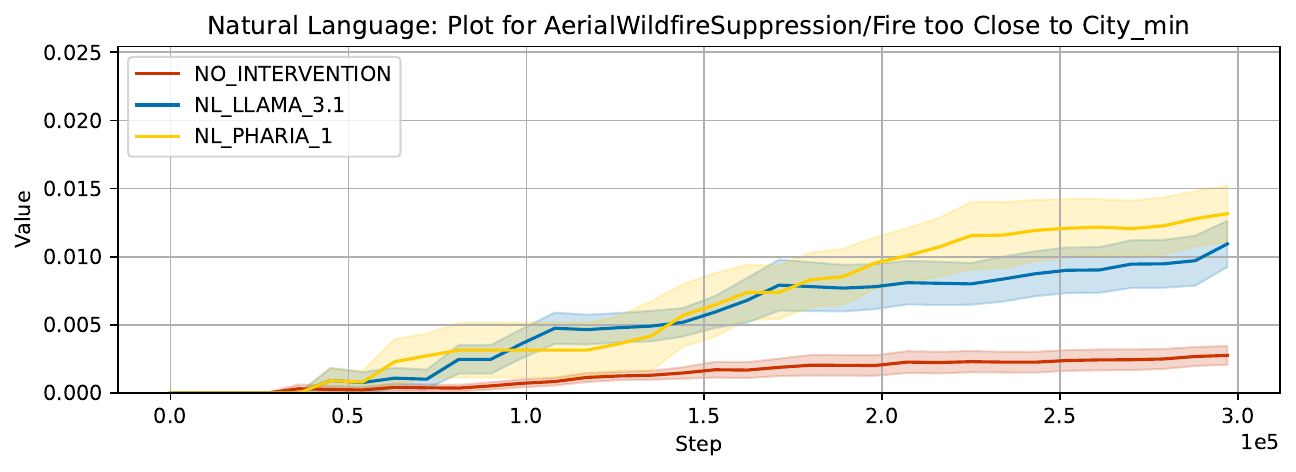}
    \end{minipage}\hfill
    \begin{minipage}{0.33\textwidth}
        \centering
        \includegraphics[width=\linewidth]{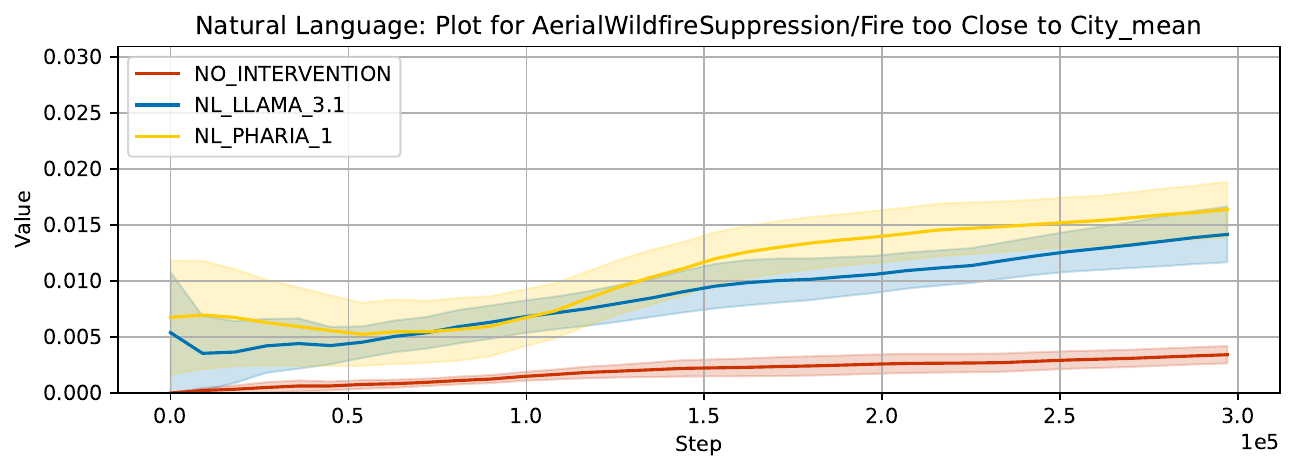}
    \end{minipage}
    \begin{minipage}{0.33\textwidth}
        \centering
        \includegraphics[width=\linewidth]{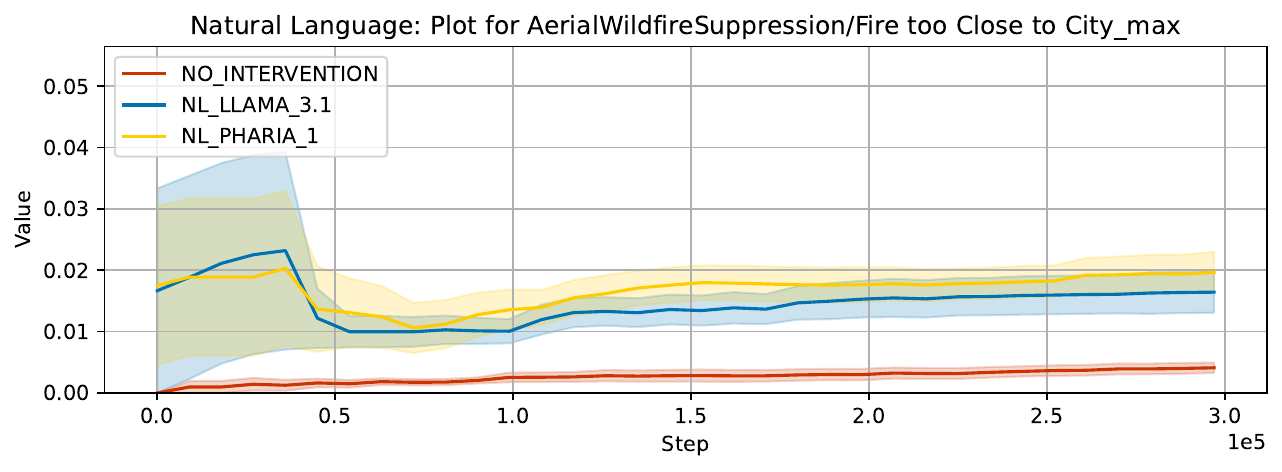}
    \end{minipage}
    \caption{Fire too Close to Village \textbf{(Natural Language)} - No controller baseline VS Natural Language Controller with Llama-3.1-8B Instruct: min, mean and max.}
\end{figure}

\begin{figure}[h!]
    \centering
    \label{results:preparing_trees_RB}
    \begin{minipage}{0.33\textwidth}
        \centering
        \includegraphics[width=\linewidth]{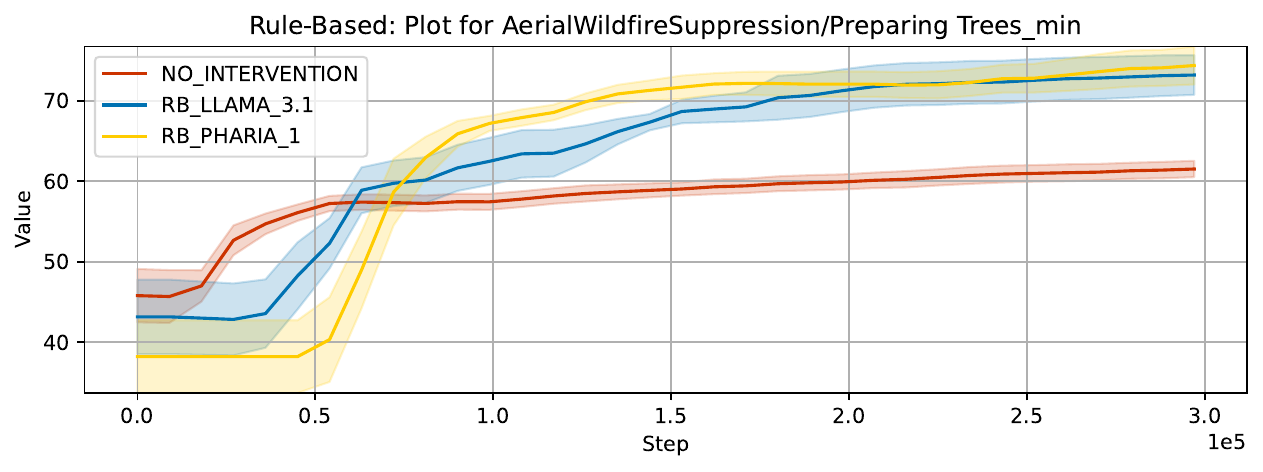}
    \end{minipage}\hfill
    \begin{minipage}{0.33\textwidth}
        \centering
        \includegraphics[width=\linewidth]{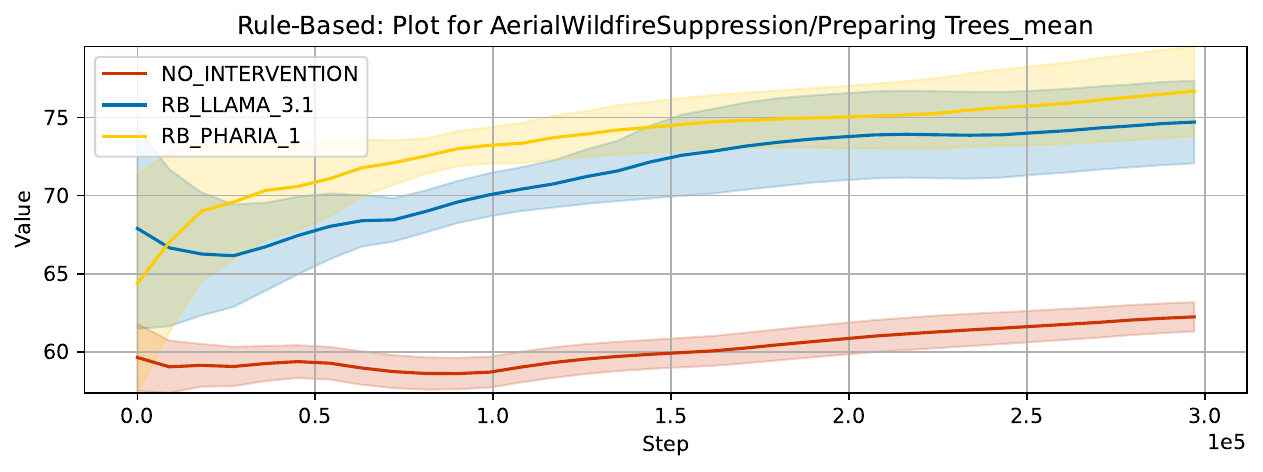}
    \end{minipage}
    \begin{minipage}{0.33\textwidth}
        \centering
        \includegraphics[width=\linewidth]{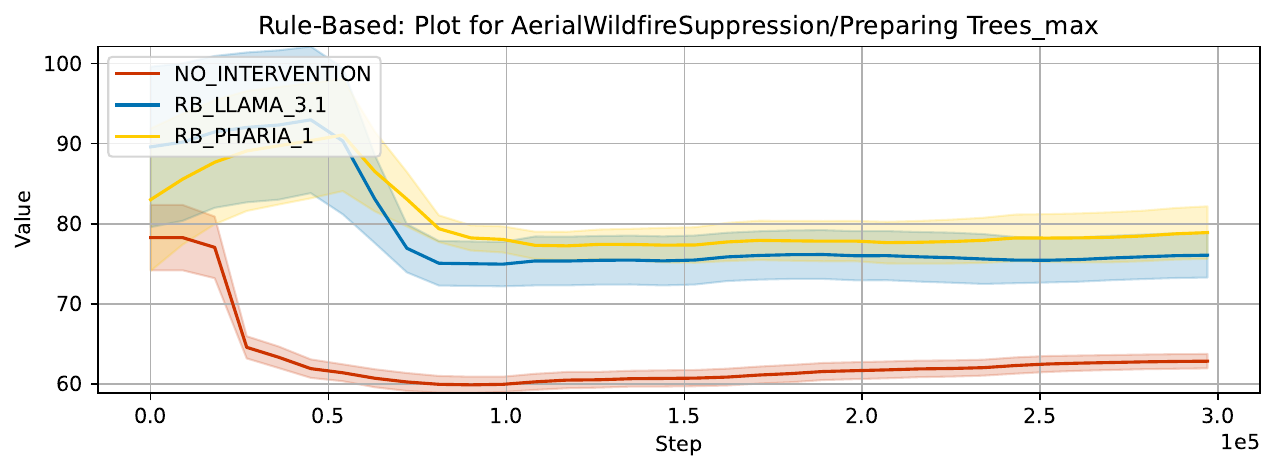}
    \end{minipage}
    \caption{Preparing Trees \textbf{(Rule-Based)} - No controller baseline VS Rule-Based Controller with Llama-3.1-8B Instruct: min, mean and max.}
\end{figure}

\begin{figure}[h!]
    \centering
    \label{results:preparing_trees_NL}
    \begin{minipage}{0.33\textwidth}
        \centering
        \includegraphics[width=\linewidth]{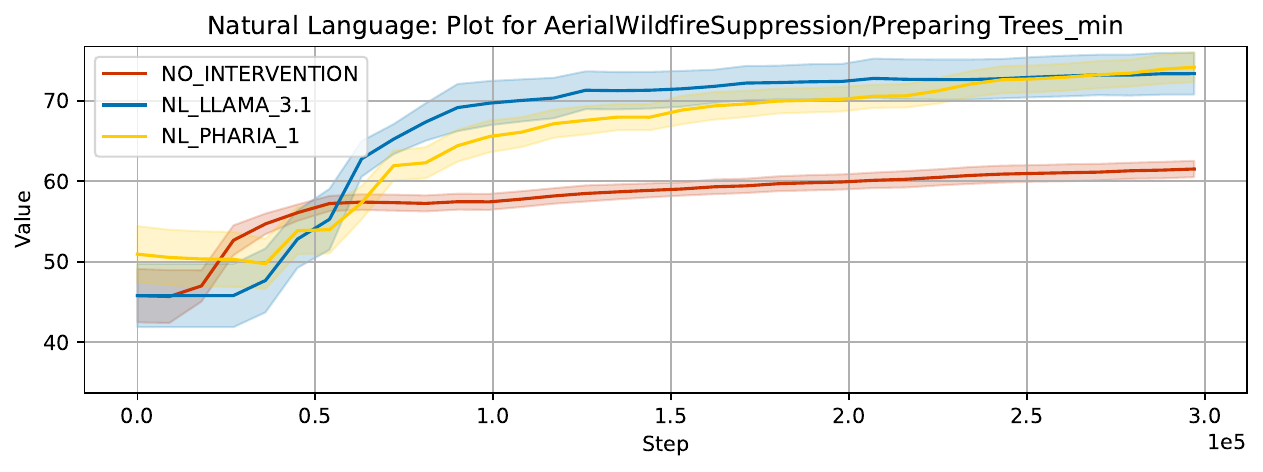}
    \end{minipage}\hfill
    \begin{minipage}{0.33\textwidth}
        \centering
        \includegraphics[width=\linewidth]{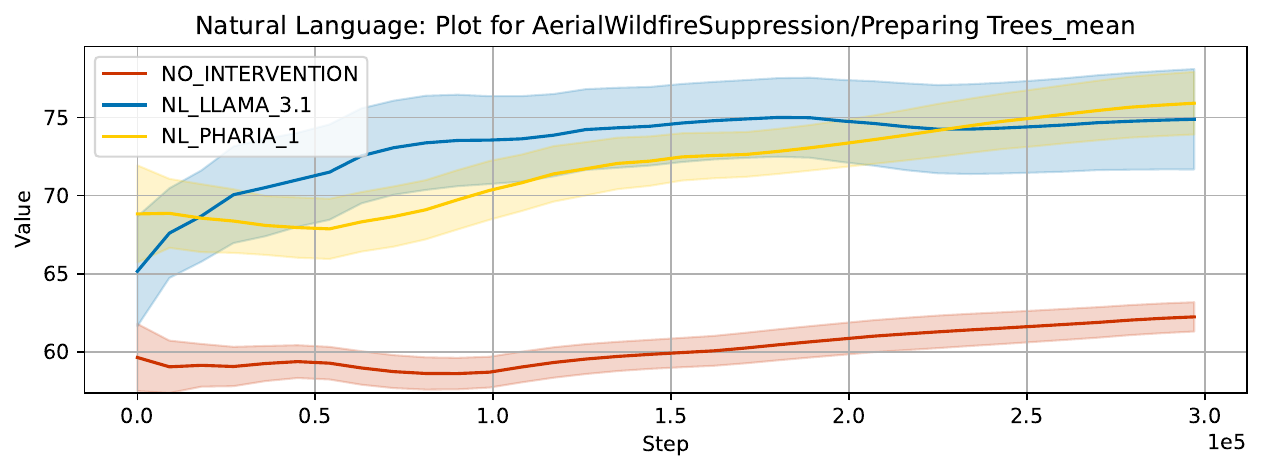}
    \end{minipage}
    \begin{minipage}{0.33\textwidth}
        \centering
        \includegraphics[width=\linewidth]{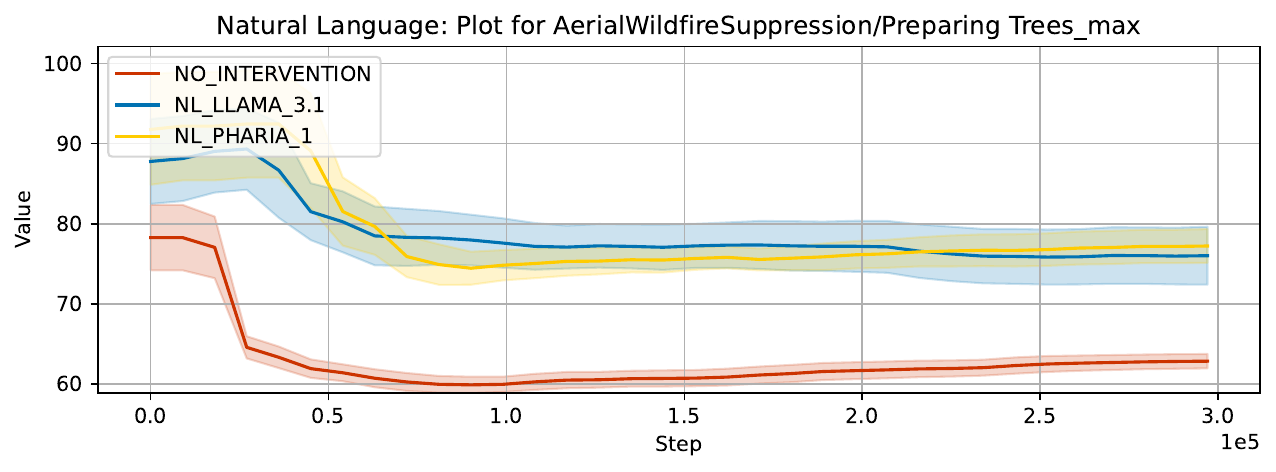}
    \end{minipage}
    \caption{Preparing Trees \textbf{(Natural Language)} - No controller baseline VS Natural Language Controller with Llama-3.1-8B Instruct: min, mean and max.}
\end{figure}

\begin{figure}[h!]
    \centering
    \label{results:preparing_trees_reward_RB}
    \begin{minipage}{0.33\textwidth}
        \centering
        \includegraphics[width=\linewidth]{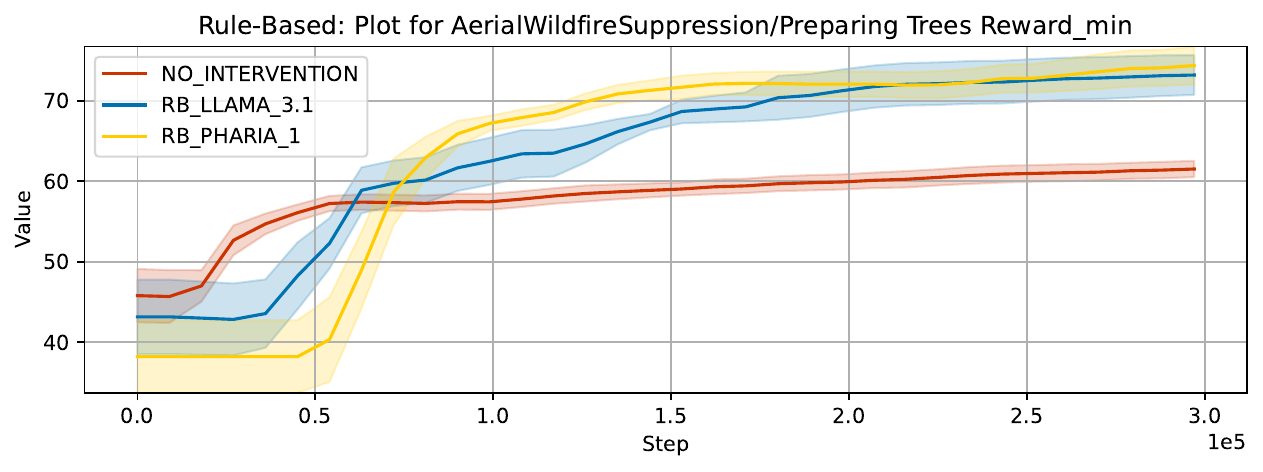}
    \end{minipage}\hfill
    \begin{minipage}{0.33\textwidth}
        \centering
        \includegraphics[width=\linewidth]{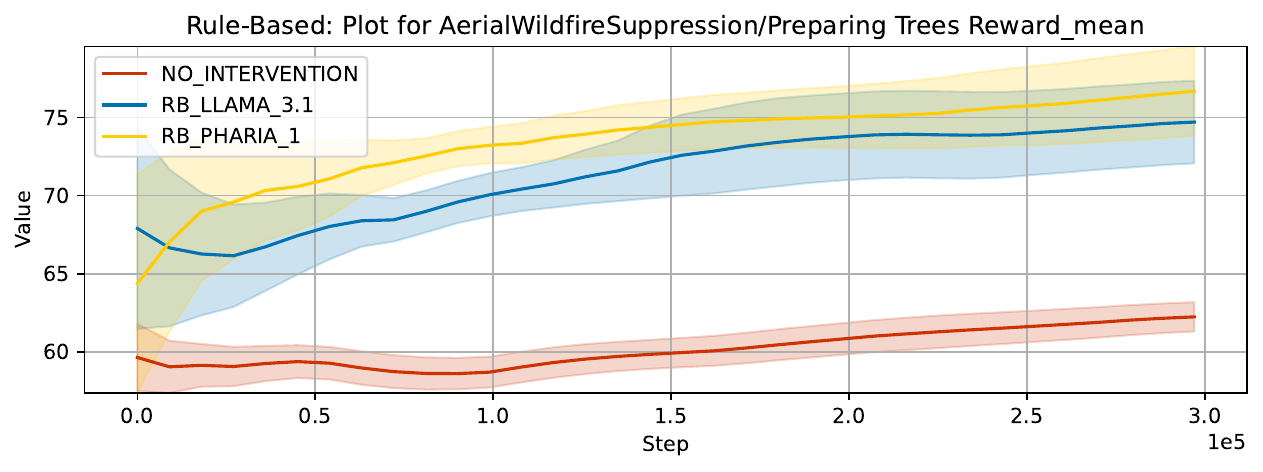}
    \end{minipage}
    \begin{minipage}{0.33\textwidth}
        \centering
        \includegraphics[width=\linewidth]{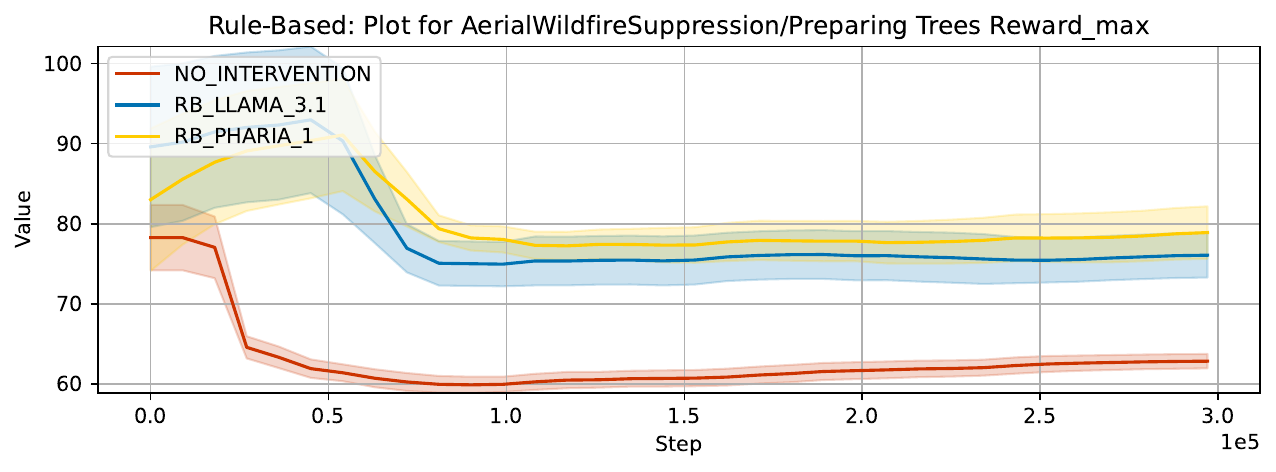}
    \end{minipage}
    \caption{Preparing Trees Reward \textbf{(Rule-Based)} - No controller baseline VS Rule-Based Controller with Llama-3.1-8B Instruct: min, mean and max.}
\end{figure}

\begin{figure}[h!]
    \centering
    \label{results:preparing_trees_reward_NL}
    \begin{minipage}{0.33\textwidth}
        \centering
        \includegraphics[width=\linewidth]{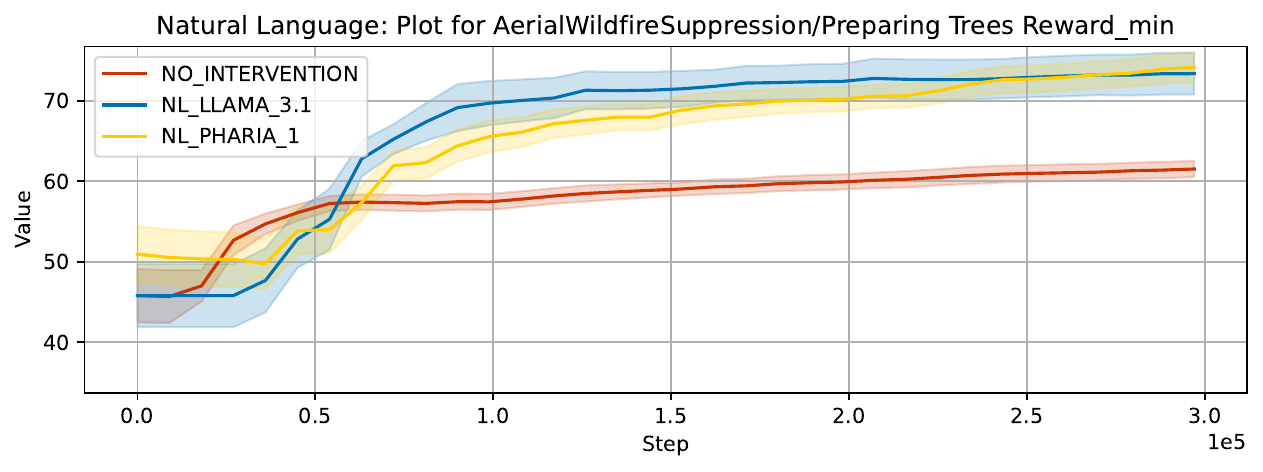}
    \end{minipage}\hfill
    \begin{minipage}{0.33\textwidth}
        \centering
        \includegraphics[width=\linewidth]{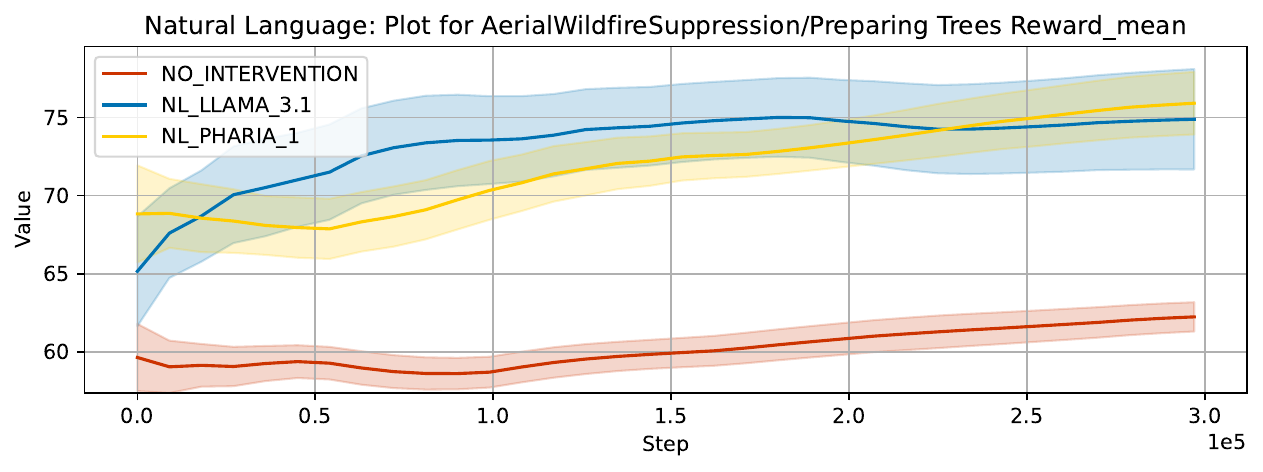}
    \end{minipage}
    \begin{minipage}{0.33\textwidth}
        \centering
        \includegraphics[width=\linewidth]{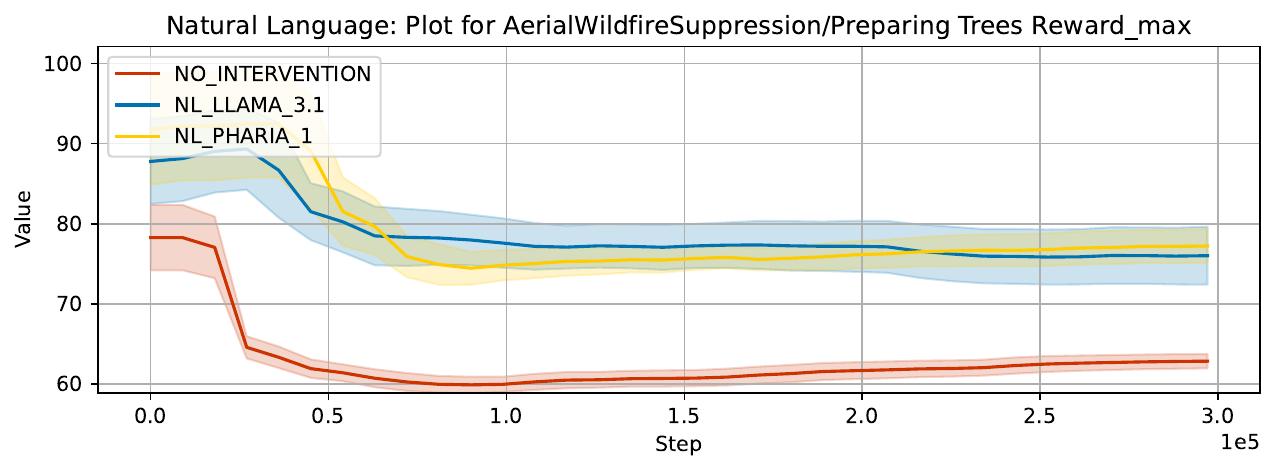}
    \end{minipage}
    \caption{Preparing Trees Reward \textbf{(Natural Language)} - No controller baseline VS Natural Language Controller with Llama-3.1-8B Instruct: min, mean and max.}
\end{figure}

\begin{figure}[h!]
    \centering
    \label{results:time_step_count_RB}
    \begin{minipage}{0.33\textwidth}
        \centering
        \includegraphics[width=\linewidth]{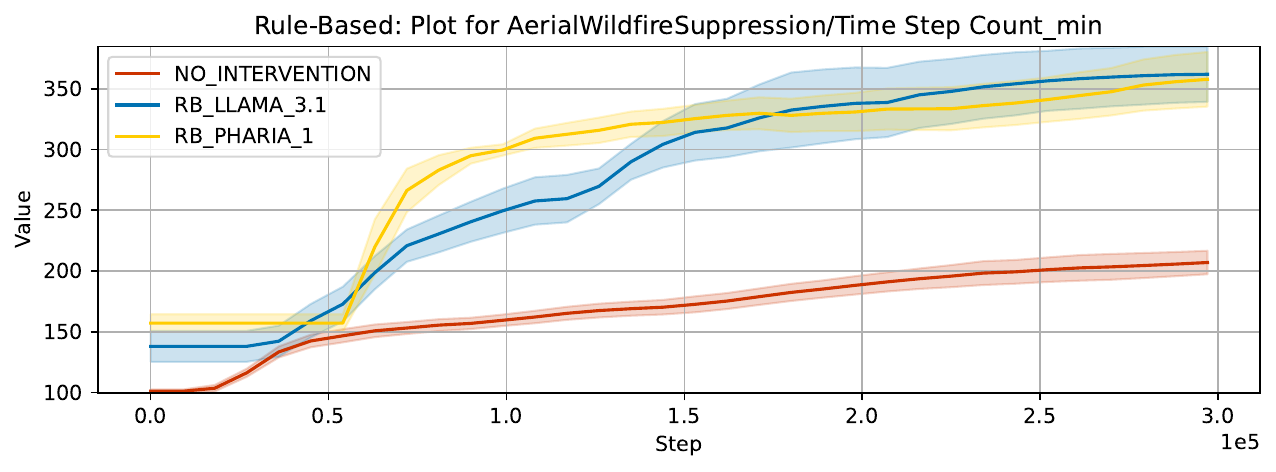}
    \end{minipage}\hfill
    \begin{minipage}{0.33\textwidth}
        \centering
        \includegraphics[width=\linewidth]{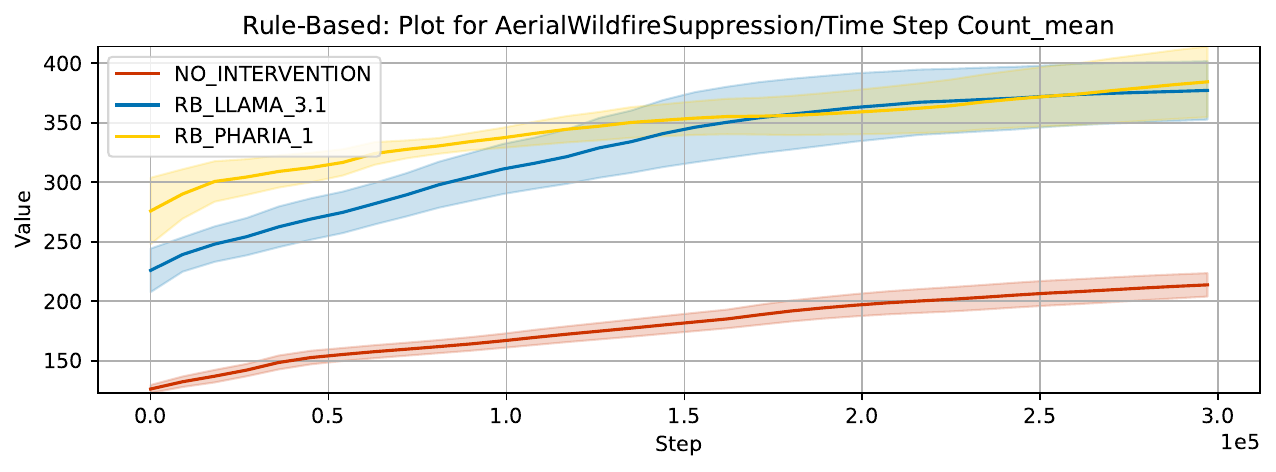}
    \end{minipage}
    \begin{minipage}{0.33\textwidth}
        \centering
        \includegraphics[width=\linewidth]{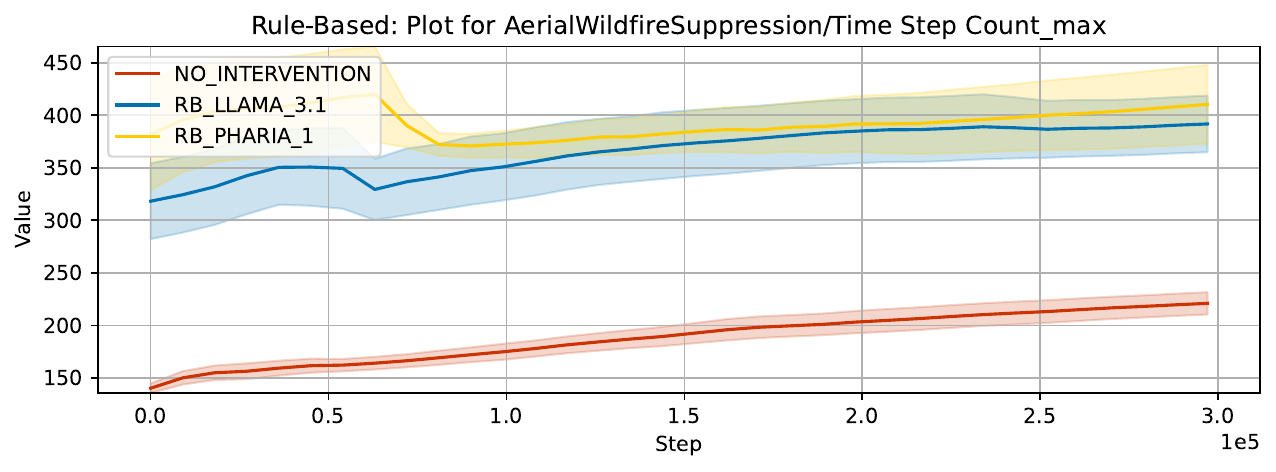}
    \end{minipage}
    \caption{Time Step Count \textbf{(Rule-Based)} - No controller baseline VS Rule-Based Controller with Llama-3.1-8B Instruct: min, mean and max.}
\end{figure}

\begin{figure}[h!]
    \centering
    \label{results:time_step_count_NL}
    \begin{minipage}{0.33\textwidth}
        \centering
        \includegraphics[width=\linewidth]{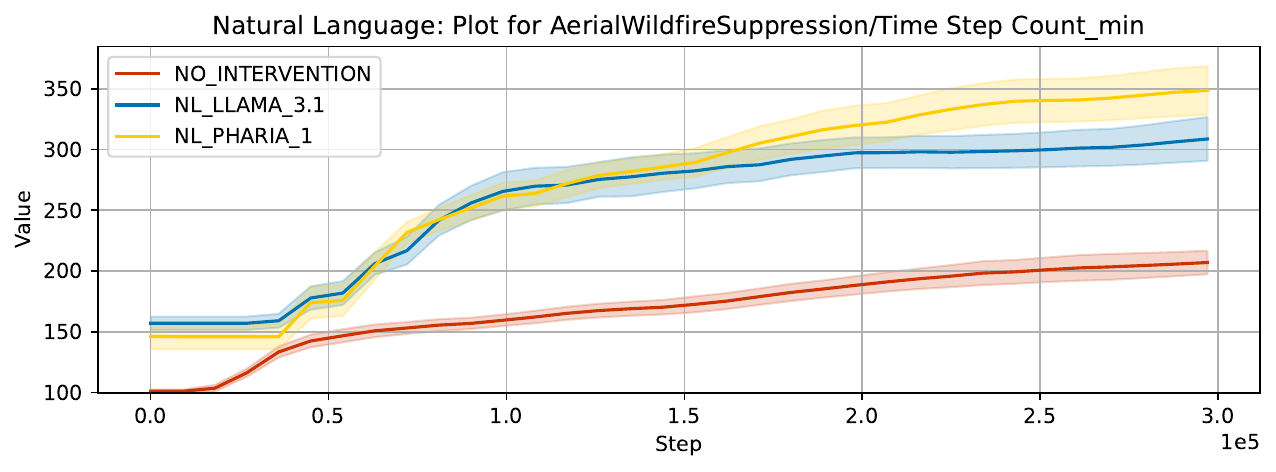}
    \end{minipage}\hfill
    \begin{minipage}{0.33\textwidth}
        \centering
        \includegraphics[width=\linewidth]{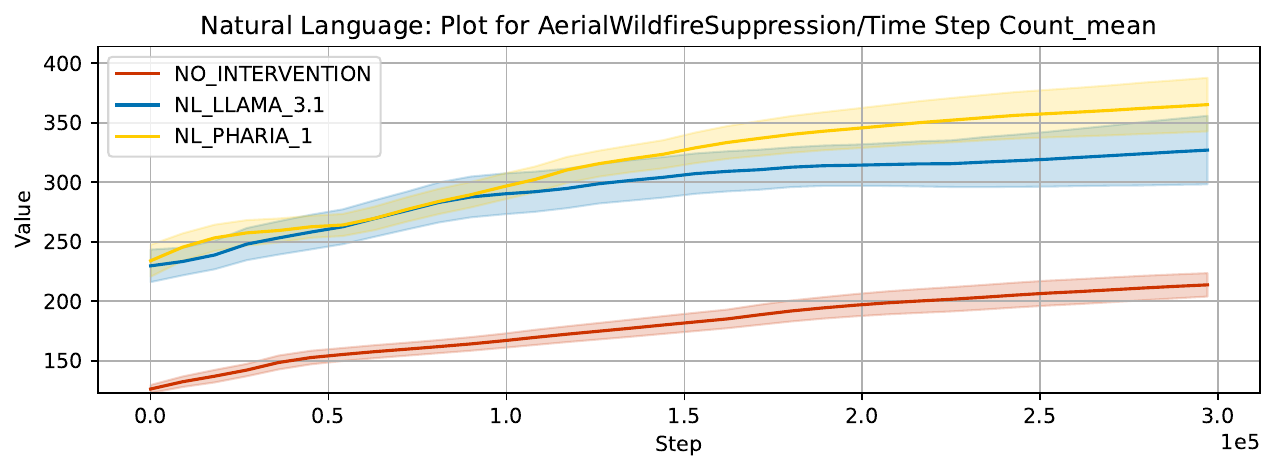}
    \end{minipage}
    \begin{minipage}{0.33\textwidth}
        \centering
        \includegraphics[width=\linewidth]{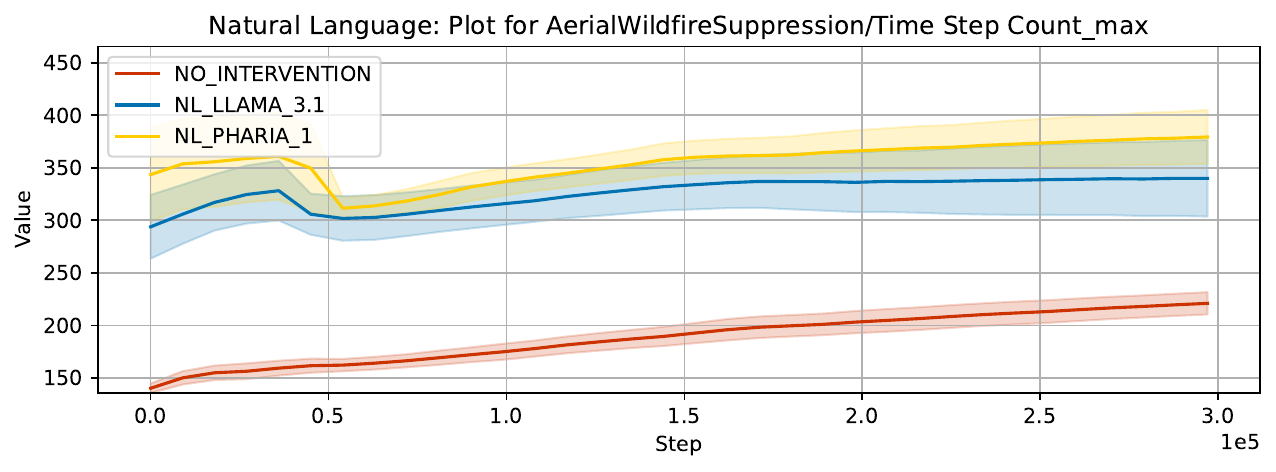}
    \end{minipage}
    \caption{Time Step Count \textbf{(Natural Language)} - No controller baseline VS Natural Language Controller with Llama-3.1-8B Instruct: min, mean and max.}
\end{figure}

\begin{figure}[h!]
    \centering
    \label{results:water_drop_count_RB}
    \begin{minipage}{0.33\textwidth}
        \centering
        \includegraphics[width=\linewidth]{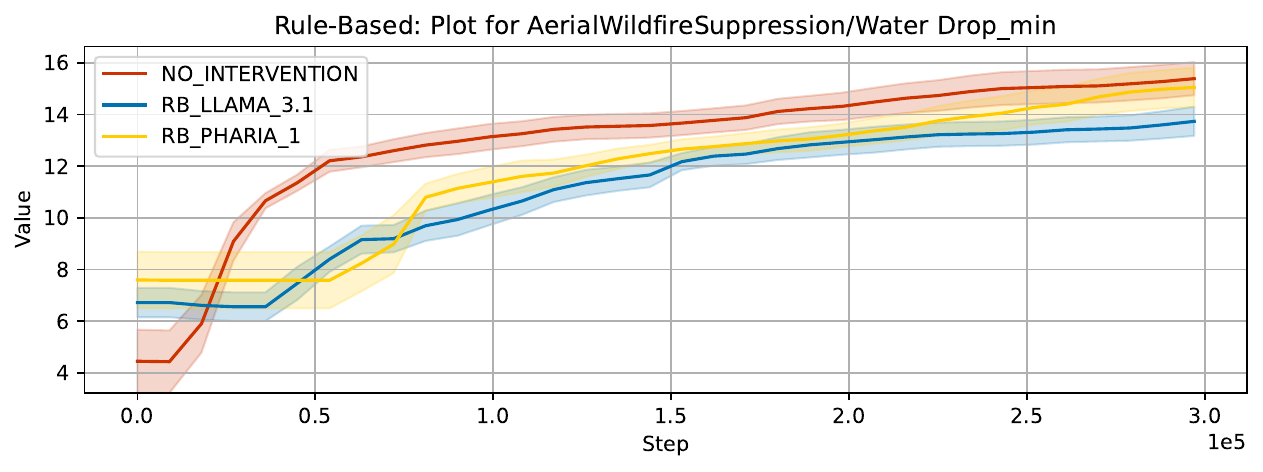}
    \end{minipage}\hfill
    \begin{minipage}{0.33\textwidth}
        \centering
        \includegraphics[width=\linewidth]{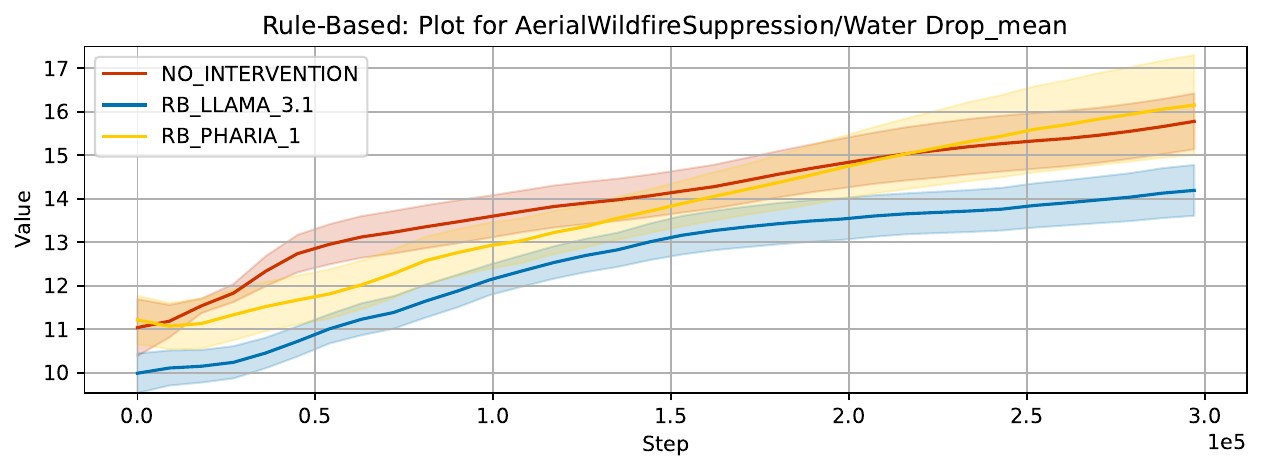}
    \end{minipage}
    \begin{minipage}{0.33\textwidth}
        \centering
        \includegraphics[width=\linewidth]{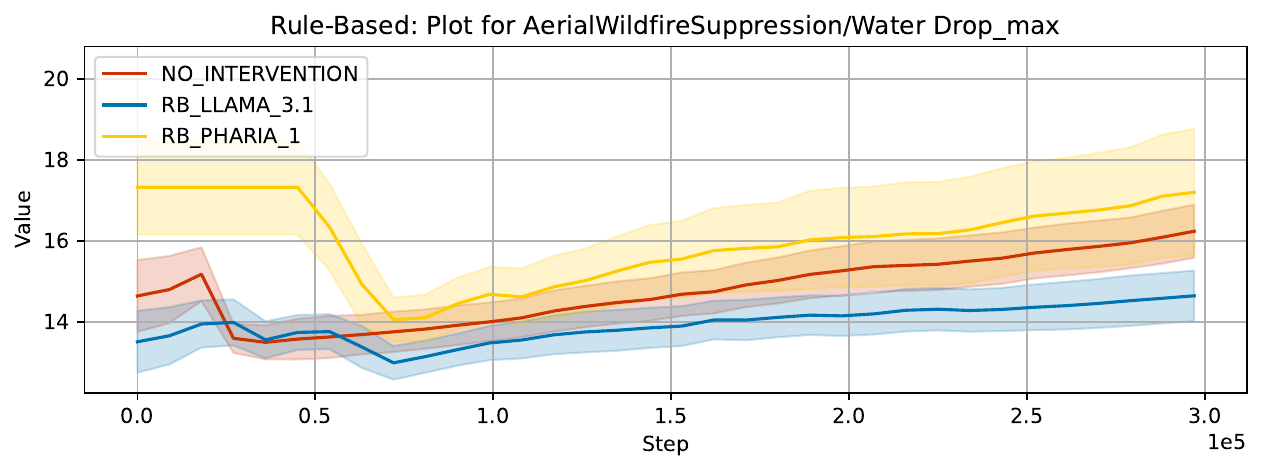}
    \end{minipage}
    \caption{Water Drop Count \textbf{(Rule-Based)} - No controller baseline VS Rule-Based Controller with Llama-3.1-8B Instruct: min, mean and max.}
\end{figure}

\begin{figure}[h!]
    \centering
    \label{results:water_drop_count_NL}
    \begin{minipage}{0.33\textwidth}
        \centering
        \includegraphics[width=\linewidth]{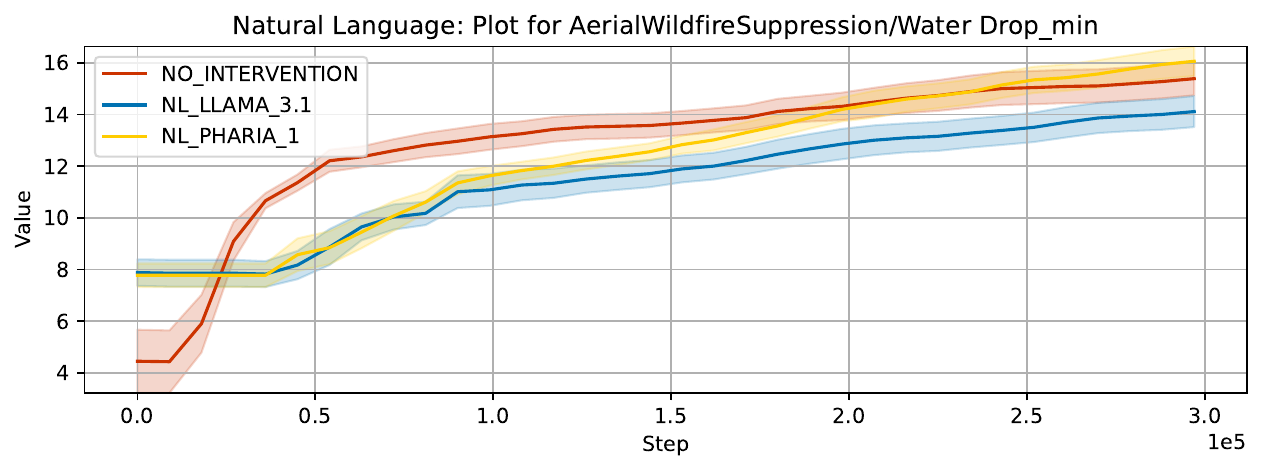}
    \end{minipage}\hfill
    \begin{minipage}{0.33\textwidth}
        \centering
        \includegraphics[width=\linewidth]{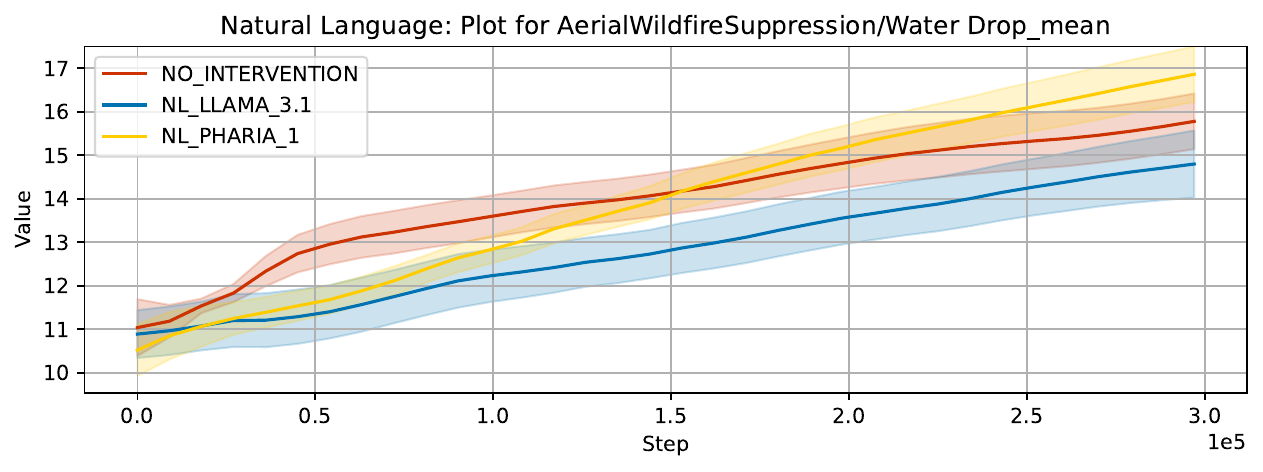}
    \end{minipage}
    \begin{minipage}{0.33\textwidth}
        \centering
        \includegraphics[width=\linewidth]{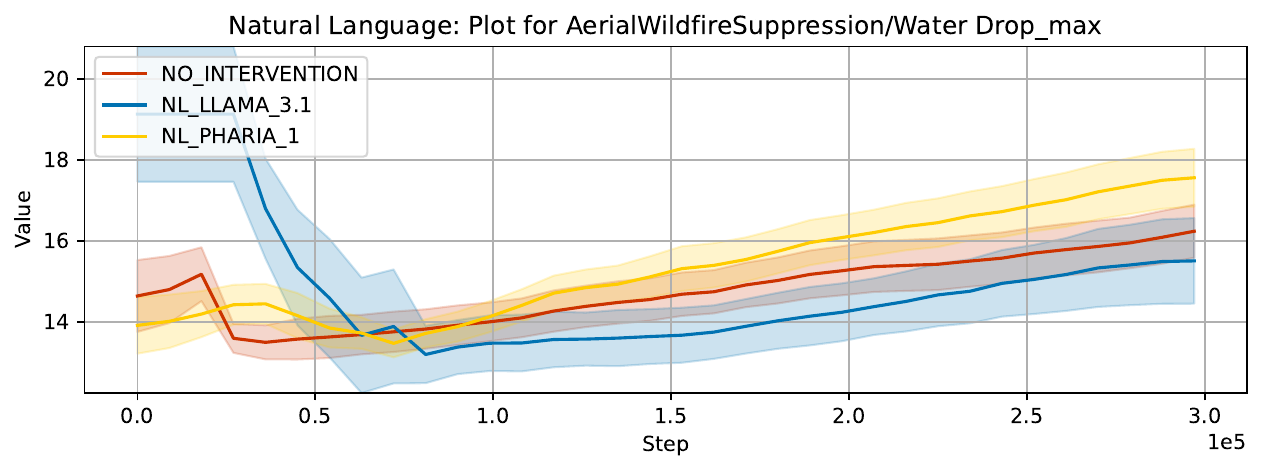}
    \end{minipage}
    \caption{Water Drop Count \textbf{(Natural Language)} - No controller baseline VS Natural Language Controller with Llama-3.1-8B Instruct: min, mean and max.}
\end{figure}

\begin{figure}[h!]
    \centering
    \label{results:water_pickup_count_RB}
    \begin{minipage}{0.33\textwidth}
        \centering
        \includegraphics[width=\linewidth]{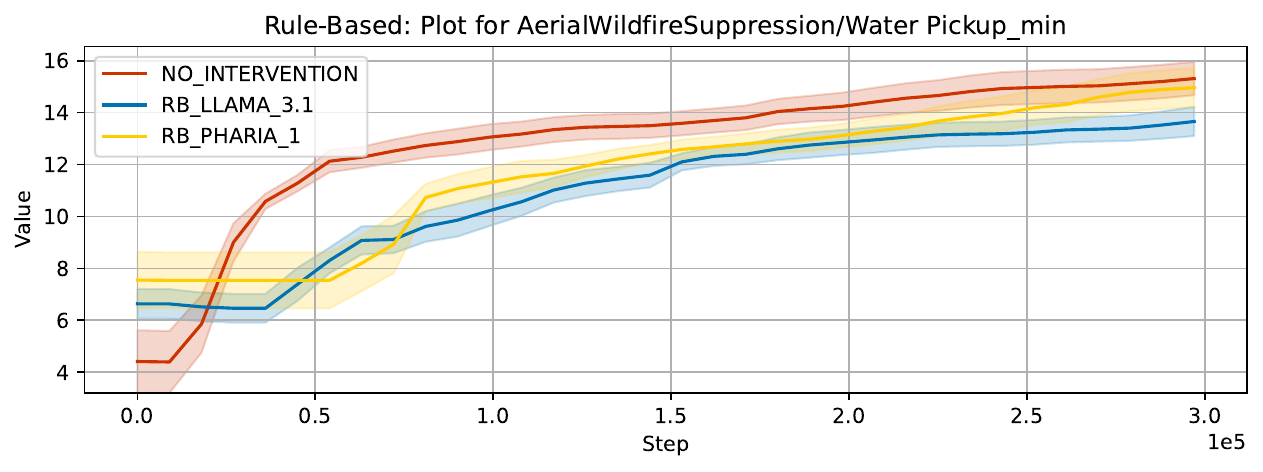}
    \end{minipage}\hfill
    \begin{minipage}{0.33\textwidth}
        \centering
        \includegraphics[width=\linewidth]{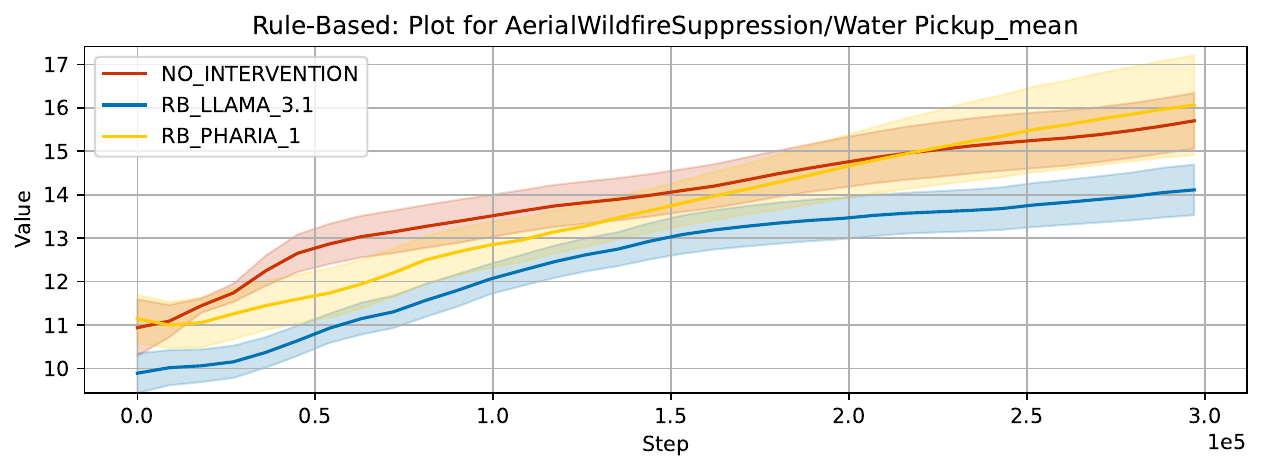}
    \end{minipage}
    \begin{minipage}{0.33\textwidth}
        \centering
        \includegraphics[width=\linewidth]{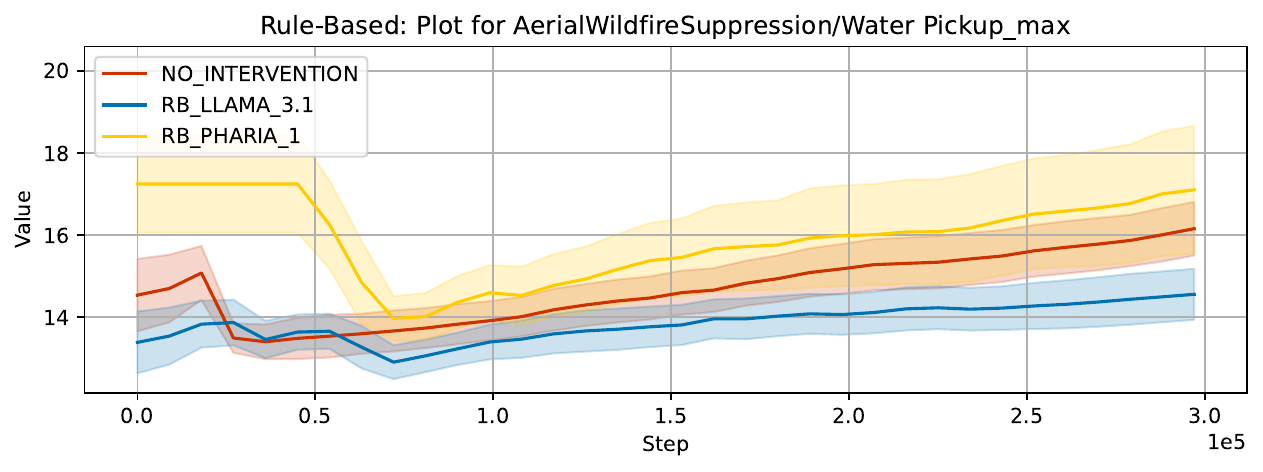}
    \end{minipage}
    \caption{Water Pickup Count \textbf{(Rule-Based)} - No controller baseline VS Rule-Based Controller with Llama-3.1-8B Instruct: min, mean and max.}
\end{figure}

\begin{figure}[h!]
    \centering
    \label{results:water_pickup_count_NL}
    \begin{minipage}{0.33\textwidth}
        \centering
        \includegraphics[width=\linewidth]{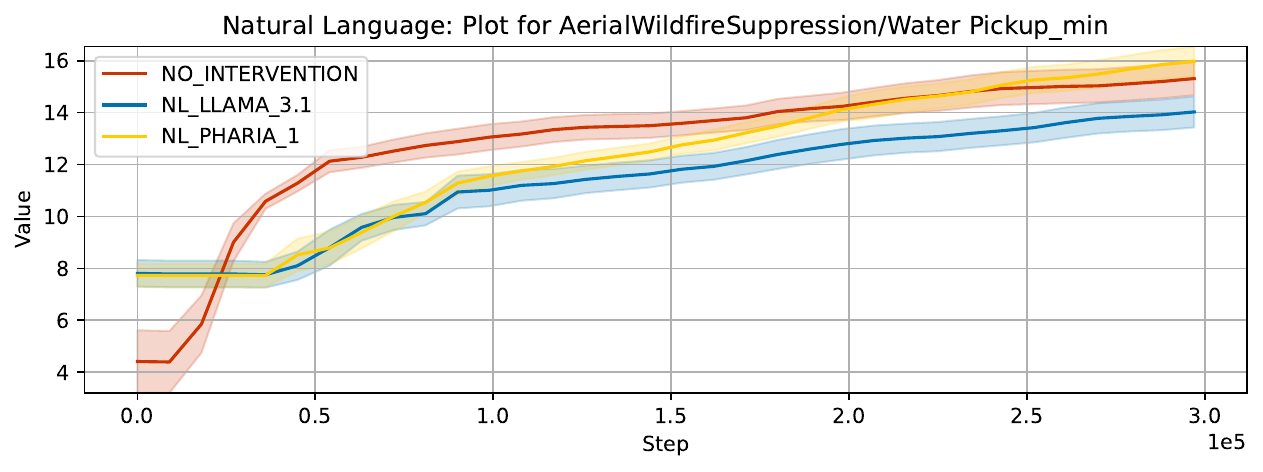}
    \end{minipage}\hfill
    \begin{minipage}{0.33\textwidth}
        \centering
        \includegraphics[width=\linewidth]{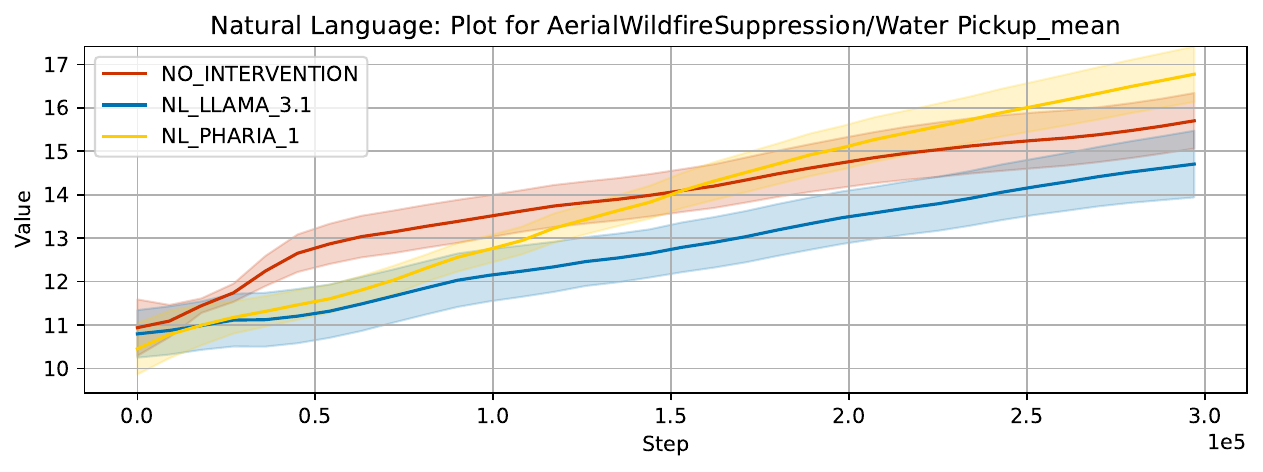}
    \end{minipage}
    \begin{minipage}{0.33\textwidth}
        \centering
        \includegraphics[width=\linewidth]{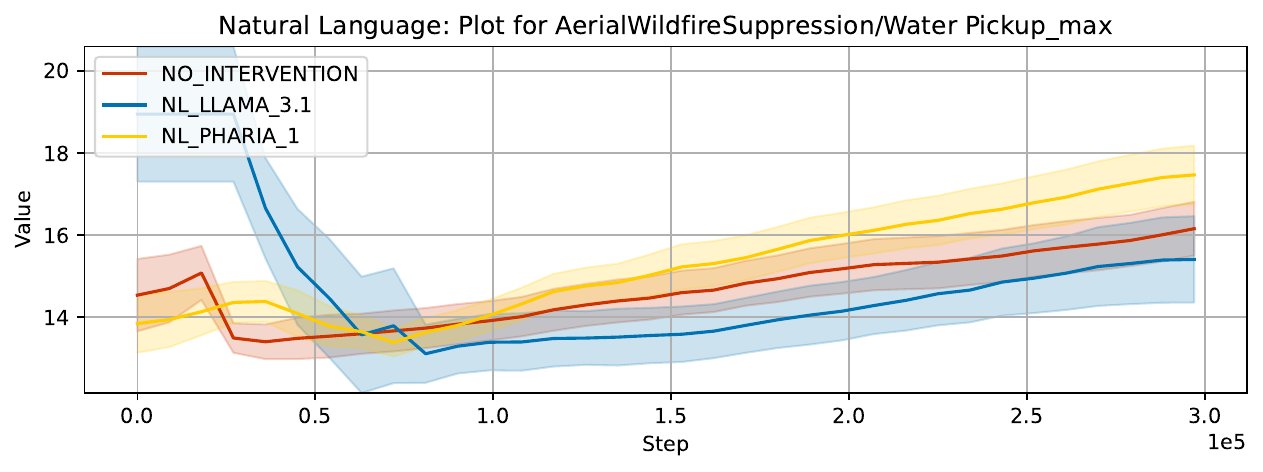}
    \end{minipage}
    \caption{Water Pickup Count \textbf{(Natural Language)} - No controller baseline VS Natural Language Controller with Llama-3.1-8B Instruct: min, mean and max.}
\end{figure}

\begin{figure}[h!]
    \centering
    \label{results:episode_return_RB}
    \begin{minipage}{0.33\textwidth}
        \centering
        \includegraphics[width=\linewidth]{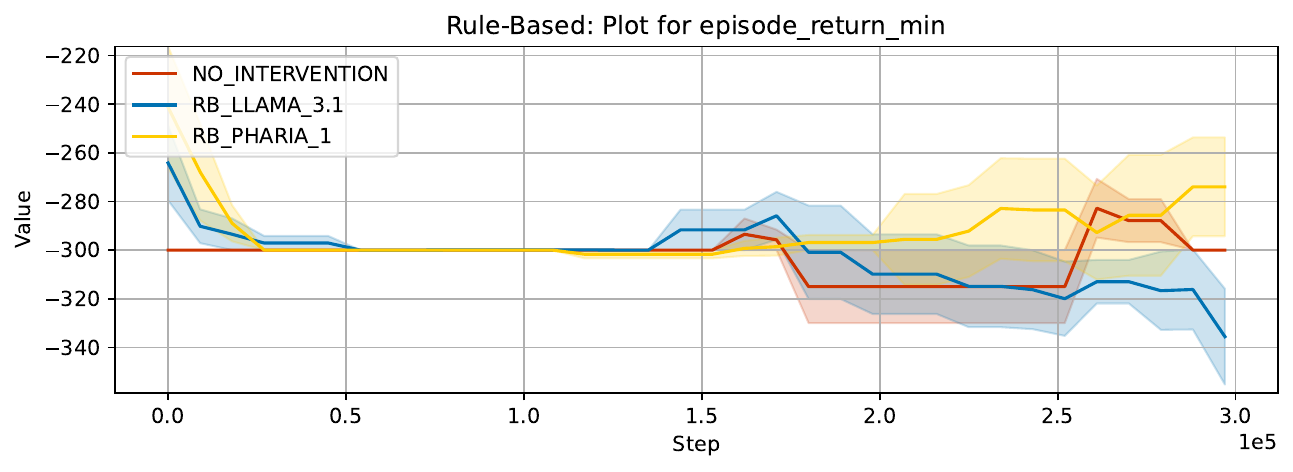}
    \end{minipage}\hfill
    \begin{minipage}{0.33\textwidth}
        \centering
        \includegraphics[width=\linewidth]{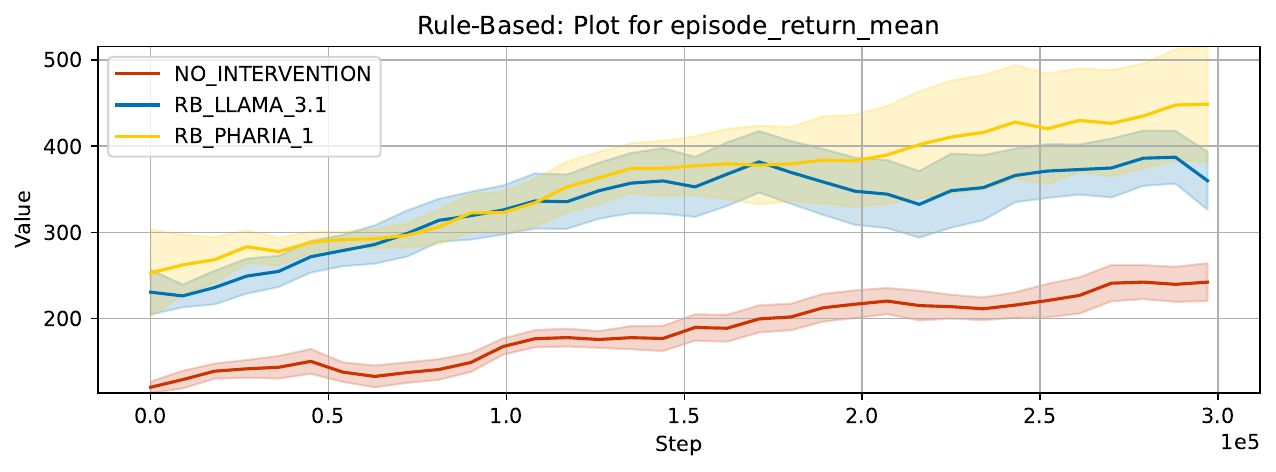}
    \end{minipage}
    \begin{minipage}{0.33\textwidth}
        \centering
        \includegraphics[width=\linewidth]{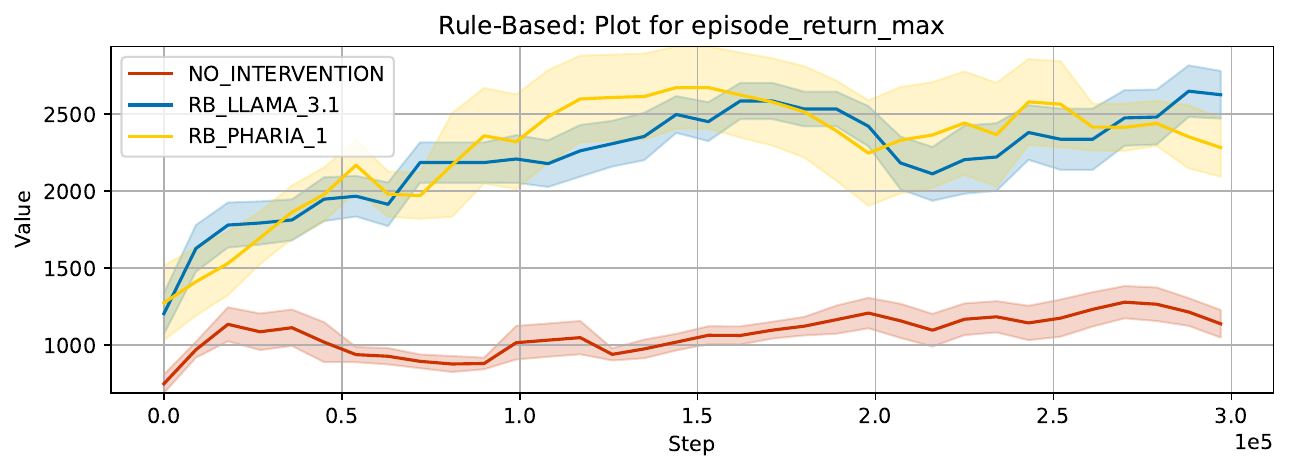}
    \end{minipage}
    \caption{Episode Return \textbf{(Rule-Based)} - No controller baseline VS Rule-Based Controller with Llama-3.1-8B Instruct: min, mean and max.}
\end{figure}

\begin{figure}[h!]
    \centering
    \label{results:episode_return_NL}
    \begin{minipage}{0.33\textwidth}
        \centering
        \includegraphics[width=\linewidth]{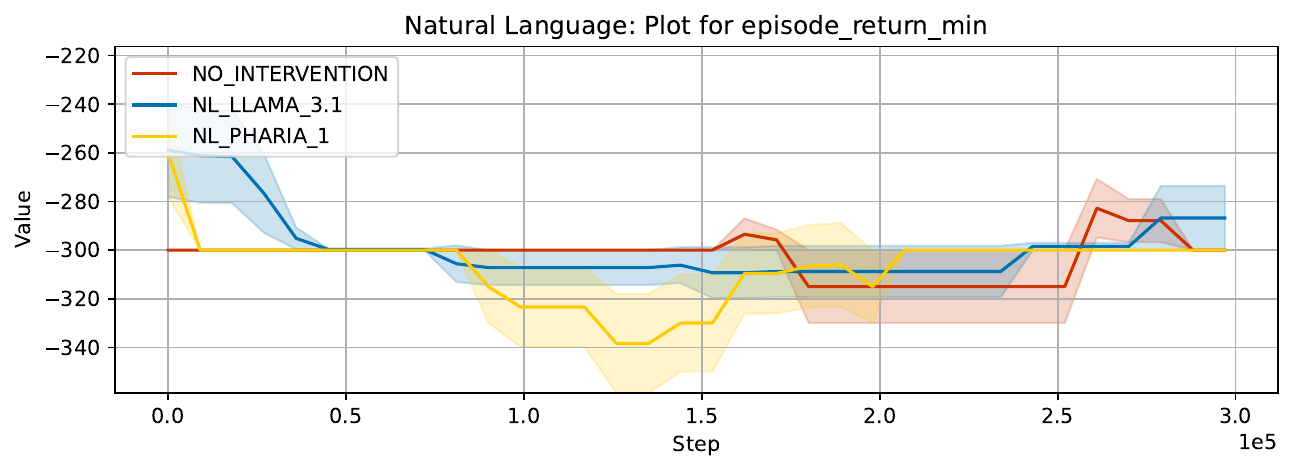}
    \end{minipage}\hfill
    \begin{minipage}{0.33\textwidth}
        \centering
        \includegraphics[width=\linewidth]{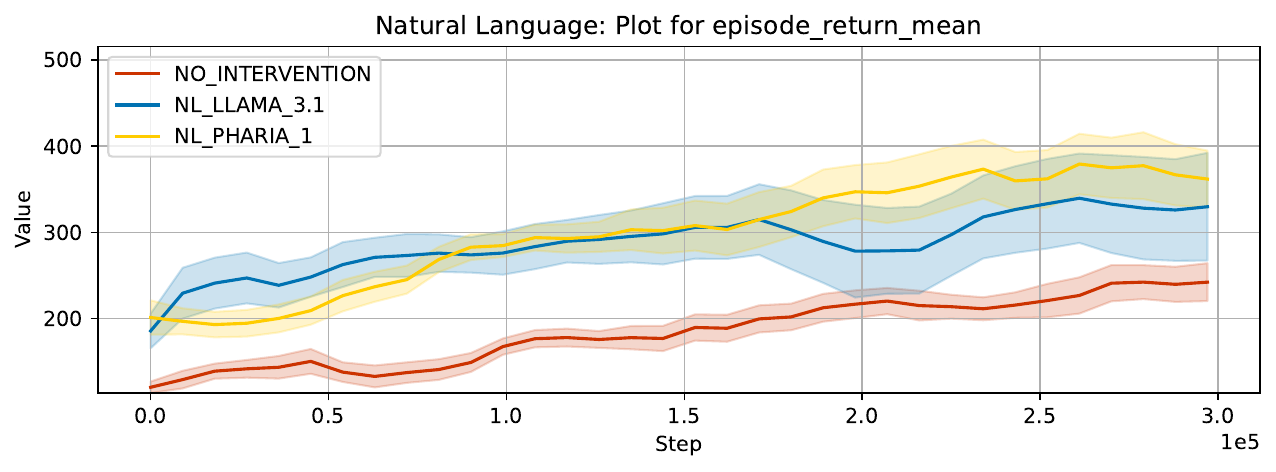}
    \end{minipage}
    \begin{minipage}{0.33\textwidth}
        \centering
        \includegraphics[width=\linewidth]{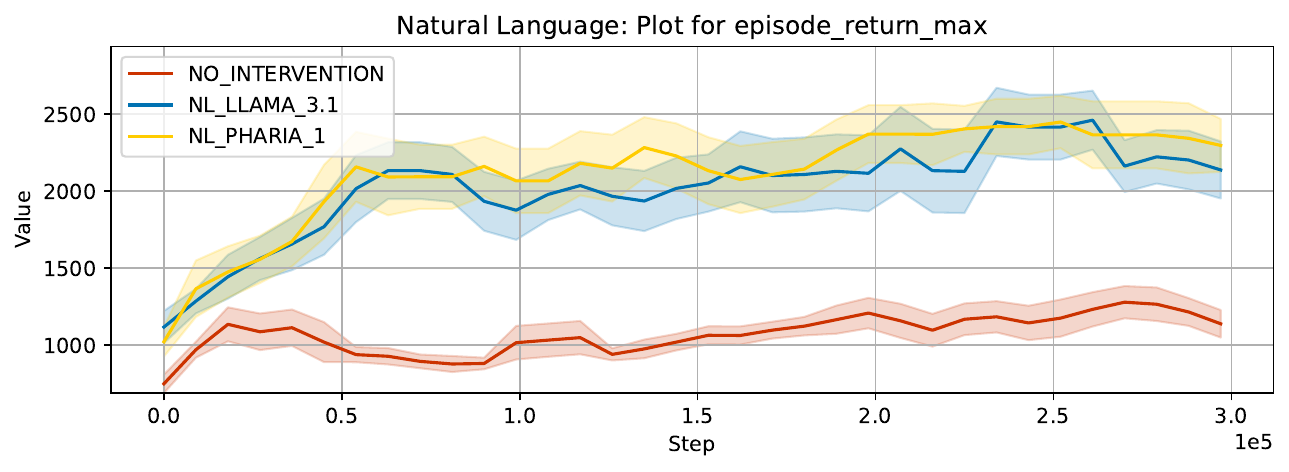}
    \end{minipage}
    \caption{Episode Return \textbf{(Natural Language)} - No controller baseline VS Natural Language Controller with Llama-3.1-8B Instruct: min, mean and max.}
\end{figure}

\begin{figure}[h!]
    \centering
    \label{results:episode_reward_RB}
    \begin{minipage}{0.33\textwidth}
        \centering
        \includegraphics[width=\linewidth]{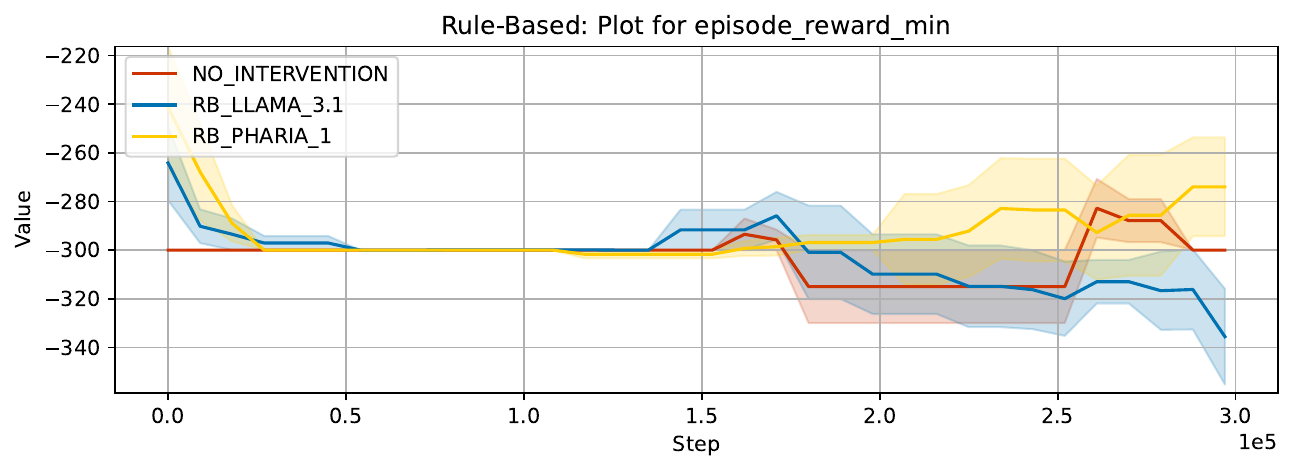}
    \end{minipage}\hfill
    \begin{minipage}{0.33\textwidth}
        \centering
        \includegraphics[width=\linewidth]{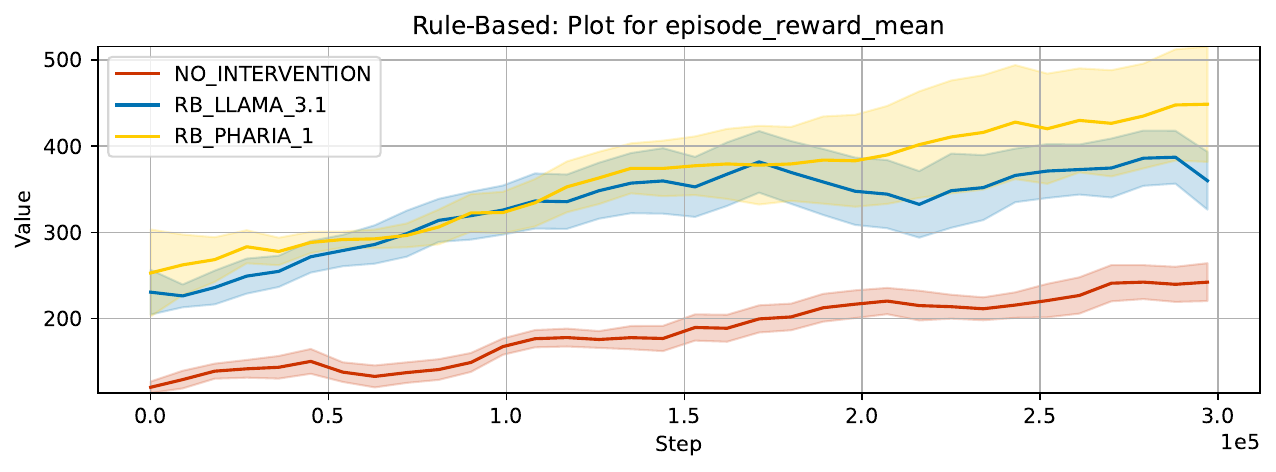}
    \end{minipage}
    \begin{minipage}{0.33\textwidth}
        \centering
        \includegraphics[width=\linewidth]{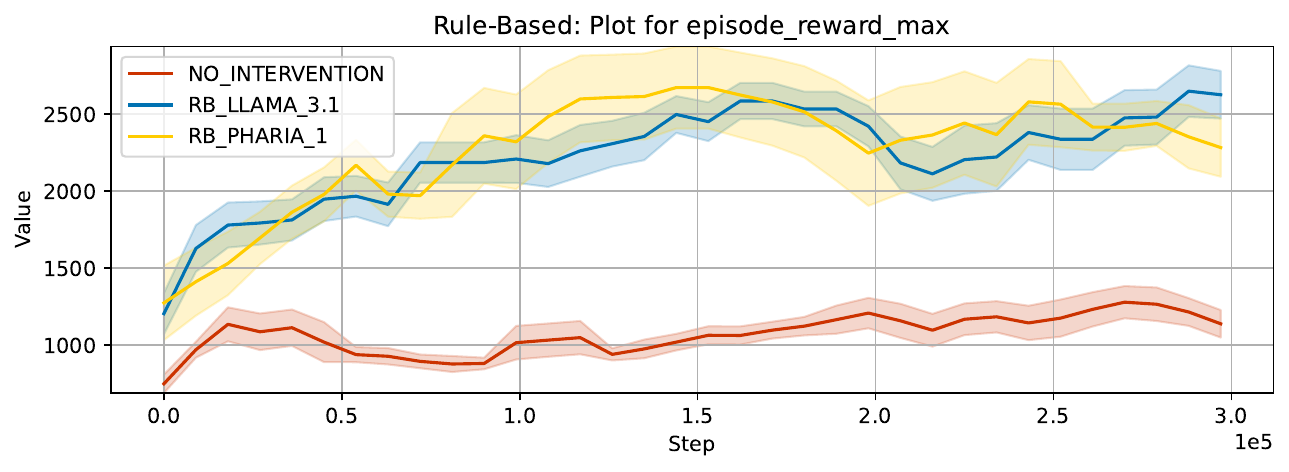}
    \end{minipage}
    \caption{Episode Reward \textbf{(Rule-Based)} - No controller baseline VS Rule-Based Controller with Llama-3.1-8B Instruct: min, mean and max.}
\end{figure}

\begin{figure}[h!]
    \centering
    \label{results:episode_reward_NL}
    \begin{minipage}{0.33\textwidth}
        \centering
        \includegraphics[width=\linewidth]{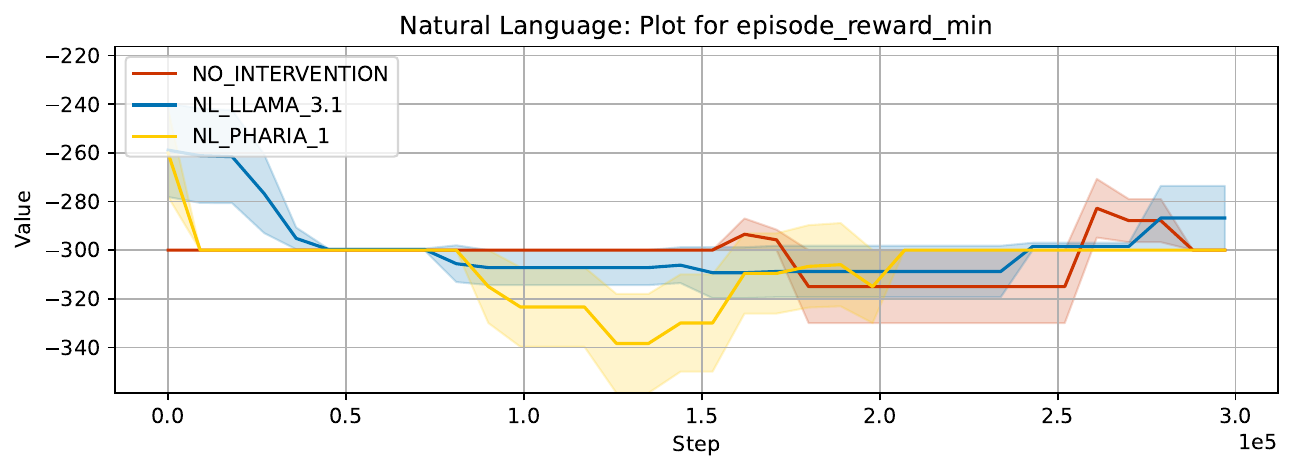}
    \end{minipage}\hfill
    \begin{minipage}{0.33\textwidth}
        \centering
        \includegraphics[width=\linewidth]{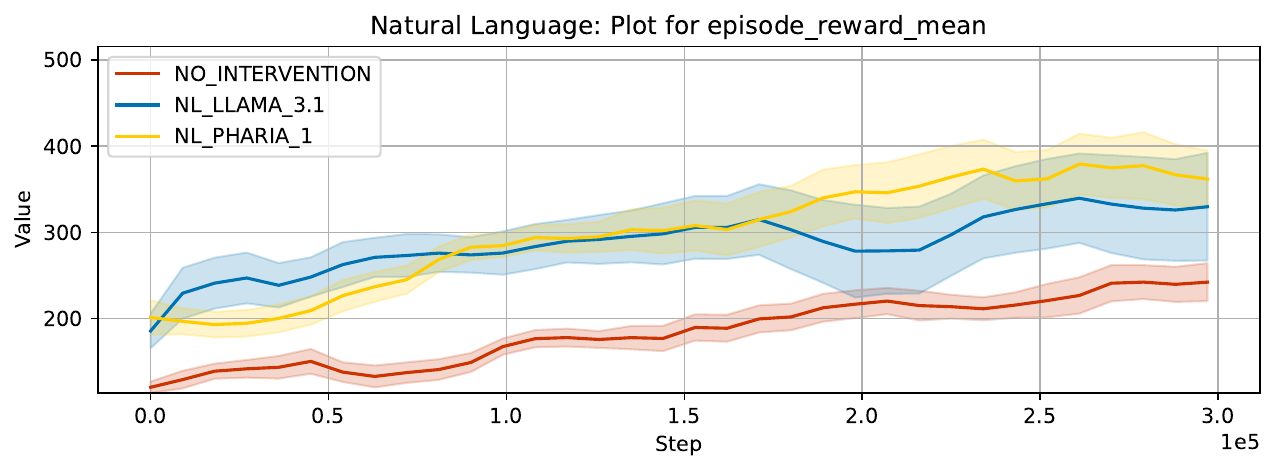}
    \end{minipage}
    \begin{minipage}{0.33\textwidth}
        \centering
        \includegraphics[width=\linewidth]{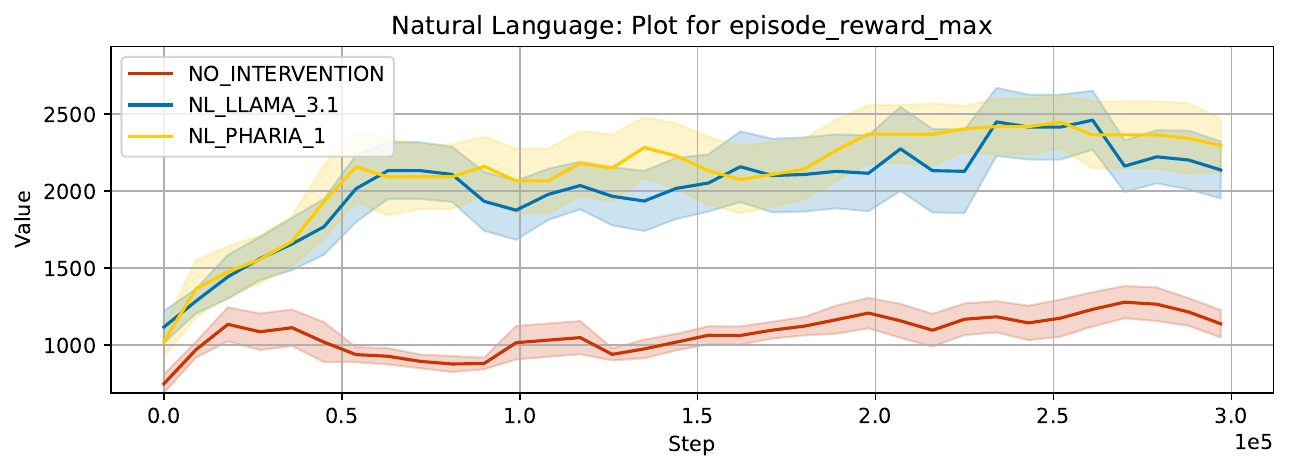}
    \end{minipage}
    \caption{Episode Reward \textbf{(Natural Language)} - No controller baseline VS Natural Language Controller with Llama-3.1-8B Instruct: min, mean and max.}
\end{figure}

\begin{figure}[h!]
    \centering
    \label{results:task_count_RB}
    \begin{minipage}{0.33\textwidth}
        \centering
        \includegraphics[width=\linewidth]{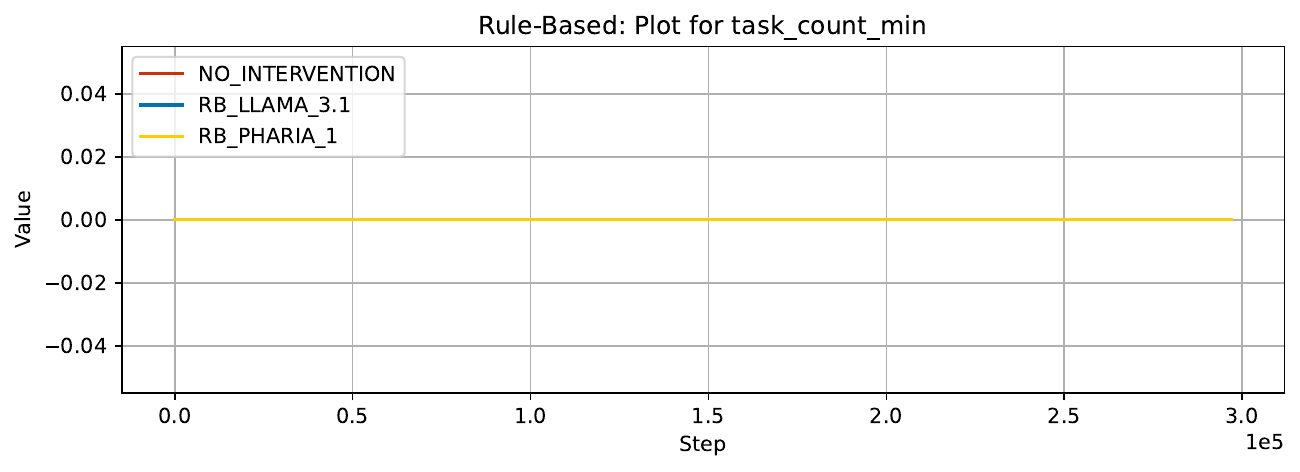}
    \end{minipage}\hfill
    \begin{minipage}{0.33\textwidth}
        \centering
        \includegraphics[width=\linewidth]{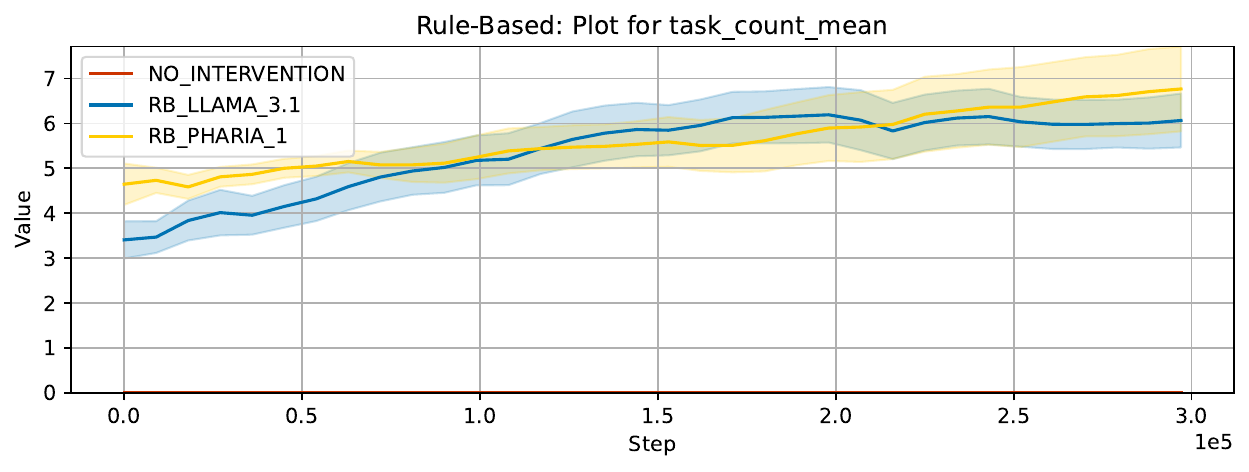}
    \end{minipage}
    \begin{minipage}{0.33\textwidth}
        \centering
        \includegraphics[width=\linewidth]{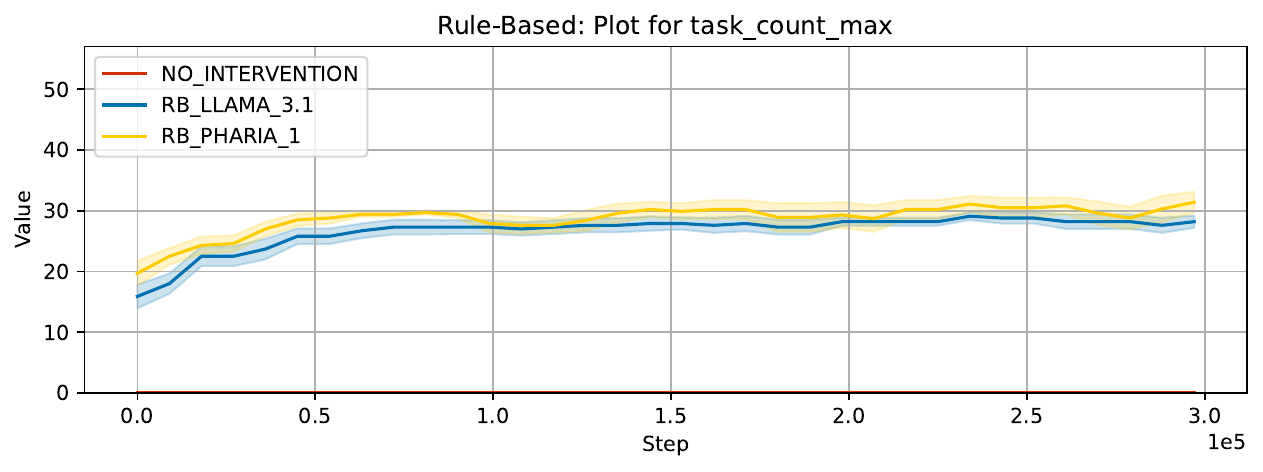}
    \end{minipage}
    \caption{Task Count \textbf{(Rule-Based)} - No controller baseline VS Rule-Based Controller with Llama-3.1-8B Instruct: min, mean and max.}
\end{figure}

\begin{figure}[h!]
    \centering
    \label{results:task_count_NL}
    \begin{minipage}{0.33\textwidth}
        \centering
        \includegraphics[width=\linewidth]{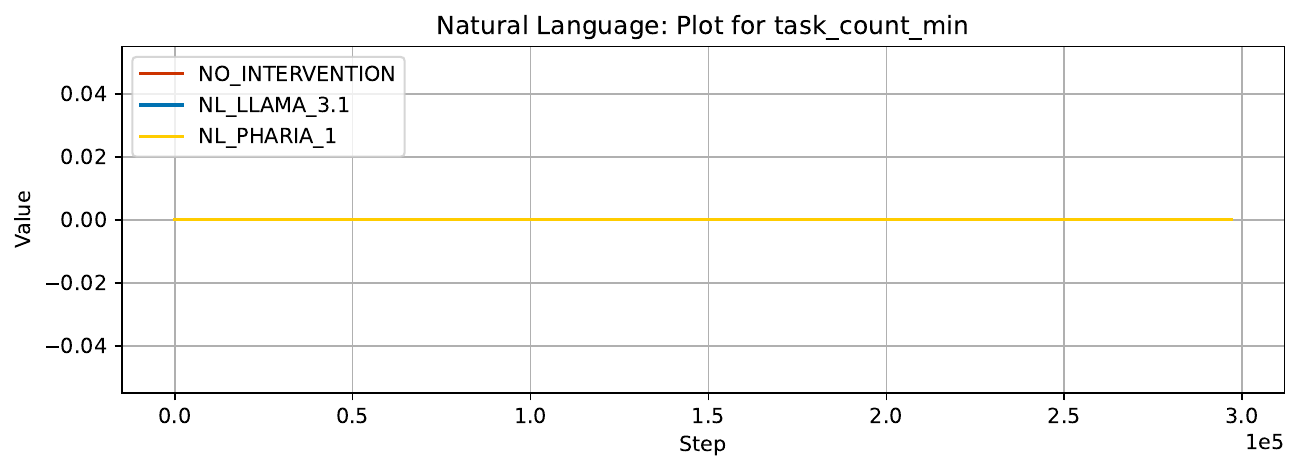}
    \end{minipage}\hfill
    \begin{minipage}{0.33\textwidth}
        \centering
        \includegraphics[width=\linewidth]{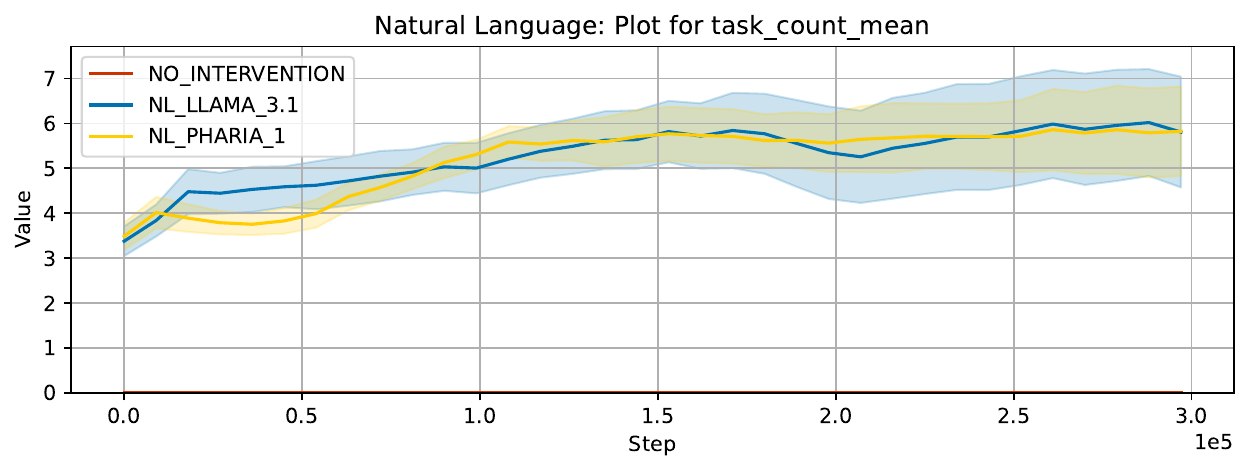}
    \end{minipage}
    \begin{minipage}{0.33\textwidth}
        \centering
        \includegraphics[width=\linewidth]{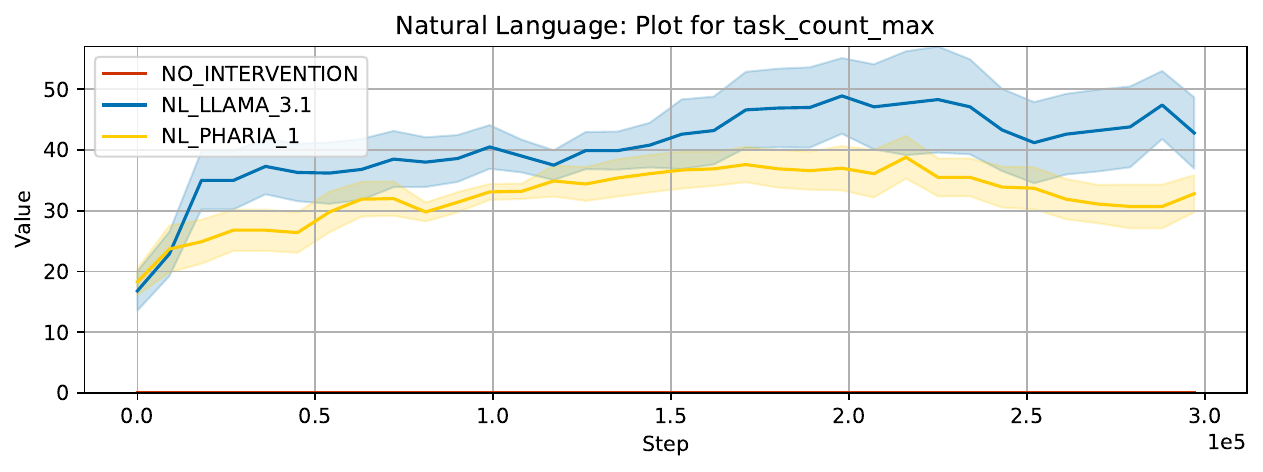}
    \end{minipage}
    \caption{Task Count \textbf{(Natural Language)} - No controller baseline VS Natural Language Controller with Llama-3.1-8B Instruct: min, mean and max.}
\end{figure}

\begin{figure}[h!]
    \centering
    \label{results:total_task_count_RB}
    \begin{minipage}{0.33\textwidth}
        \centering
        \includegraphics[width=\linewidth]{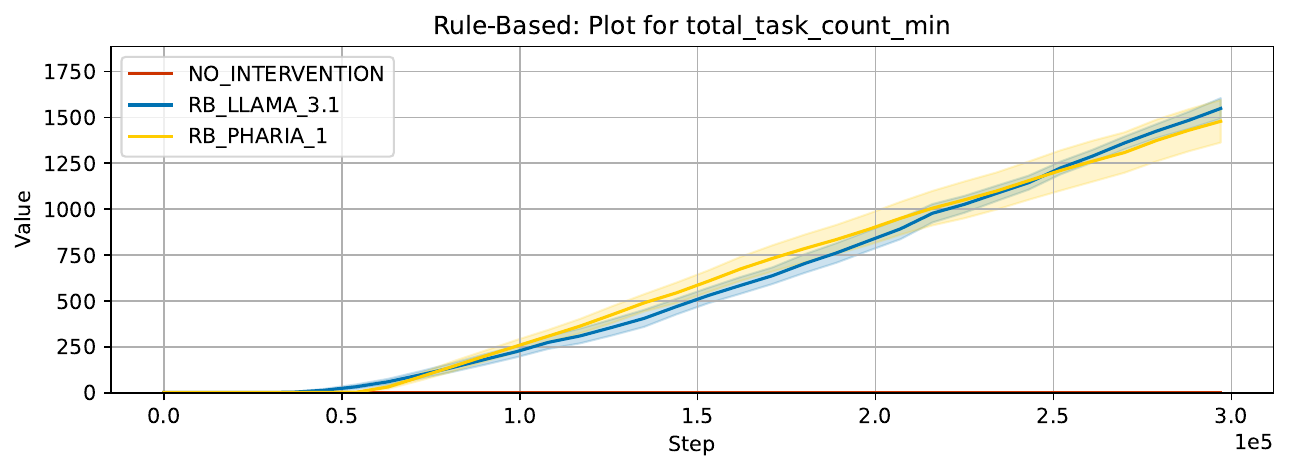}
    \end{minipage}\hfill
    \begin{minipage}{0.33\textwidth}
        \centering
        \includegraphics[width=\linewidth]{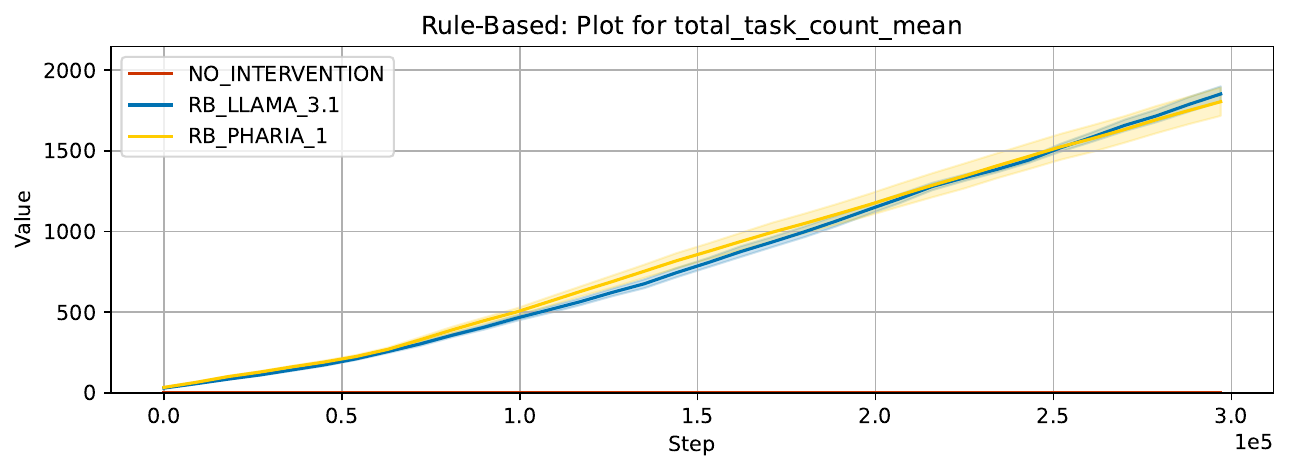}
    \end{minipage}
    \begin{minipage}{0.33\textwidth}
        \centering
        \includegraphics[width=\linewidth]{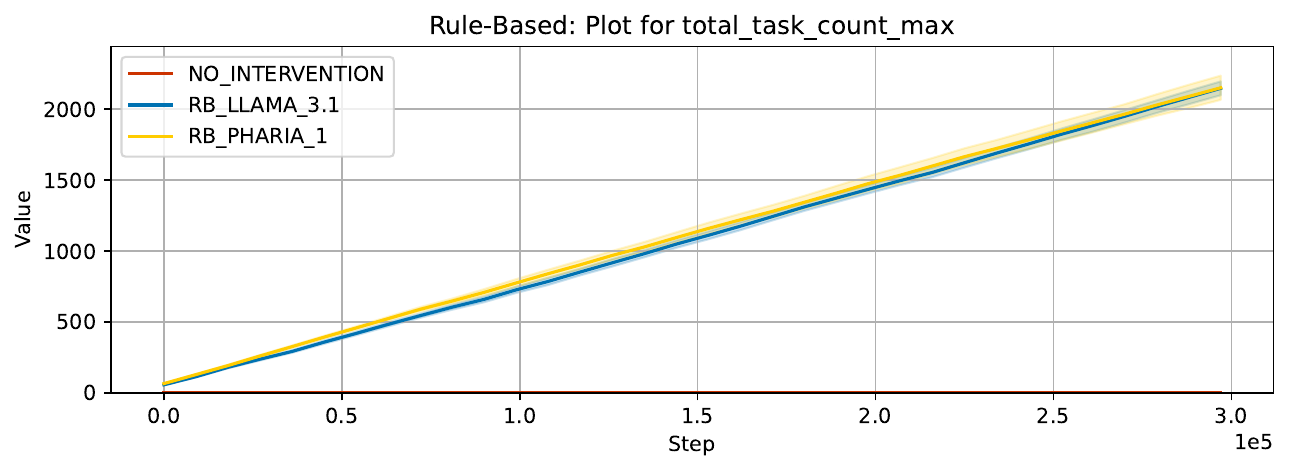}
    \end{minipage}
    \caption{Total Task Count \textbf{(Rule-Based)} - No controller baseline VS Rule-Based Controller with Llama-3.1-8B Instruct: min, mean and max.}
\end{figure}

\begin{figure}[h!]
    \centering
    \label{results:total_task_count_NL}
    \begin{minipage}{0.33\textwidth}
        \centering
        \includegraphics[width=\linewidth]{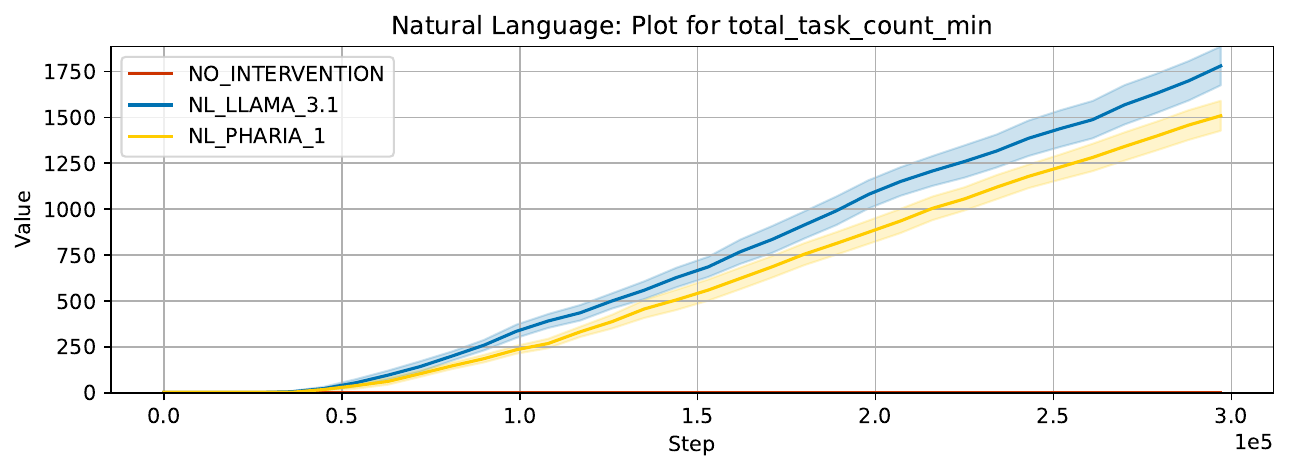}
    \end{minipage}\hfill
    \begin{minipage}{0.33\textwidth}
        \centering
        \includegraphics[width=\linewidth]{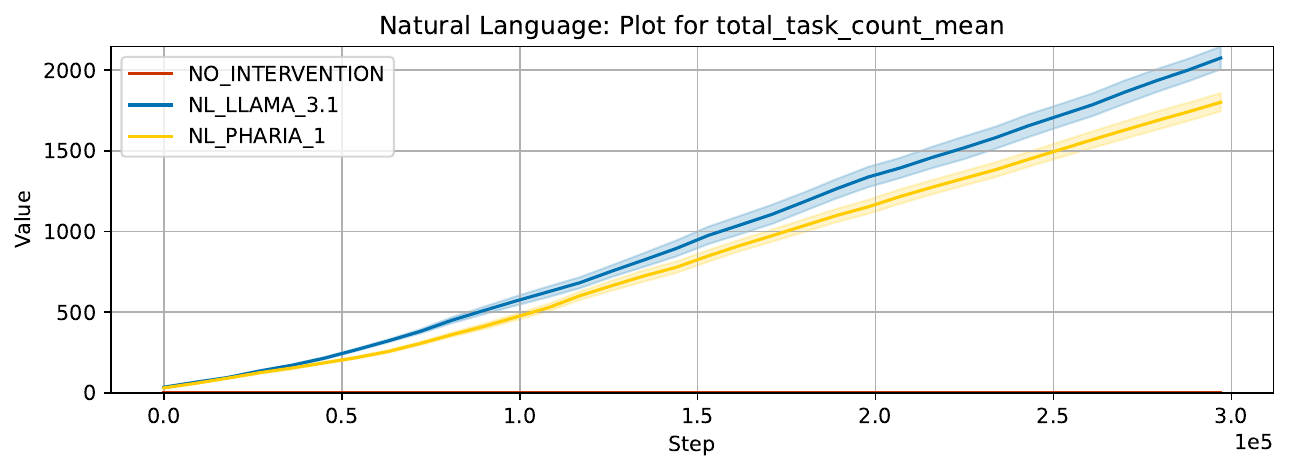}
    \end{minipage}
    \begin{minipage}{0.33\textwidth}
        \centering
        \includegraphics[width=\linewidth]{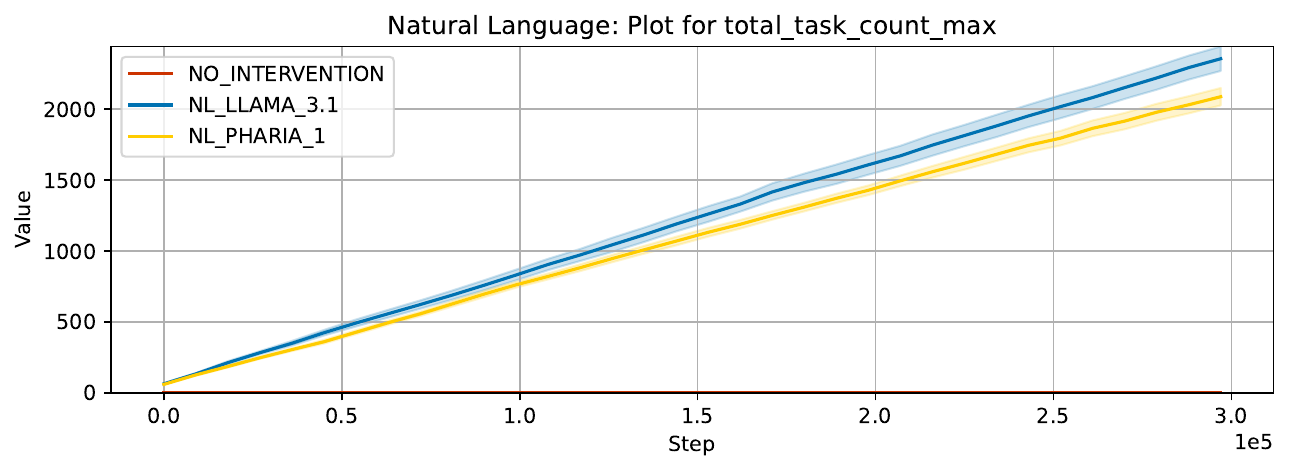}
    \end{minipage}
    \caption{Total Task Count \textbf{(Natural Language)} - No controller baseline VS Natural Language Controller with Llama-3.1-8B Instruct: min, mean and max.}
\end{figure}

\begin{figure}[h!]
    \centering
    \label{results:episode_len_RB_NL}
    \begin{minipage}{0.5\textwidth}
        \centering
        \includegraphics[width=\linewidth]{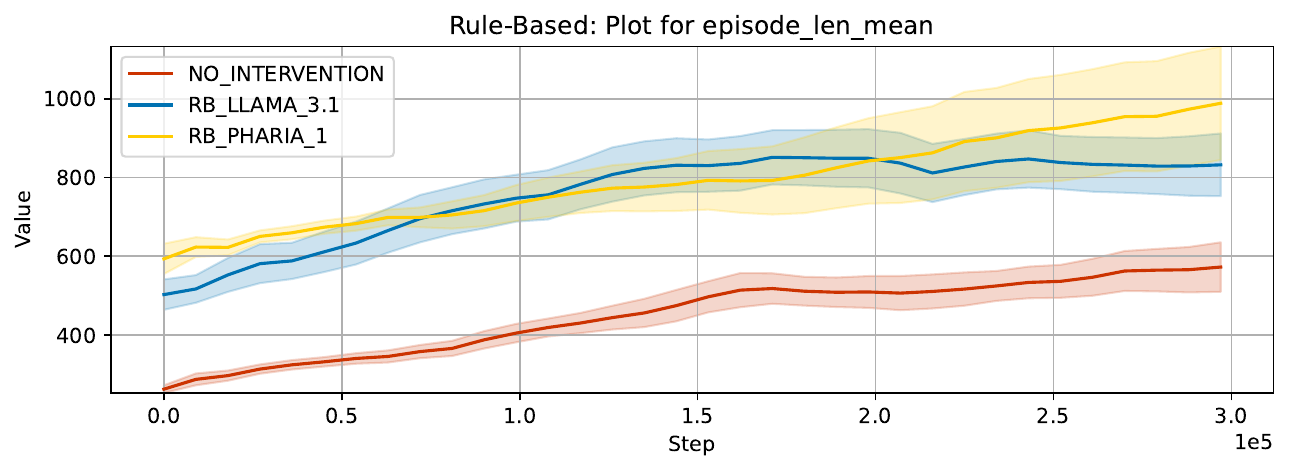}
    \end{minipage}\hfill
    \begin{minipage}{0.5\textwidth}
        \centering
        \includegraphics[width=\linewidth]{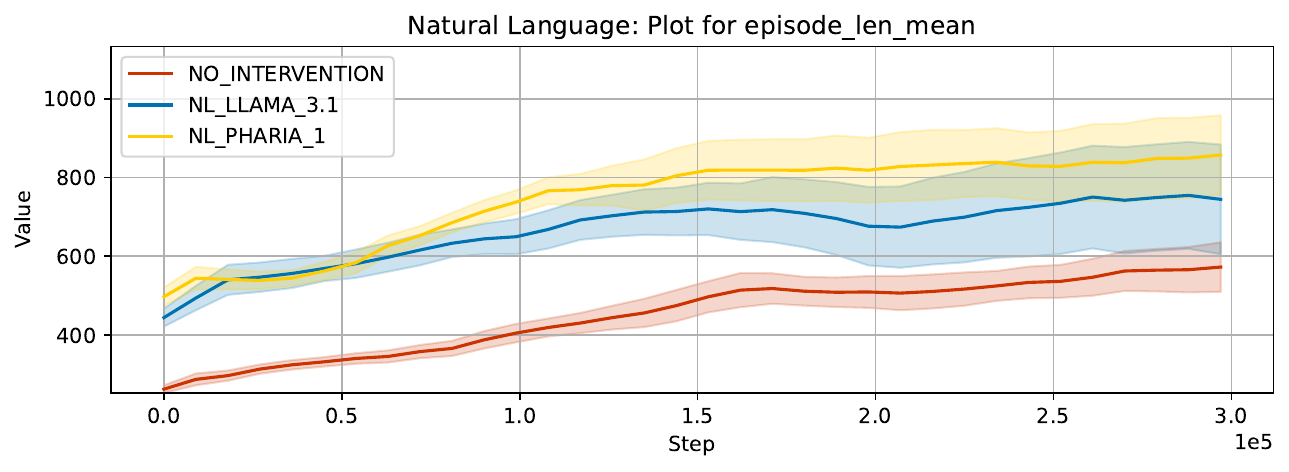}
    \end{minipage}
    \caption{Episode Length - No controller baseline VS Rule-Based (left) and Natural Language (right) Controller with Llama-3.1-8B Instruct: min, mean and max.}
\end{figure}

\begin{figure}[h!]
    \centering
    \label{results:episodes_this_iter_RB_NL}
    \begin{minipage}{0.5\textwidth}
        \centering
        \includegraphics[width=\linewidth]{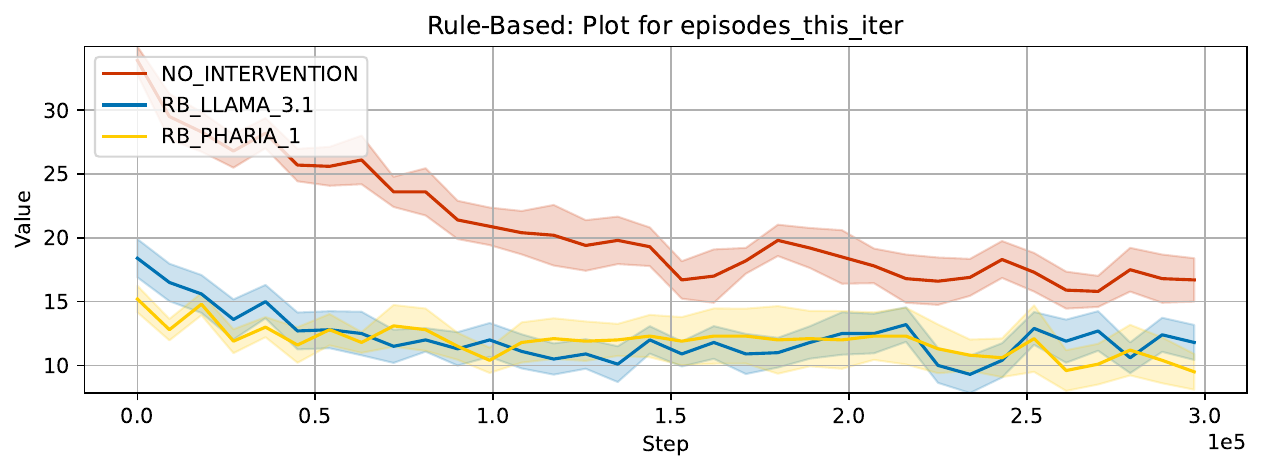}
    \end{minipage}\hfill
    \begin{minipage}{0.5\textwidth}
        \centering
        \includegraphics[width=\linewidth]{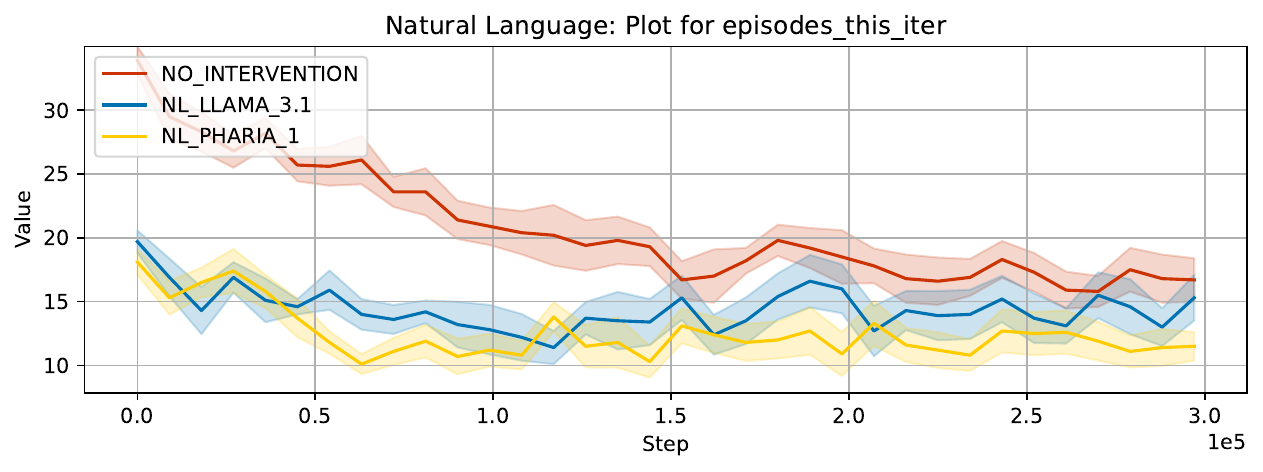}
    \end{minipage}
    \caption{Episodes This Iteration - No controller baseline VS Rule-Based (left) and Natural Language (right) Controller with Llama-3.1-8B Instruct: min, mean and max.}
\end{figure}

\begin{figure}[h!]
    \centering
    \label{results:episodes_timesteps_total_RB_NL}
    \begin{minipage}{0.5\textwidth}
        \centering
        \includegraphics[width=\linewidth]{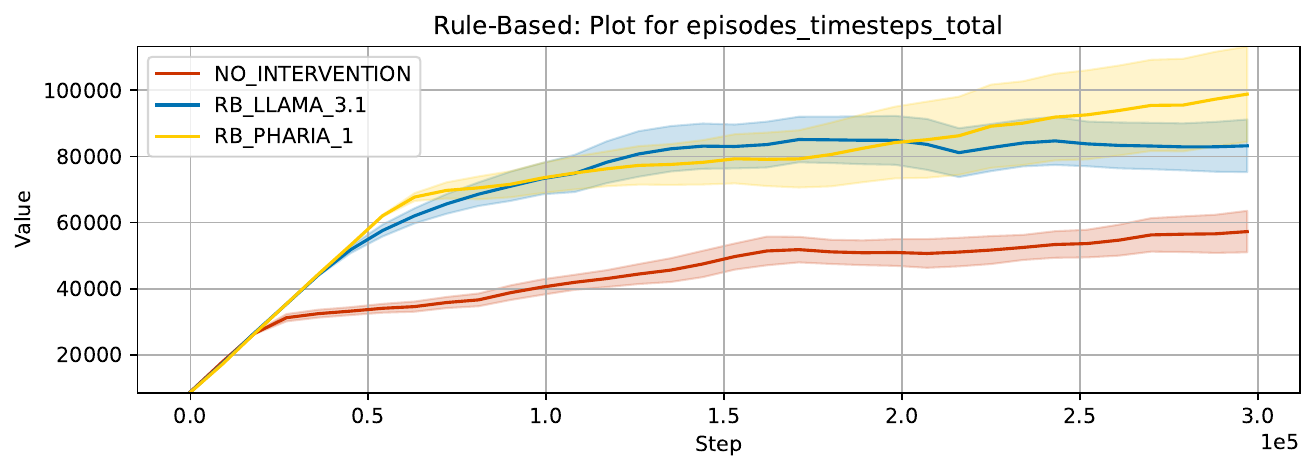}
    \end{minipage}\hfill
    \begin{minipage}{0.5\textwidth}
        \centering
        \includegraphics[width=\linewidth]{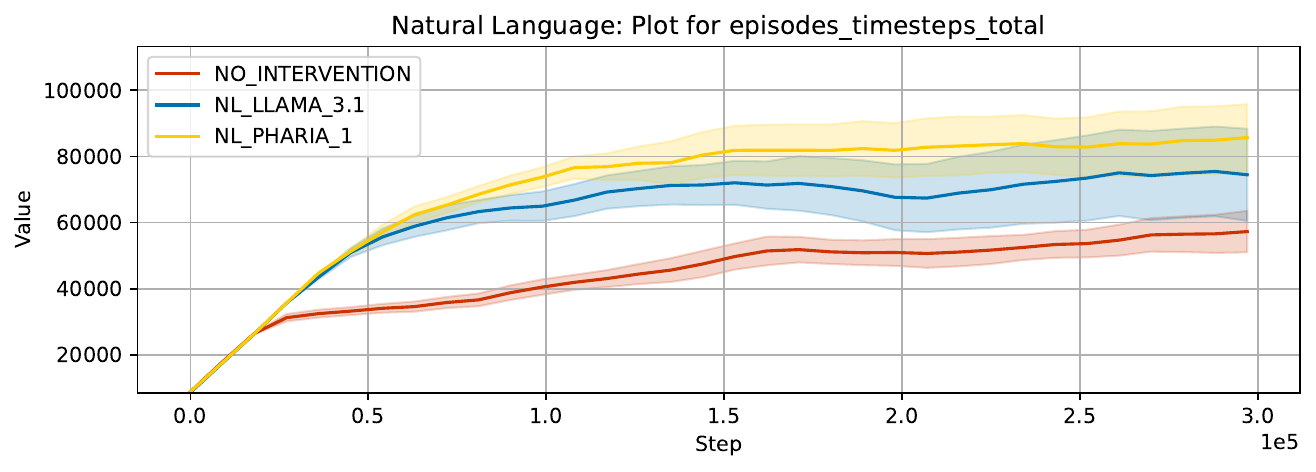}
    \end{minipage}
    \caption{Episodes Timesteps Total - No controller baseline VS Rule-Based (left) and Natural Language (right) Controller with Llama-3.1-8B Instruct: min, mean and max.}
\end{figure}

\begin{figure}[h!]
    \centering
    \label{results:num_episodes_RB_NL}
    \begin{minipage}{0.5\textwidth}
        \centering
        \includegraphics[width=\linewidth]{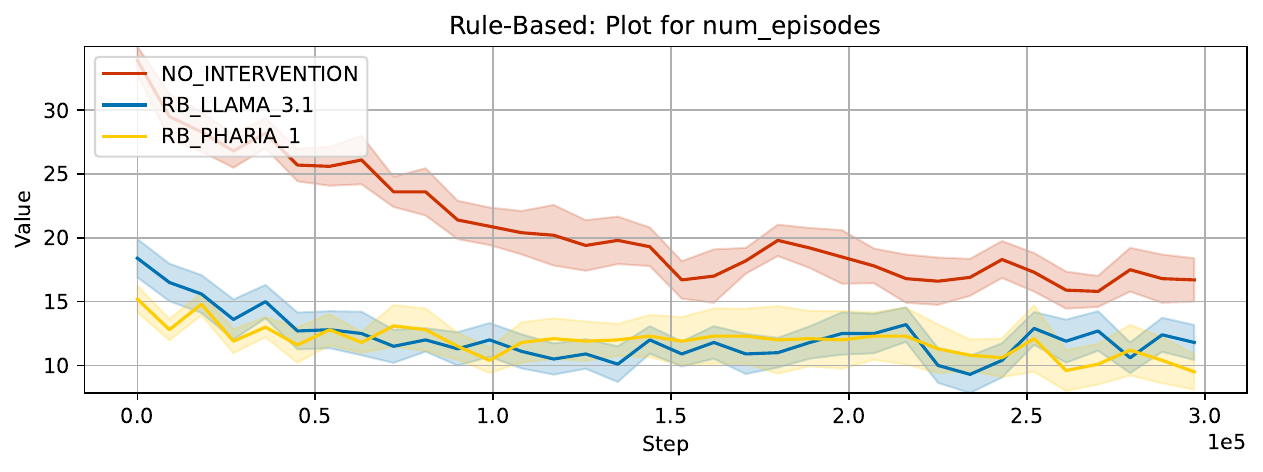}
    \end{minipage}\hfill
    \begin{minipage}{0.5\textwidth}
        \centering
        \includegraphics[width=\linewidth]{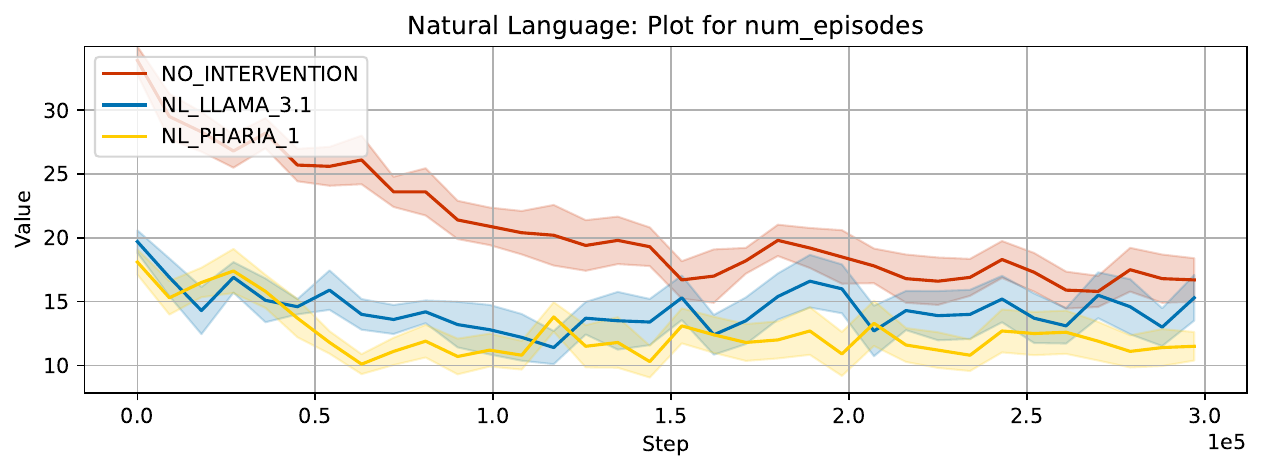}
    \end{minipage}
    \caption{Number Episodes - No controller baseline VS Rule-Based (left) and Natural Language (right) Controller with Llama-3.1-8B Instruct: min, mean and max.}
\end{figure}

\end{document}